\documentclass[journal=jacsat,manuscript=article]{achemso}

\usepackage[version=3]{mhchem} % Formula subscripts using \ce{}
\usepackage{paralist}
\usepackage{booktabs,siunitx}
\usepackage{array}
\usepackage{dcolumn}
\newcommand{\mc}{\multicolumn}
\newcolumntype{H}{>{\setbox0=\hbox\bgroup}c<{\egroup}@{}}
\usepackage{amssymb}

\usepackage[flushleft]{threeparttable}
\usepackage[T1]{fontenc}
\usepackage{booktabs,braket,cleveref,mathptmx,xcolor,siunitx,dcolumn,graphicx,bm,graphics,caption,subcaption,multirow,array,paralist,rotating}
\usepackage[normalem]{ulem}%

\newcommand{\abinitio}{\emph{ab initio}}
\newcommand{\cm}{cm$^{-1}$}

\newcolumntype{d}{D{.}{.}{-1}}

\graphicspath{{Figs/}}

\SectionNumbersOn

\author{Juan C. Zapata Trujillo}
\author{Laura K. McKemmish}
\email{l.mckemmish@unsw.edu.au}
\affiliation[University of New South Wales]
{School of Chemistry, University of New South Wales, 2052, Sydney}
\title[An \textsf{achemso} demo]
  {Model Chemistry Recommendations for Scaled Harmonic Frequency Calculations: A Benchmark Study}

\usepackage[utf8]{inputenc}
\DeclareUnicodeCharacter{2212}{-}
\begin{document}

% \begin{tocentry}

% Some journals require a graphical entry for the Table of Contents.
% This should be laid out ``print ready'' so that the sizing of the

% Inside the \texttt{tocentry} environment, the font used is Helvetica
% 8\,pt, as required by \emph{Journal of the American Chemical

% The surrounding frame is 9\,cm by 3.5\,cm, which is the maximum
% permitted for  \emph{Journal of the American Chemical Society}
% graphical table of content entries. The box will not resize if the
% content is too big: instead it will overflow the edge of the box.

% This box and the associated title will always be printed on a
% separate page at the end of the document.

% \end{tocentry}

\begin{abstract}
    Despite the widespread popularity of scaled harmonic frequency calculations to predict experimental fundamental frequencies in chemistry, sparse benchmarking is available to guide users on appropriate level of theory and basis set choices (model chemistry) or deep understanding of expected errors. An updated assessment of the best approach for scaling to minimise errors is also overdue. 

    Here, we assess the performance of over 600 popular, contemporary, and robust model chemistries in the calculation of scaled harmonic frequencies, evaluating different scaling factors types and their implications in the scaled harmonic frequencies and model chemistry performance.

    We can summarise our results into three main findings: (1) using model chemistry-specific scaling factors optimised for three different frequency regions (low ($<$\,1,000\,\cm{}), mid- (1,000--2,000\,\cm{}), and high-frequency ($>$\,2,000\,\cm{})) results in substantial improvements in the agreement between the scaled harmonic and experimental frequencies compared to other choices; (2) larger basis sets and more robust levels of theory generally lead to superior performance; however, the particular model chemistry choice matters and poor choices lead to significantly reduced accuracies; and (3) outliers are expected in routine calculations regardless of the model chemistry choice. 

    Our benchmarking results here do not consider the intensity of vibrational transitions; however, we draw upon previous benchmarking results for dipole moments which highlight the importance of diffuse functions (i.e. augmented basis sets) in high-quality intensity predictions.

    In terms of specific recommendations, overall, the highest accuracy model chemistries are double-hybrid density functional approximations with a non-Pople augmented triple zeta basis set, which can produce median frequency errors of down to 7.6\,\cm{} (DSD-PBEP86/def2-TZVPD) which is very close to the error in the harmonic approximation, i.e., the anharmonicity error. Double-zeta basis sets should not be used with double-hybrid functionals as there is no improvement compared to hybrid functionals (unlike for double-hybrid triple-zeta model chemistries). Note that 6-311G* and 6-311+G* basis sets perform like a double-zeta basis set for vibrational frequencies.

    After scaling, all studied hybrid functionals with non-Pople triple-zeta basis sets will produce median errors of less than 15\,\cm{}, with the best result of 9.9\,\cm{} with B97-1/def2-TZVPD. Appropriate matching of double-zeta basis sets with hybrid functionals can produce high quality results, but the precise choice of functional and basis set is more important. The B97-1, TPSS0-D3(BJ) or $\omega$B97X-D hybrid density functionals with 6-31G*, pc-1 or pcseg-1 are recommended for fast routine calculations, all delivering median errors of 11-12 \cm{}. Note that dispersion corrections are not easily available for B97-1; given its strong performance here, we recommend these be added to major programs in coming updates. 
\end{abstract}


\section{Introduction}

The evaluation of different level of theory and basis set pairs (a.k.a. model chemistries) for the calculation of chemical and physical properties is essential to guarantee the appropriate use of the powerful tools of computational quantum chemistry. Extensive and critically evaluated benchmarking studies have been performed to provide reliable model chemistry recommendations for the calculation of many important general-chemistry properties such as dipole moments \cite{18HaHe1,20ZaMc} and polarisabilities \cite{18HaHe2}, isomerisation energies and, thermochemical and kinetic properties \cite{17MaHe,17GoHaBa}, amongst others. These benchmark studies allow users --particularly non-experts-- to easily incorporate more accurate quantum-chemistry calculations in their research, highlighting estimates for the uncertainties in the computed values, as well as key limitations in their procedures.

However, despite the widespread popularity of using scaled harmonic frequency calculations in chemistry \cite{13GrMaBl,20ZhYuZh,21LoWaXi,21NiTr,22LoMaLa} (as evidenced by, for example, the thousands of citations for \citet{96ScRa}), sparse benchmarking is available to guide users on appropriate model chemistry recommendations. 

Instead of comparing the quality of theoretical predictions against experimental values, the assessment of model chemistries for harmonic frequency calculations has largely focused on the optimisation of scaling factors for individual model chemistries \cite{21ZaMc}. The often-produced root-mean-square-error (RMSE) between the scaled harmonic and experimental fundamental frequencies is therefore widely used as a metric to judge and compare model chemistry performance, allowing potential recommendations as is done in \citet{21ZaMc}. Indeed, this procedure has allowed the comparison of nearly 1,500 model chemistries combining hundreds of levels of theory and basis sets. However, recommendations from this analysis were only preliminary as (1) the model chemistry space covered mostly corresponds to traditional levels of theory and basis sets, disregarding contemporary options and, most importantly, (2) the differences in the benchmark databases used when optimising and assessing the performance of the scaling factors means that the RMSEs cannot be fairly compared across different publications.

Benchmark databases play a crucial role when optimising scaling factors and, consequently, assessing model chemistry performance. With the most extensive and frequently used database from \citet{93PoScWo} (also known as the F1 set as used in \citet{96ScRa}) collated almost thirty years ago, we recently compiled a new and updated benchmark database for vibrational frequency calculations (VIBFREQ1295) storing 1,295 experimental fundamental frequencies and CCSD(T)(F12*)\cite{10HaTeKo}/cc-pVDZ-F12\cite{08PeAdWe,10HiPe} \abinitio{} harmonic frequencies for 141 organic-like molecules \cite{22ZaMc_VIBFREQ}. The development of VIBFREQ1295 was based on an extensive compilation of contemporary experimental data (usually at high resolution), providing general updates to previous molecular frequency assignments. 

Comparisons between the experimental and high-level \abinitio{} data in VIBFREQ1295 allowed us to define and further understand the best approach to assess model chemistry performance through the use of harmonic frequency scaling factors. We found \cite{22ZaMc_VIBFREQ} that a frequency-range-specific scaling approach, specifically dividing the frequency range into three different frequency regions (low ($<$\,1,000\,\cm{}), mid (1,000--2,000\,\cm{}), and high ($\geq$\,2,000\,\cm{})), each one with a corresponding scaling factor, leads to significantly improved performance compared to the traditional global scaling approach (corresponding to a single value to scale all frequencies in the spectrum) or a high- and low-frequency division. We also found no benefit in optimising scaling factors for different molecular groups (e.g., halogen- and non-halogen-containing molecules), or vibrational mode types (e.g., stretches and non-stretches) \cite{22ZaMc_VIBFREQ}. Moreover, our results showed that median errors are more appropriate than RMSEs in describing the distribution of errors and predicting performance of model chemistries in new calculations given that RMSEs are dominated by outliers rather than typical expected results.

Using VIBFREQ1295 as our reference set of experimental fundamental frequencies and noting the aspects mentioned above to ensure the appropriate comparison of different model chemistries performance, here we perform an extensive benchmarking of a wide cross-section of model chemistry choices in harmonic frequency calculations. Not only we are interested in providing solid level of theory and basis sets recommendations, but we aim to provide users with a deep understanding of the expected errors and dependencies in the calculations.

\begin{figure*}
    \centering
    \includegraphics[width=1\textwidth]{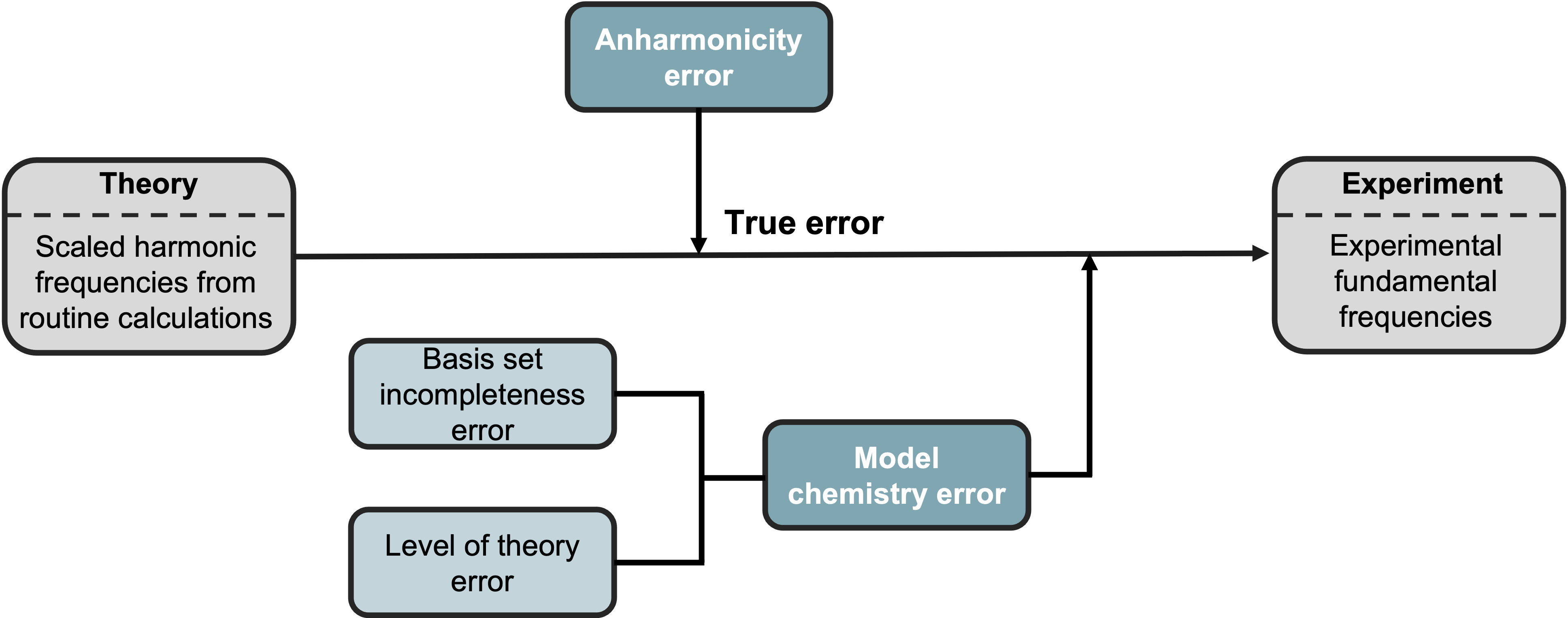}
    \caption{Schematic representation of the two different errors (anharmonicity and model chemistry error) contributing towards the observed deviations (true error) between the scaled harmonic frequencies from routine quantum-chemistry calculations, and the experimental fundamental frequencies for a given molecule.}
    \label{fig:summary}
\end{figure*}

\Cref{fig:summary} schematically presents the two main sources of error that contribute towards the observed deviation (henceforth True error as displayed in the figure) between the scaled harmonic frequencies from routine calculations (i.e., calculations involving moderate-quality model chemistries, such as those assessed in this study) and the experimental fundamental frequencies for a given molecule:

\begin{enumerate}
    \item \textbf{The Model Chemistry Error:} Error introduced in the calculations when using a finite basis set and approximate treatment of Schrodinger equation (i.e. level of theory).  We can divide the model chemistry error into the \textbf{Level of Theory Error (LoTE)} (estimated by comparing large basis sets results for the approximate level of theory --usually DFT-- against a benchmark-quality high-level theory calculations such as the CCSD(T)(F12*)/cc-pVDZ-F12 harmonic frequencies in VIBFREQ1295) and the \textbf{Basis Set Incompleteness Error (BSIE)} (estimated by comparing results for a given density functional theory with the moderate size basis set against near complete basis set limit results, e.g. def2-QZVP or cc-pV5Z). The relative magnitude of the LoTE and BSIE can be used to determine the sensitivity of the results to level of theory and basis set choice; and
    \item \textbf{The Anharmonicity Error:} Error introduced by using a scaled harmonic frequency rather than a true anharmonic result. This error is irreducible within the double harmonic approximation. We recently estimated the median anharmonicity error as 7.5\,\cm{} based on comparisons between the CCSD(T)(F12*)/cc-pVDZ-F12 scaled harmonic and experimental fundamental frequencies in VIBFREQ1295, thus setting a lower bound for reliable model chemistry performance \cite{22ZaMc_VIBFREQ} (ignoring potential cancellation of errors as unreliable).
\end{enumerate}

Given that the anharmonicity error (at 7.5\,\cm{}) represents a lower bound for reliable model chemistry performance for fundamental frequency predictions using scaled harmonic frequency calculations, here our goal is to provide model chemistry recommendations that can produce results as close as possible to this threshold, i.e. having negligible model chemistry error. Results from our scaling factors review paper demonstrate that model chemistry error becomes very small relative to the anharmonicity error at quite modest model chemistry levels, probably hybrid or double-hybrid functionals and double- or triple-zeta basis sets \cite{21ZaMc}; this paper's results allow rigorous quantification of these errors (i.e., calculation of uncertainties). 

It is important to note that our assessment is solely based on scaled harmonic frequencies, given their widespread use in the literature and affordable computational timings for larger molecular systems and high-throughput methodologies. No evaluation of non-scaled raw harmonic frequencies is considered here, though we expect our recommendations may follow similar performance in the non-scaled raw harmonic frequency case. Recommendations from this analysis will set the base level performance that higher anharmonic vibrational treatments must exceed to be useful \cite{20BoCeTa,20BaCeFu}. 

This publication is organised as follows. Section \ref{sec:comp_details} describes the methodology of this benchmark study,  outlining the VIBFREQ1295 benchmark database, explaining our selection of levels of theory and basis sets, presents the computational details for our calculations, and discusses the procedure we use to optimise and evaluate the scaling factors. This section concludes with an analysis of global vs frequency-range-specific scaling factors as well as universal vs model-chemistry-specific scaling factors, justifying the consideration of only frequency-range-specific and model-chemistry-specific scaling factors for the rest of the manuscript. In Section \ref{sec:uncert_harmcalcs} we delve into understanding the different sources of error and uncertainties in harmonic frequency calculations, demonstrating that anharmonicity error dominates over model chemistry error at modest computational levels. The section considers the source of model chemistry error by evaluating level of theory (LoTE) and basis set incompleteness errors (BSIE) to further understand the dependency of the calculations to the level of theory and basis set choice. Section \ref{sec:modechem_perfm} is dedicated to the complete evaluation of all model chemistries considered, highlighting median error performance, the prevalence of outliers, and the model chemistry reliability. We also discuss the effect of dispersion corrections in the density functional and diffuse functions in the basis set has in the overall model chemistry performance. For ease of reference, our model chemistry recommendations outlining expected errors and performance are presented in Section \ref{sec:recommendations}. Finally, we summarise our findings in Section \ref{sec:conclusions}, discussing limitations in our procedure and future directions.

We understand that the terminology and assessment of the different calculations and errors in this publication can be overwhelming and challenging for the inexperienced reader in computational quantum chemistry. We outline crucial concepts relevant to our discussion here, but refer the reader to \citet{19GoMe}  and \citet{17NaJe}  for accessible introductions to density functional approximations and basis sets respectfully. In-depth knowledge of the differences between computational quantum chemistry methodologies is not essential for understanding the recommendations in this paper.

\section{Computational Details and Methodology}
\label{sec:comp_details}

\subsection{Benchmark Database}

To assess the performance of all model chemistries featured in this study, we used the recently developed VIBFREQ1295 benchmark database for vibrational frequency calculations collating 1,295 gas-phase experimental fundamental frequencies for 141 common organic-like molecules \cite{22ZaMc_VIBFREQ}. The data collated in VIBFREQ1295 represent a robust and extensive compilation of contemporary experimental frequencies, providing general updates in previous molecular frequency assignments. \Cref{tab:all_mol} presents an overview of all molecules considered in the VIBFREQ1295 database. A more self-contained version of \Cref{tab:all_mol} can be found in \citet{22ZaMc_VIBFREQ} with references to the original publications outlining the collected experimental fundamental frequencies for each molecule.

\begin{table*}
    \centering
    \caption{Summary of molecular species in VIBFREQ1295\cite{22ZaMc_VIBFREQ}. Cyclic molecules are indicated with $c-$ before the molecular formula. Isomeric molecules (except for $cis$ and $trans$) are distinguished by an underscore and number after the molecular formula, e.g., \ce{C3H4} and \ce{C3H4}$\_$1. The electronic state for each molecular species is indicated as a superscript in the left-hand side of the molecular formula whenever the electronic state is not a singlet. Only $^{\textit{1}}$\ce{CH2} (singlet state) is explicitly presented in the table to distinguish it from $^{\textit{3}}$\ce{CH2} (triplet state).}
    \label{tab:all_mol}
    \scalebox{0.7}{
    \begin{tabular}{llllll}
    \toprule
    \mc{1}{l}{Formula}         & \mc{1}{l}{Name}           & \mc{1}{l}{Formula}      & \mc{1}{l}{Name}                  & \mc{1}{l}{Formula}      & \mc{1}{l}{Name}  \\
    \midrule
        \ce{AlCl3}             & Aluminium chloride        & \ce{CCl4}               & Tetrachloromethane               & \ce{HNCO}               & Isocyanic acid                            \\
        \ce{BH}                & $\lambda^{1}$-borane      & \ce{CF4}                & Tetrafluoromethane               & \ce{HNO}                & Nitroxyl                                  \\
        \ce{BH3CO}             & Borane carbonyl           & \ce{$^{\textit{2}}$CH}  & Methylidyne radical              & \ce{HNO3}               & Nitric acid                               \\
        \ce{$^{\textit{2}}$BO} & Oxoboron                  & \ce{CH2Cl2}             & Dichloromethane                  & \ce{$^{\textit{2}}$HO2} & Hydroperoxy radical                       \\
        \ce{c-C2H4S3}          & 1,2,4-trithiolane         & \ce{CH2N2}              & Diazomethane                     & \ce{HOCl}               & Hypochlorous acid                         \\
        \ce{c-C3H3NO}          & 1,3-oxazole               & \ce{CH2O2}              & Formic acid                      & \ce{HOF}                & Hypofluorous acid                         \\
        \ce{c-C3H3NO}\_1       & 1,2-oxazole               & \ce{CH2S}               & Thioformaldehyde                 & \ce{N2F2}               & (E)-difluorodiazene                       \\
        \ce{c-C3H6}            & Cyclopropane              & \ce{CH3Cl}              & Chloromethane                    & \ce{N2H4}               & Hydrazine                                 \\
        \ce{c-C3H6S}           & Thietane                  & \ce{CH3F}               & Fluoromethane                    & \ce{N2O}                & Nitrous oxide                             \\
        \ce{c-C4H4N2}          & Pyrazine                  & \ce{CH3N}               & Methanimine                      & \ce{NCl2F}              & Diclhorofluoroamine                       \\
        \ce{c-C4H4O}           & Furan                     & \ce{CH3NO}              & Formamide                        & \ce{NClF2}              & Chlorodifluoroamine                       \\
        \ce{c-C4H5N}           & 1H-pyrrole                & \ce{CH4}                & Methane                          & \ce{NF3}                & Nitrogen trifluoride                      \\
        \ce{c-C4H8O2}          & 1,4-dioxane               & \ce{CH4O}               & Methanol                         & \ce{$^{\textit{3}}$NH}  & $\lambda^{1}$-azane                       \\
        \ce{c-C5H5N}           & Pyridine                  & \ce{CH4S}               & Methanethiol                     & \ce{NH3}                & Amonia                                    \\
        \ce{c-CH2N4}           & 1H-tetrazole              & \ce{CH5N}               & Methanamine                      & \ce{NHF2}               & Difluoroamine                             \\
        \ce{C2Cl2}             & 1,2-dichloroethyne        & \ce{CH5P}               & Methylphosphine                  & \ce{$^{\textit{2}}$NO2} & Nitrogen dioxide                          \\
        \ce{C2H2}              & Acetylene                 & \ce{CH6Si}              & Methylsilane                     & \ce{NSCl}               & Azanylidyne(chloro)-$\lambda^{4}$-sulfane \\
        \ce{C2H2O}             & Ketene                    & \ce{CHCl3}              & Chloroform                       & \ce{NSF}                & Azanylidyne(fluoro)-$\lambda^{4}$-sulfane \\
        \ce{C2H2O2}            & Oxaldehyde                & \ce{CHF3}               & Fluoroform                       & \ce{O3}                 & Ozone                                     \\
        \ce{C2H3Cl}            & Chloroethene              & \ce{Cl2}                & Molecular chlorine               & \ce{OCS}                & Carbonyl sulfide                          \\
        \ce{C2H3N}             & Acetonitrile              & \ce{Cl2O}               & Chloro hypochlorite              & \ce{ONF}                & Nitrosyl fluoride                         \\
        \ce{C2H3N}\_1          & Isocyanomethane           & \ce{ClCN}               & Carbononitridic chloride         & \ce{P4}                 & Tetraphosphorus                           \\
        \ce{C2H3OF}            & Acetyl fluoride           & \ce{ClF}                & Chlorine fluoride                & \ce{PCl3}               & Trichlorophosphane                        \\
        \ce{C2H4O}             & Acetaldehyde              & \ce{ClF3}               & Trifluoro-$\lambda^{3}$-chlorane & \ce{PF3}                & Trifluorophosphane                        \\
        \ce{C2H4O2}            & Methyl formate            & \ce{ClNO}               & Nitrosyl chloride                & \ce{$^{\textit{3}}$PH}  & $\lambda^{1}$-phosphane                   \\
        \ce{C2H4O2}\_1         & Acetic acid               & \ce{ClNO2}              & Chloro nitrite                   & \ce{PH3}                & Phosphine                                 \\
        \ce{C2H5F}             & Fluoroethane              & \ce{$^{\textit{2}}$ClO} & Chlorosyl                        & \ce{PN}                 & Azanylidynephosphane                      \\
        \ce{C2H6}              & Ethane                    & \ce{$^{\textit{2}}$CN}  & Cyano radical                    & \ce{S2}                 & Disulfur                                  \\
        \ce{C2H6O}             & Methoxymethane            & \ce{CO}                 & Carbon monoxide                  & \ce{S2F2}               & Fluorosulfanyl thiohypofluorite           \\
        \ce{C2H6S}             & Methylsulfanylmethane     & \ce{CO2}                & Carbon dioxide                   & \ce{S2O}                & Disulfur monoxide                         \\
        \ce{C2HCl}             & Chloroethyne              & \ce{COCl2}              & Carbonyl dichloride              & \ce{SCl2}               & Chloro thiohypochlorite                   \\
        \ce{C2HF}              & Fluoroethyne              & \ce{COClF}              & Carbonyl chloride fluoride       & \ce{$^{\textit{2}}$SH}  & $\lambda^{1}$-sulfane                     \\
        \ce{C2N2}              & Oxalonitrile              & \ce{COF2}               & Carbonyl difluoride              & \ce{Si2H6}              & Disilane                                  \\
        \ce{C3H3Cl}            & 3-chloroprop-1-yne        & \ce{CS}                 & Methanidylidynesulfanium         & \ce{SiH2}               & Silylene                                  \\
        \ce{C3H3F}             & 3-fluoroprop-1-yne        & \ce{CS2}                & Carbon disulfide                 & \ce{SiH2Cl2}            & Dichlorosilane                            \\
        \ce{C3H3N}             & Acrylonitrile             & \ce{CSCl2}              & Thiocarbonyl dichloride          & \ce{SiH3Cl}             & Chlorosilane                              \\
        \ce{C3H4}              & Propa-1,2-diene           & \ce{CSF2}               & Difluoromethanethione            & \ce{SiH3F}              & Fluorosilane                              \\
        \ce{C3H4}\_1           & Prop-1-yne                & \ce{F2}                 & Molecular fluorine               & \ce{SiH4}               & Silane                                    \\
        \ce{C3H4O}             & Prop-2-enal               & \ce{F2O}                & Fluoro hypofluorite              & \ce{SiHCl3}             & Trichlorosilane                           \\
        \ce{C3H6}              & Prop-1-ene                & \ce{F2SO}               & Thionyl fluoride                 & \ce{SiHF3}              & Trifluorosilane                           \\
        \ce{C3H8}              & propane                   & \ce{FCN}                & Carbononitridic fluoride         & \ce{$^{\textit{1}}$CH2} & Methylene                                 \\
        \ce{C3O2}              & Carbon suboxide           & \ce{H2CO}               & Formaldehyde                     & \ce{SiO}                & Oxosilicon                                \\
        \ce{C4H2}              & Buta-1,3-diyne            & \ce{H2O}                & Water                            & \ce{SO2}                & Sulfur dioxide                            \\
        \ce{C4H6}              & Buta-1,3-diene            & \ce{H2S}                & Sulfane                          & \ce{SO3}                & Sulfur trioxide                           \\
        \ce{C4N2}              & Dicyanoacetylene          & \ce{HCl}                & Hydrochloric acid                & \ce{SOCl2}              & Thionyl dichloride                        \\
        \ce{C6H8}              & (3E)-hexa-1,3,5-triene    & \ce{HCN}                & Hydrogen cyanide                 & \ce{trans-C2H4Cl2}      & 1,2-dichloroethane                        \\
        \ce{CCl2F2}            & Dichloro(difluoro)methane & \ce{$^{\textit{2}}$HCO} & Formyl radical                   & \ce{$^{\textit{3}}$CH2} & Methylene                                 \\
    \bottomrule
    \end{tabular}}
\end{table*}

The experimental frequencies in VIBFREQ1295 are complemented with \abinitio{} harmonic frequencies computed at the CCSD(T)(F12*)\cite{10HaTeKo}/cc-pVDZ-F12\cite{08PeAdWe,10HiPe} level for all 141 molecules. Comparison against this high-level \abinitio{} data will be crucial in understanding the computational errors associated with routine and less-demanding harmonic frequency calculations. 

\subsection{Levels of Theory and Basis Sets}

Given the wide diversity of levels of theory \cite{16HaBaMa,19HaBaFr} and basis sets \cite{19PrAlDi} currently available, assessing the performance of a wide cross-section of model chemistries for routine harmonic frequency calculations could represent, in principle, a challenging and altogether unfeasible task. Here, however, we constrained our analysis to primarily consider model chemistries involving hybrid and double-hybrid functionals, together with double- and triple-zeta basis sets, given their expected superior performance yet affordable computational timings in routine harmonic frequency calculations based on the recent \citet{21ZaMc} review study. For comparison and completeness, we also included  strongly performing functionals from the GGA and meta-GGA classes, as recommended from thorough benchmark studies \cite{17GoHaBa,17MaHe}. The HF and MP2 wavefunction methods were also considered for historical reference.

\begin{table*}
    \centering
    \caption{Overview of the levels of theory considered here, along with the level of theory class (in increasing order of complexity and computational time), reference to original publications, and reasoning for inclusion in this study.}
    \label{tab:methods}
    \scalebox{0.8}{
    \begin{tabular}{p{0.2\linewidth}  p{0.05\linewidth}  p{0.1\linewidth}  p{0.8\linewidth}}
    \toprule
        \mc{1}{c}{Level of Theory}  & \mc{1}{c}{Class$^{a}$}  & \mc{1}{c}{Ref.} & \mc{1}{c}{Description} \\
        \midrule
        HF                          & HF                & \cite{51Ro}                                           & Historical reference  \\  [2mm]
        B97-D3(BJ)                  & GGA               & \cite{06Gr,11GrEhGo}                                  & Superior performance in main group thermochemistry, kinetics and noncovalent interaction calculations \cite{17GoHaBa} \\
        BLYP-D3(BJ)                 & GGA               & \cite{88Be,88LeYaPa,89MiSaSt,06Gr}                    & Superior performance for main group thermochemistry, kinetics and noncovalent interaction calculations \cite{17GoHaBa} \\
        PBE                         & GGA               & \cite{96PeBuEr}                                       & Popular choice in computational quantum chemistry \\ [2mm]
        M06L-D3(0)                  & mGGA              & \cite{06ZhTr,11GoGr}                                  & Superior performance for main group thermochemistry, kinetics and noncovalent interaction calculations \cite{17GoHaBa} \\ [2mm]
        B3LYP                       & Hybrid            & \cite{93Be,94StDeCh}                                  & Popular choice in computational quantum chemistry \\
        B3LYP-D3(BJ)                & Hybrid            & \cite{93Be,94StDeCh,06Gr}                             & Popular choice in computational quantum chemistry \\
        B97-1                       & Hybrid            & \cite{98HaCoTo}                                       & Popular choice in computational quantum chemistry \cite{17GoHaBa} \\
        B97-2                       & Hybrid            & \cite{01WiBrTo}                                       & Accurate dipole moment calculations \cite{18HaHe1} \\
        M052X                       & Hybrid            & \cite{06ZhScTr}                                       & Popular choice in computational quantum chemistry \\
        M052X-D3(0)                 & Hybrid            & \cite{06ZhScTr,11GoGr}                                & Superior performance for main group thermochemistry, kinetics and noncovalent interaction calculations \cite{17GoHaBa} \\
        SOGGA11-X                   & Hybrid            & \cite{11PeTr}                                         & Accurate dipole moment \cite{18HaHe1,20ZaMc} and polarisability calculations \cite{18HaHe2} \\
        PBE0                        & Hybrid            & \cite{99AdBa,99ErSc}                                  & Popular choice in computational quantum chemistry \\
        PBE0-D3(BJ)                 & Hybrid            & \cite{99AdBa,99ErSc,06Gr}                             & Accurate dipole moment \cite{18HaHe1} and polarisability calculations \cite{18HaHe2} \\
        TPSS0                       & Hybrid            & \cite{05Gr}                                           & Popular choice in computational quantum chemistry \\
        TPSS0-D3(BJ)                & Hybrid            & \cite{05Gr,06Gr}                                      & Popular choice in computational quantum chemistry \\
        $\omega$B97X-D              & Hybrid            & \cite{08ChHe}                                                     & Superior performance for main group thermochemistry, kinetics and noncovalent interaction calculations  \cite{17GoHaBa,17MaHe} \\
        $\omega$B97X-V              & Hybrid            & \cite{14MaHe}                                         & Accurate dipole moment \cite{18HaHe1,20ZaMc} and polarisability calculations \cite{18HaHe2}, with superior performance in main group thermochemistry, kinetics and noncovalent interactions \cite{17GoHaBa,17MaHe} \\ [2mm]
        MP2                         & MP2               & \cite{88HePoFr,89SaAl,90FrHePo_1,90FrHePo_2,94HeHe}   & Historical reference  \\  [2mm] 
        B2PLYP                      & DH                & \cite{06Gr}                                           & Adequate performance in harmonic frequency calculations \cite{10BiPaSc,20BaCeFu} \\
        B2PLYP-D3(BJ)               & DH                & \cite{06Gr,11GoGr}                                    & Superior performance in main group thermochemistry, kinetics and noncovalent interactions \cite{11GoGr,17GoHaBa}\\
        DSD-PBEP86                  & DH                & \cite{11KoMa}                                         & Superior performance for main  group thermochemistry, kinetics and noncovalent interaction calculations \cite{17GoHaBa} \\
        PBE0DH                      & DH                & \cite{11BrAd}                                         & Popular choice in computational quantum chemistry \\
    \bottomrule
    \end{tabular}}
        \begin{tablenotes}
        \scriptsize{\item[] (a) GGA: generalised gradient approximation, mGGA: meta-GGA, and DH: Double-hybrid.}
        \end{tablenotes}
\end{table*}

Our specific selection of levels of theory is detailed in \Cref{tab:methods} and focused on density functionals with strong performance for general-purpose chemistry \cite{17GoHaBa,17MaHe}, though we also included popular choices, e.g., B3LYP, for comparison.  Levels of theory in \Cref{tab:methods} are organised into different classes according to Jacob's ladder classification as HF $<$ GGA (generalised gradient approximation) $<$ mGGA (meta-GGA) $<$ Hybrid $<$ MP2 $<$ DH (double-hybrid) in terms of time and predicted accuracy (see \citet{19GoMe} for further details).  
We recognise, however, the absence of some common functionals in our analysis. As our goal is to provide suitable and reliable recommendations for harmonic frequency calculations, we decided to exclude selected time-consuming and problematic functionals. For instance, we didn't assess further the performance of double-hybrid functionals like B2GPLYP and DSD-BLYP, as only numerical second-order derivatives were available for these functionals in common quantum-chemistry packages, significantly increasing computational timings. We also excluded hybrid functionals like PW6B95 and PW6B95-D3(BJ) due the large number of unresolved convergence problems. 

\begin{table*}
    \centering
    \caption{Basis sets considered in this publication. Table lists references to the original publication, the basis set family,  zeta-quality, and the augmentation in the basis set (i.e. polarisation and diffuse functions). Augmentation is for both H and non-H atom unless otherwise noted.}
    \label{tab:bases}
    \scalebox{0.83}{
    \begin{tabular}{lcccl}
    \toprule
        \mc{1}{c}{Basis Set}  & \mc{1}{c}{Basis Set Family}  & \mc{1}{c}{Zeta-quality} & \mc{1}{c}{Augmentation} & \mc{1}{c}{Ref.}  \\
        \midrule
                  6-31G       & Pople-style                  &  Double                  & None &  \cite{71DiHePo,72HeDiPo,75DiPo,77BiPo,82FrPiHe}                                                                             \\
                                    6-31+G      & Pople-style                  & Double                  & Diffuse functions on non-H atoms& \cite{83ClChSp, 87SpClRa, 92GiJoPo}                                                        \\
                  
                  6-31G*      & Pople-style                  &  Double                  & Polarisation functions on non-H atoms & \cite{73HaPo, 82FrPiHe}                                                                                    \\
6-31+G*     & Pople-style                  & Double                 & Polarisation \& diffuse functions on non-H atoms  &\cite{71DiHePo, 72HeDiPo, 73HaPo, 75DiPo, 77BiPo, 82FrPiHe, 83ClChSp, 87SpClRa, 92GiJoPo}     \\
                  6-311G      & Pople-style                  & Triple                  & None & \cite{80KrBiSe,80McCh}                                                                                                     \\
                  6-311+G     & Pople-style                  & Triple                  & Diffuse functions on non-H atoms &\cite{83ClChSp, 87SpClRa, 92GiJoPo}                                                         \\
                  6-311G*     & Pople-style                  & Triple                  &  Polarisation functions on non-H atoms & \cite{80KrBiSe, 82FrPiHe}                                                                               \\
                  6-311+G*    & Pople-style                  & Triple                  & Polarisation \& diffuse functions on non-H atoms&   \cite{80KrBiSe, 80McCh, 82FrPiHe, 83ClChSp, 87SpClRa, 92GiJoPo}                               \\ [2mm]
                  def2-SVP    & Ahlrichs-Karlsruhe           & Double                  &  Polarisation functions  &  \cite{05WeAh}                                                                                          \\
                  def2-TZVP   & Ahlrichs-Karlsruhe           & Triple                  &  Polarisation functions   & \cite{05WeAh}                                                                                          \\
                  def2-TZVPP  & Ahlrichs-Karlsruhe           & Triple                  & Polarisation functions  & \cite{05WeAh}                                                                                            \\
                  def2-SVPD   & Ahlrichs-Karlsruhe           & Double      & Polarisation \& diffuse functions            & \cite{05WeAh, 10RaFu}                                                                          \\
                  def2-TZVPD  & Ahlrichs-Karlsruhe           & Triple                  & Polarisation \& diffuse functions  & \cite{05WeAh, 10RaFu}                                                                        \\
                  def2-TZVPPD & Ahlrichs-Karlsruhe           & Triple          & Polarisation \& diffuse functions         & \cite{05WeAh, 10RaFu}                                                                         \\ [2mm]
                  cc-pVDZ     & Dunning                      & Double         & Polarisation functions              & \cite{89Du,93WoDu,94WoDu,11PrWoPe}                                                                   \\
                  cc-pVTZ     & Dunning                      & Triple            & Polarisation functions          & \cite{89Du,93WoDu,94WoDu,11PrWoPe}                                                                    \\
                  aug-cc-pVDZ & Dunning                      & Double          & Polarisation \& diffuse functions        & \cite{89Du,92KeDuHa,93WoDu,94WoDu,11PrWoPe}                                                    \\
                  aug-cc-pVTZ & Dunning                      & Triple          & Polarisation \& diffuse functions          & \cite{89Du,92KeDuHa,93WoDu,94WoDu,11PrWoPe}                                                  \\ [2mm]
                  pc-1        & Jensen pc-$n$                & Double     & Polarisation functions                  & \cite{01Je,02Je,04JeHe,07Je}                                                                         \\
                  pc-2        & Jensen pc-$n$                & Triple     & Polarisation functions                  & \cite{01Je,02Je,04JeHe,07Je}                                                                         \\
                  aug-pc-1    & Jensen pc-$n$                & Double       & Polarisation \& diffuse functions               & \cite{01Je,02Je,02Je_diffuse,04JeHe,07Je}                                                 \\
                  aug-pc-2    & Jensen pc-$n$                & Triple   & Polarisation \& diffuse functions               & \cite{01Je,02Je,02Je_diffuse,04JeHe,07Je}                                                     \\ [2mm]
                  pcseg-1     & Jensen pcseg-$n$             & Double        & Polarisation functions              & \cite{14Je}                                                                                          \\
                  pcseg-2     & Jensen pcseg-$n$             & Triple       & Polarisation functions              & \cite{14Je}                                                                                           \\
                  aug-pcseg-1 & Jensen pcseg-$n$             & Double      & Polarisation \& diffuse functions            & \cite{14Je}                                                                                   \\
                  aug-pcseg-2 & Jensen pcseg-$n$             & Triple     & Polarisation \& diffuse functions               & \cite{14Je}                                                                                 \\ [2mm]
                  SNSD        & Barone                       & Triple   & Polarisation \& diffuse functions                  & \cite{13BaBiBl}                    \\
    \bottomrule
    \end{tabular}}
\end{table*}

Our consideration of basis sets is fairly comprehensive, as detailed in \Cref{tab:bases}, as insufficient benchmarking is available to enable recommendation of particular basis sets over others. It is worth noting that modern basis set theoretical understanding recommends inclusion of polarisation for all calculations in line with zeta level, e.g. modern carbon double  zeta basis sets always include 1 d basis function while triple zeta basis sets always include 2 d functions and at least 1 f function, meaning that 6-31G and 6-311G would not fit the modern quality definition of double and triple zeta basis sets. Given the widespread use of Pople basis sets and the common understanding of the term double and triple zeta, we choose to maintain the original definitions and specify polarisation explicitly as an augmentation to the basis set. 

Modern use of the term augmented basis functions generally refers to the addition of diffuse basis functions with low Gaussian exponents to the basis set, enabling treatment of extended electron wavefunctions such as in excited states and anions. Diffuse basis functions are not recommended for typical calculations as they often cause SCF convergence issues and are more expensive computationally. However, when predicting infrared intensities of vibrational frequencies, we do recommend inclusion of diffuse basis functions because of their benchmarked importance in accurate dipole moment predictions \cite{20ZaMc}.

\subsection{Computational Details}

Initial molecular geometries for all 141 molecules were obtained from their SMILES identifiers through an automated approach using the RDKit \cite{10La} and ChemCoord \cite{17We} libraries in python. We optimised the initial geometries at the corresponding model chemistry level using an ultrafine integration grid (99 radial shells and 590 angular points per shell) and a tight convergence criterion with a maximum force and maximum displacements smaller than 2.0x10$^{-6}$\,Hartree/Bohr and 6.0x10$^{-6}$\,\AA, respectively. We investigated the use of tighter convergence criteria (thresholds of 1.5x10$^{-5}$\,Hartree/Bohr and 6.0x10$^{-5}$\,\AA 
 for the maximum force and maximum displacements, respectively) but this led to a substantial increase of unconverged calculations in our approach. For a selection of molecules where both convergence criteria converged, we found average differences in frequencies of less than 1.0\,\cm{}, which should be taken as an approximate error on our results.

For a small number of calculations ($\sim$\,5\,\%), however, loose convergence criteria in addition to implementing further functionality in the input file (e.g., changing the SCF default algorithm, turning symmetry off, etc.) were carried out to ensure the collection of all data necessary for our analysis. These cases are outlined in more detail in the supplemental material for this paper.

Most calculations were carried out using the Gaussian quantum chemistry package \cite{g16}, with a smaller number of jobs ran with the Q-Chem package \cite{15ShGaEp}.

\subsection{Computing Scaling Factors and Statistical Metrics}

In line with good-practice procedures in data science, here we computed harmonic frequency scaling factors for all model chemistries by splitting the frequencies in VIBFREQ1295 into a training (70\% of the data) and test (30\% of the data) set \cite{22ZaMc_VIBFREQ}. We ensured convergence in the reported values by randomising the partition of the training and test sets over 100 repetitions. Average values are reported throughout this publication along with their corresponding uncertainties (in brakets) as one standard deviation of the computed value across all 100 repetitions.

We computed all predicted fundamental frequencies, $\nu_i^\textrm{calc}$, using an optimised  scaling factor $\lambda$ defined as $\lambda = (\sum_{i}^{N} \omega_{i}^\textrm{calc}\nu_{i}^\textrm{exp})/(\sum_{i}^{N}\omega_{i}^{\textrm{calc}^{2}})$, where $\omega_{i}^\textrm{calc}$ and $\nu_{i}^\textrm{exp}$ represent the calculated raw harmonic and experimental fundamental frequencies, respectively, with both summations running over the total number of frequencies $N$ considered in the training set. This equation is found through a least-squares procedure minimising the root-mean-squared-error (RMSE) between the scaled harmonic $\lambda \omega_{i}^\textrm{calc}$ and experimental fundamental frequencies $\nu_{i}^\textrm{exp}$ \cite{96ScRa}.

Traditionally, model chemistry performance in harmonic frequency calculations has been largely assessed through the RMSE between the scaled harmonic and experimental fundamental frequencies \cite{21ZaMc}. However, RMSEs are highly influenced by the presence of outliers, meaning that the reported metrics could potentially overestimate the expected errors in the scaled harmonic frequencies from a given model chemistry. 

Instead, we find \cite{22ZaMc_VIBFREQ} that median errors are generally a more appropriate statistical metric to judge model chemistry performance, providing a single number description of the distribution of errors in the scaled harmonic frequencies. To further assess the distribution of errors for important model chemistries, we also use  box-and-whisker plots and quantify the first, second (median error), and third quartiles. 

Outliers are considered separately and explicitly. Though there are many possible definitions, here we use the simple criteria of a deviation $>$\,50\,\cm{} as these predictions will usually be unhelpful for prediction of vibrational frequencies or analysis of experimental spectra. 

\subsection{Types of Scaling Factors}

It is not enough to simply say that a scaling factor is used to convert from computed raw harmonic frequencies to predicted fundamental frequencies; we need to specify whether the scaling factor is (1) global vs frequency-range-specific and (2) universal or model-chemistry-specific.

Due to their simplicity, global scaling factors are  commonly used  despite their lower accuracy \cite{21ZaMc,22ZaMc_VIBFREQ}. In the global scaling approach, a single value is used across the whole frequency range to predict fundamental frequencies by scaling raw harmonic frequencies from a given quantum-chemistry calculation. However, as thoroughly demonstrated \cite{01HaVeSc,04SiBoGu,05AnUv,07MeMoRa,08AnGoJo,11LaBoHa,12LaCaWi,16ChRa,17Cha,22ZaMc_VIBFREQ}, implementing a frequency-range-specific scaling approach can substantially improve upon the accuracy of predicted fundamental frequencies, by allowing a more targeted scaling of similar vibrational modes together. Specifically, based on considerations of different potential frequency thresholds, we found that superior performance can be achieved by dividing the frequency range into three different frequency regions, each one of them with a corresponding scaling factor, i.e., low-frequency: $<$\,1,000\,\cm{}, mid-frequency: 1,000--2,000\,\cm{}, and high-frequency: $\geq$\,2,000\,\cm{}. 

The literature on scaled harmonic frequencies strongly preferences the use of a model-chemistry-specific scaling factors over a single universal value used for all model chemistries; indeed most of the methodology assessment papers in this field have been focused on finding model-chemistry-specific scaling factors rather than assessing the relative performance of individual model chemistries. However, our recent review paper \cite{21ZaMc} showed that the scaling factor value converged at quite moderate computational levels, i.e., hybrid functionals with double-zeta basis sets, thus suggesting that a universal scaling factor for all model chemistries at or above hybrid/double-zeta level might give sufficient accuracy \cite{21ZaMc}. Comparison of high-level theory and experiment in \citet{22ZaMc_VIBFREQ} allowed quantification of these universal scaling factors for users of both global and frequency-range-specific approaches; see \Cref{tab:frange_ccsdt}.

\begin{table}[h!]
    \centering
    \caption{Global and frequency-range-specific scaling factors for the CCSD(T)(F12*)/cc-pVDZ-F12 model chemistry as provided in \citet{22ZaMc_VIBFREQ}. These scaling factors correct for anharmonicity error, and can be used for model chemistries in which the model-chemistry-specific scaling factor is not available. Numbers in parenthesis are one standard deviation in the last digit of the reported value, calculated from 100 different training/test partitions of the full database.}
    \label{tab:frange_ccsdt}
    \scalebox{0.95}{
    \begin{tabular}{lcc}
    \toprule
        \mc{1}{c}{Type} & \mc{1}{c}{Frequency Coverage (\cm{})} & \mc{1}{c}{Scaling Factor} \\
        \midrule
        Global & All freq. range & 0.9617(3) \\
        \\
        Low-frequency & $<$\,1,000 & 0.987(1) \\
        Mid-frequency & 1,000 -- 2,000 & 0.9727(6) \\
        High-frequency & $\geq$\,2,000 & 0.9564(4) \\
    \bottomrule
    \end{tabular}}
\end{table}

In practice, raw harmonic frequencies are scaled using a combination of these scaling factor types, e.g., using a global scaling approach with model-chemistry-specific scaling factors. In this section, we aim to compare the performance of the four options --global universal scaling factor, global model-chemistry-specific scaling factor, frequency-range-specific universal scaling factors and frequency-range-specific and model-chemistry-specific scaling factors-- and determine the influence of this choice in performance. 

\begin{figure}[h!]
    \centering
    \includegraphics[width=1\textwidth]{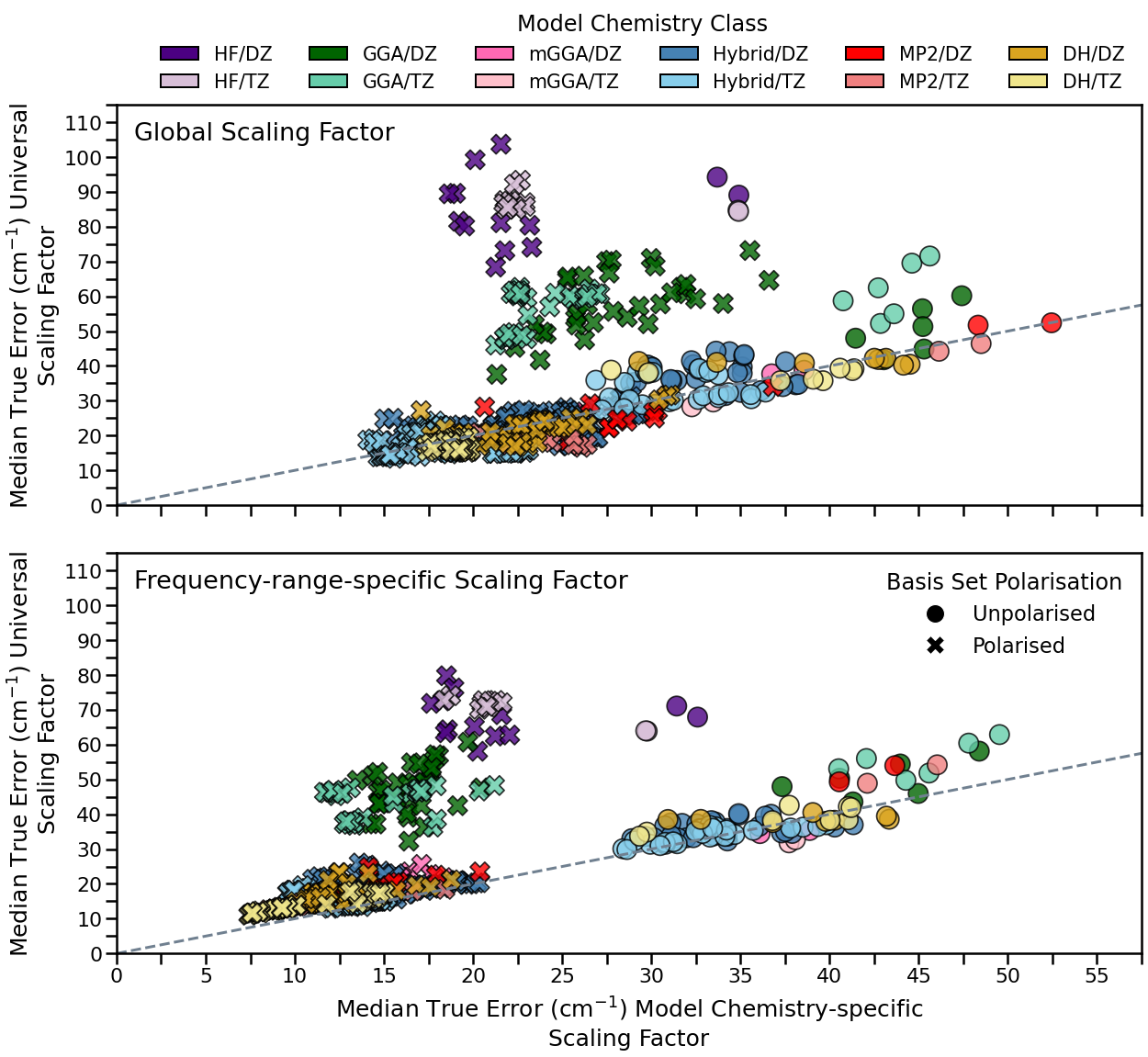}
    \caption{Comparison between the median true errors (in \cm{}) for the universal ($y$-axis) and model chemistry-specific ($x$-axis) scaling factors using a global (top) and frequency-range-specific (bottom) scaling approach for all model chemistries featured in this publication. The grey dashed line in the figure corresponding to the diagonal $x$ = $y$ is used as a guiding reference. Model chemistry classes are hued by different colours. The shapes indicate the presence of polarisation functions in the basis sets.}
    \label{fig:sf_comparison}
\end{figure}

\Cref{fig:sf_comparison} presents the correlation between the median true errors (henceforth median errors) in \cm{} for all model chemistries featured in this study using a global (top) and frequency-range-specific (bottom) scaling approach. The median errors along the $x$-axis in the figure correspond to using the model chemistry-specific scaling factors computed here, whereas the median errors in the $y$-axis come from implementing the universal scaling factors presented in \Cref{tab:frange_ccsdt}. Data points are hued depending on the model chemistry class with the different shapes indicating whether the basis set used in the calculations is complemented with polarisation functions. 

Poor performing model chemistries are immediately apparent in \Cref{fig:sf_comparison}. Unpolarised basis sets (circles) are far inferior to polarised basis sets (crosses) regardless of the scaling factor type used and will not be considered further in this paper. HF and GGA functionals are generally inferior to hybrid and double-hybrid functionals, though the performance of some GGAs with model-chemistry and frequency-range-specific scaling factors can be surprisingly impressive (down to $\sim$\,12\,\cm{} median error) and might be worth further consideration for larger molecule applications. However, there is substantial error cancellation of model chemistry error associated with GGA fundamental frequency predictions, as evidenced by the big differences between the results using the universal vs model-chemistry-specific scaling factor. This error cancellation may be unreliable. Therefore, for this rest of this paper, we focus on the higher reliability and accuracy achievable with hybrid and double-hybrid functionals. 

For these stronger-performing model chemistries in \Cref{fig:sf_comparison},  frequency-range-specific scaling factors usually produce  median true errors about half as large as those using the global scaling approach, aligning with the results of the high level theoretical results in \citet{22ZaMc_VIBFREQ}; this improvement in accuracy will, in most cases, be worth the minimal additional data processing step. 

The relative performance of model-chemistry-specific scaling factors vs universal scaling factors depends on whether a global or frequency-range-specific scaling approach is used. In \Cref{fig:sf_comparison}, the top plot (global scaling) shows no significant difference between the median errors for the universal and model chemistry-specific scaling factors (there is a fair alignment of the data-points along the $x = y$ line), thus suggesting that when using global scaling, universal scaling factors can be used without a significant loss in accuracy. Conversely, for the frequency-range-specific scaling (bottom plot), significant improvement is achieved when using model chemistry-specific instead of universal scaling factors. It is worth mentioning that, as the median errors are not what is optimised when calculating scaling factors, it is possible for the universal scaling factor median error to be lower than the model-chemistry-specific scaling factor median error.

Based on these results, we strongly encourage the reader to use frequency-range and model-chemistry-specific scaling factors to achieve the best accuracy when predicting fundamental frequencies by scaling computed harmonic frequencies. Further analysis in this publication will focus on this recommendation unless otherwise noted.

\section{Understanding Errors in Harmonic Frequency Calculations}
\label{sec:uncert_harmcalcs}

\subsection{Anharmonicity vs Model Chemistry Error: Components of the True Error}

\begin{figure}
    \centering
    \includegraphics[width=1\textwidth]{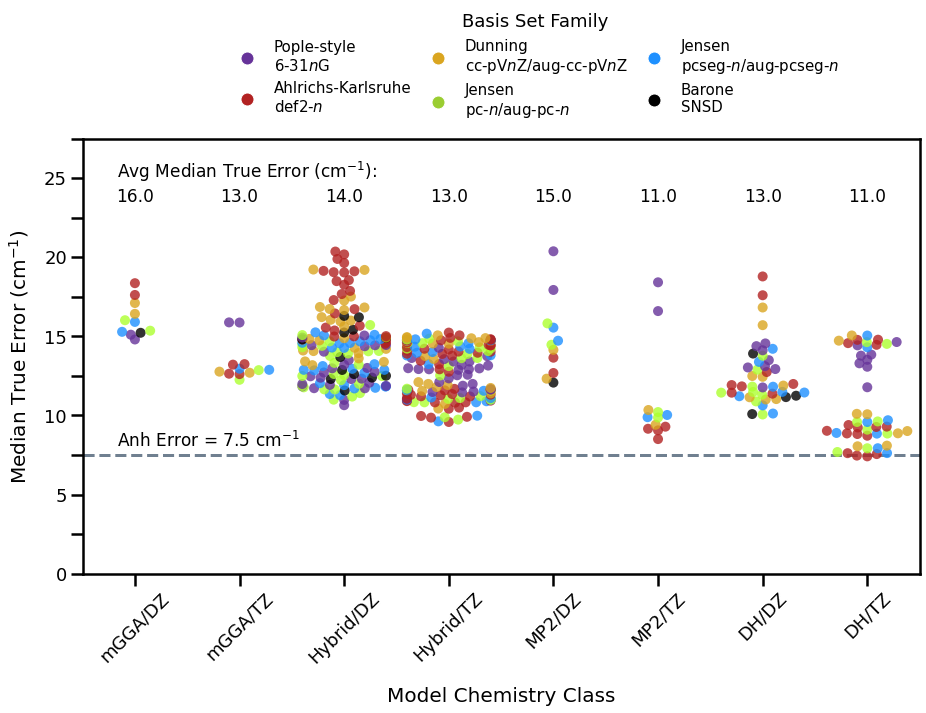}
    \caption{Distribution of median errors between scaled harmonic frequencies from routine model chemistries and the experimental fundamental frequencies in VIBFREQ1295, grouped by different model chemistry classes. Basis set families are hued by different colours. The average median true errors for each model chemistry class (in \cm{}) are presented at top of the figure. The anharmonicity error is represented with the dashed grey line at 7.5\,\cm{}.}
    \label{fig:anhmc_errors}
\end{figure}

Based on a meta-analysis of RMSEs for vibrational frequency predictions in the literature for a variety of model chemistries, \citet{21ZaMc} concluded that when predicting vibrational frequencies within the double harmonic approximation, it was likely that anharmonicity error dominated model chemistry error in determining the true error, i.e. the difference between the scaled harmonic frequency and experimental fundamental frequency. 

To explore this, \Cref{fig:anhmc_errors} shows new results for the median true error for vibrational frequencies for a variety of model chemistries across the VIBFREQ1295 database. The anharmonicity error of 7.5 \cm{} (computed by \citet{22ZaMc_VIBFREQ} by comparing experimental vs scaled CCSD(T)(F12*)/cc-pVDZ-F12 \abinitio{} harmonic frequencies) is visualised as a horizontal dashed line. From \Cref{fig:anhmc_errors}, it is clear that judicious choice of model chemistry can result in the true error approaching the anharmonicity error lower bound, i.e. make the model chemistry error negligible --or at least that use of scaling factors can enable good cancellation of model chemistry and anharmonicity errors so the true error approaches the anharmonicity error limit.  Similarly, however, even within a particular model chemistry class, some density functional approximations and basis sets had much larger model chemistry errors than others. This result justifies the need for this benchmark paper to support users in choosing optimal model chemistries, rather than just finding a scaling factor for the model chemistry of their choice. 

We defer detailed consideration of model chemistry choice to \Cref{sec:modechem_perfm} where we produce more detailed visualisations, but note the richness of information already available in \Cref{fig:anhmc_errors}; e.g. double-hybrids can outperform hybrid density functionals but only when paired with triple-zeta basis sets. 

\subsection{Level of Theory Error (LoTE) and Basis Set Incompleteness Error (BSIE): Components of the Model Chemistry Error}

We can understand the model chemistry error by separating it into the level of theory error LoTE (error in the level of theory chosen) and the basis set incompleteness error BSIE (error in the basis set choice). Here, we estimate both errors for a sub-section of our model chemistries to understand the sensitivity of harmonic frequency calculations  to the particular basis set and level of theory choice.

For ease of analysis, we excluded any scaling of the computed frequencies and consider the raw harmonic frequencies obtained from the calculations. This is the only sub-section in the paper that uses raw instead of scaled harmonic frequencies, as this choice allows us to more cleanly delineate LoTE and BSIE without anharmonicity error complicating the interpretation. 

\begin{table}[h!]
    \centering
    \caption{Median error (Med. Error), root-mean-squared-error (RMSE), and mean signed error (MSE) describing the level of theory error (LoTE) for all levels of theory considered by comparing the corresponding large basis set/DFT harmonic frequencies against our reference high-quality CCSD(T)(F12*)/cc-pVDZ-F12 harmonic frequencies. All statistical metrics are given in \cm{}.}
    \label{tab:lot_error}
    \scalebox{1}{
    \begin{tabular}{lcccHc}
        \toprule
        \mc{1}{c}{Level of Theory}  & \mc{1}{c}{Class} & \mc{1}{c}{Med. Error} & \mc{1}{c}{RMSE} & & \mc{1}{c}{MSE}   \\
        \midrule
            HF              & HF     & 101        & 119  & 400  & -101.89 \\ [2mm]
            B97-D3(BJ)      & GGA    & 39         & 56   & 299  & 43.09   \\
            BLYP-D3(BJ)     & GGA    & 49         & 72   & 304  & 58.19   \\
            PBE             & GGA    & 51         & 64   & 293  & 51.64   \\ [2mm]
            M06L-D3(0)      & mGGA   & 11         & 29   & 386  & -2.07   \\ [2mm]
            B3LYP           & Hybrid & 15         & 26   & 191  & 8.93    \\
            B3LYP-D3(BJ)    & Hybrid & 15         & 25   & 191  & 8.54    \\
            B97-1           & Hybrid & 16         & 24   & 182  & 9.85    \\
            B97-2           & Hybrid & 10         & 23   & 192  & -4.28   \\
            M052X           & Hybrid & 18         & 41   & 233  & -27.10  \\
            M052X-D3(0)     & Hybrid & 18         & 41   & 233  & -27.05  \\
            SOGGA11X        & Hybrid & 20         & 36   & 201  & -25.07  \\
            $\omega$B97X-D  & Hybrid & 9          & 28   & 197  & -11.09  \\
            PBE0            & Hybrid & 13         & 26   & 195  & -4.39   \\
            PBE0-D3(BJ)     & Hybrid & 13         & 26   & 195  & -4.52   \\ [2mm]
            MP2             & MP2    & 8          & 159  & 5246 & -14.17  \\ [2mm]
            B2PLYP          & DH     & 6          & 16   & 179  & 0.94    \\
            B2PLYP-D3(BJ)   & DH     & 6          & 16   & 179  & 0.79    \\
        \bottomrule
    \end{tabular}}
\end{table}

\begin{figure}
    \centering
    \includegraphics[width=0.7\textwidth]{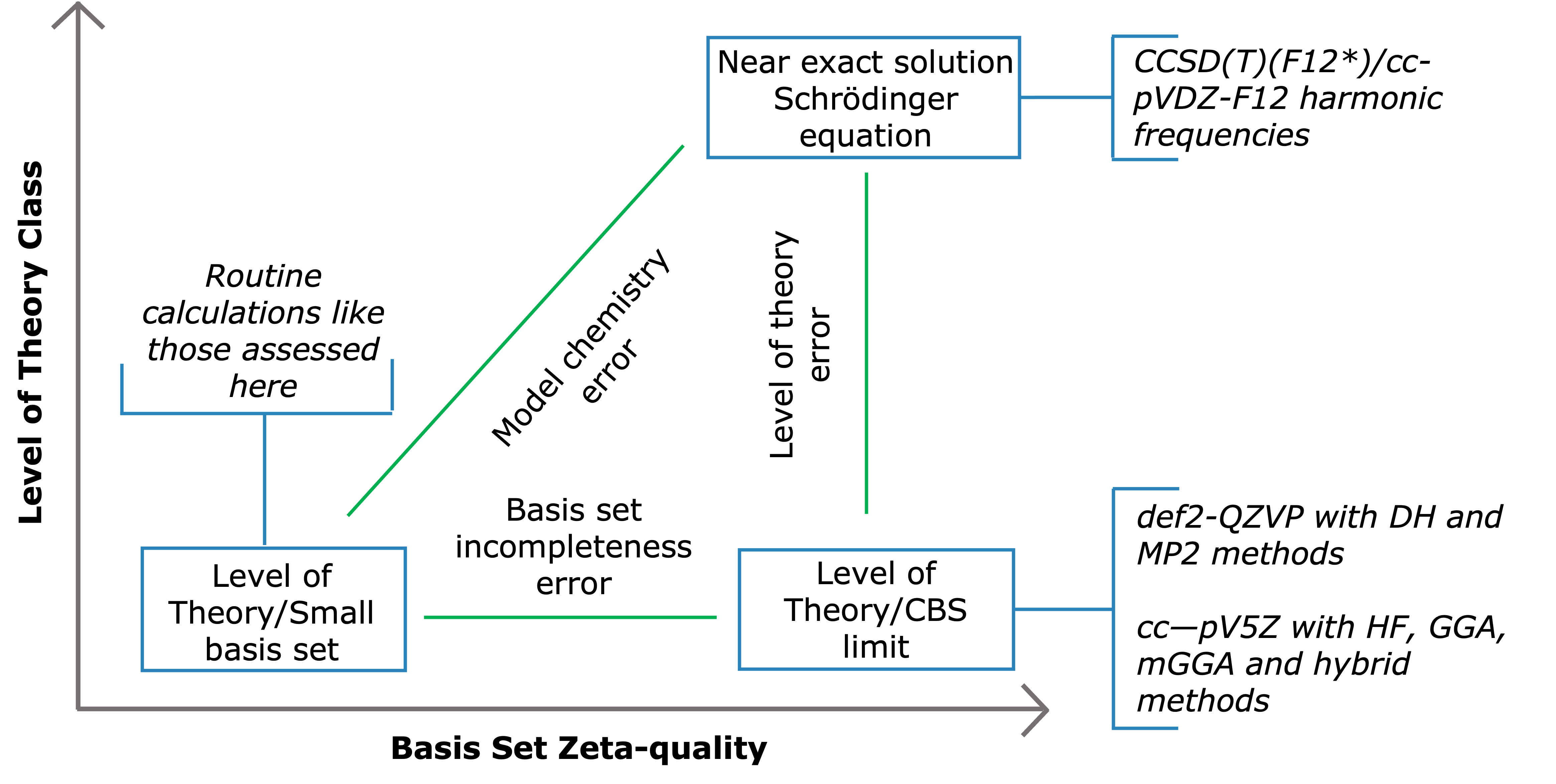}
    \caption{Combined Pople diagram and Jacob's ladder for computational quantum chemistry calculations. The boxes represent the sources of data (in blue) and the lines the type of error evaluated (in green). The squared brackets near the boxes encapsulate the type of data considered.}
    \label{fig:mc_diagram}
\end{figure}

\Cref{fig:mc_diagram} diagrammatically represents the way in which we evaluated both the level of theory LoTE and basis set incompleteness errors BSIE. The \textit{y}-axis orders the different levels of theory classes in increasing computational time and accuracy, thus merging the traditional Pople diagram for wavefunction methods \cite{65Po} and Jacob's ladder for DFT methods \cite{19GoMe}, while the \textit{x}-axis shows the different basis set zeta-qualities, also increasing in size.

In the context of this work, we define the level of theory error (LoTE) as the wavenumber difference between harmonic frequencies computed at the near-exact solution to the Schrodinger equation (the CCSD(T)(F12*)/cc-pVDZ-F12 harmonic frequencies in the database) and the harmonic frequencies from near complete basis set limit calculations using moderate levels of theory (def2-QZVP calculations using double-hybrid functionals and MP2, and cc-pV5Z calculations using HF, GGAs, mGGAs, and hybrid methods). This aligns with the density functional approximation (DFA) error metric generally quantified in large-scale DFA benchmarking studies (e.g., \cite{17GoHaBa,17MaHe}). The LoTE is independent of basis set choice (assuming these very large basis sets are sufficiently large for negligible error).

\Cref{tab:lot_error} presents the statistical distribution of the level of theory error, specifically median error (Med. Error), root-mean-squared error (RMSE), and the mean signed error (MSE). HF, GGA and mGGA present a very poor performance overall. The choice of hybrid functional is important as there is substantial variability in performance (e.g. consider the MSE).  MP2 and double-hybrid functionals significantly outperform hybrid functionals, with median level of theory errors similar to the median anharmonicity error. Given that the median true error in \Cref{fig:anhmc_errors} can lie very close to the anharmonicity error for some model chemistries, our results in \Cref{tab:lot_error} imply that there is significant cancellation of the level of theory and anharmonicity errors when evaluating the true errors.

\begin{sidewaystable}
    \centering
    \caption{Median error (Med. Error), root-mean-squared-error (RMSE), and mean signed error (MSE) describing the basis set incompleteness error (BSIE) for all basis sets considered by comparing our routine small basis set/DFT harmonic frequencies against the corresponding large-basis DFT harmonic frequencies. Note that the BSIE estimation is done across the different level of theory classes featured in this study. All statistical metrics are given in \cm{}.}
    \label{tab:bsie}
    \scalebox{0.65}{
    \begin{tabular}{lccHcccHcccHcccHcccHcccHc}
    \toprule
Basis Set      & \multicolumn{4}{c}{HF}             & \multicolumn{4}{c}{GGA}            & \multicolumn{4}{c}{mGGA}           & \multicolumn{4}{c}{Hybrid}         & \multicolumn{4}{c}{MP2}            & \multicolumn{4}{c}{Double-hybrid}  \\
\cmidrule(r){2-5} \cmidrule(r){6-9} \cmidrule(r){10-13} \cmidrule(r){14-17} \cmidrule(r){18-21} \cmidrule(r){22-25}
               & Med. Error & RMSE & MAX & MSE    & Med. Error & RMSE & MAX & MSE    & Med. Error & RMSE & MAX & MSE    & Med. Error & RMSE & MAX & MSE    & Med. Error & RMSE & MAX & MSE    & Med. Error & RMSE & MAX & MSE    \\
               
    \midrule
\textit{Double-zeta} &            &      &       &        &            &      &       &        &            &      &       &        &            &      &       &        &            &      &       &        &            &      &       &        \\ [2mm]
6-31G           & 41         & 90   & 847   & -11.62 & 34         & 65   & 569   & -9.68  & 33         & 74   & 619   & -15.37 & 36         & 69   & 623   & -9.17  & 44         & 163  & 4776  & -34.56 & 37         & 72   & 472   & -18.90 \\
6-31+G          & 39         & 94   & 842   & -16.55 & 33         & 70   & 612   & -15.10 & 32         & 78   & 610   & -19.29 & 35         & 74   & 703   & -14.28 & 41         & 169  & 4848  & -45.97 & 35         & 78   & 550   & -27.31 \\
6-31G*          & 18         & 31   & 109   & 18.17  & 17         & 27   & 167   & 9.74   & 8          & 26   & 430   & 0.21   & 17         & 25   & 123   & 12.23  & 24         & 140  & 4655  & 8.72   & 21         & 29   & 157   & 13.34  \\
6-31+G*         & 14         & 29   & 111   & 14.35  & 11         & 22   & 147   & 5.09   & 7          & 25   & 426   & -3.01  & 12         & 22   & 129   & 7.95   & 19         & 139  & 4619  & -2.35  & 15         & 25   & 178   & 5.95   \\
def2-SVP        & 12         & 23   & 139   & 7.19   & 11         & 23   & 254   & 3.03   & 13         & 26   & 350   & -5.94  & 11         & 21   & 122   & 4.24   & 13         & 136  & 4597  & 7.26   & 10         & 20   & 106   & 6.51   \\
def2-SVPD       & 11         & 19   & 98    & -0.23  & 9          & 21   & 264   & -1.98  & 15         & 29   & 440   & -10.55 & 10         & 19   & 129   & -1.38  & 8          & 134  & 4583  & -8.61  & 9          & 18   & 128   & -2.86  \\
cc-pVDZ         & 13         & 20   & 97    & -2.18  & 11         & 22   & 192   & -6.00  & 14         & 28   & 404   & -12.22 & 11         & 19   & 158   & -4.60  & 8          & 134  & 4555  & -8.44  & 10         & 19   & 112   & -5.07  \\
aug-cc-pVDZ     & 13         & 21   & 152   & -9.31  & 12         & 24   & 296   & -10.52 & 15         & 35   & 462   & -18.39 & 12         & 25   & 349   & -9.95  & 17         & 132  & 4433  & -24.66 & 13         & 25   & 310   & -14.12 \\
pc-1            & 8          & 26   & 109   & 8.51   & 7          & 16   & 174   & -0.92  & 10         & 25   & 439   & -5.34  & 8          & 16   & 72    & 1.65   & 20         & 132  & 4454  & 7.20   & 9          & 18   & 89    & 4.27   \\
aug-pc-1        & 9          & 20   & 89    & -0.72  & 8          & 15   & 159   & -5.20  & 11         & 27   & 433   & -10.61 & 8          & 20   & 204   & -4.54  & 13         & 128  & 4307  & -8.21  & 8          & 17   & 106   & -4.98  \\
pcseg-1         & 8          & 26   & 107   & 8.74   & 7          & 16   & 188   & -1.23  & 10         & 24   & 438   & -4.77  & 7          & 16   & 71    & 1.66   & 19         & 131  & 4396  & 6.18   & 8          & 18   & 92    & 3.59   \\
aug-pcseg-1     & 9          & 20   & 96    & -1.86  & 8          & 16   & 160   & -6.14  & 11         & 27   & 431   & -11.78 & 8          & 16   & 90    & -5.02  & 13         & 127  & 4284  & -9.17  & 8          & 18   & 114   & -6.01  \\
SNSD            & 7          & 16   & 112   & -3.40  & 7          & 15   & 156   & -3.53  & 11         & 24   & 408   & -11.74 & 7          & 14   & 112   & -3.67  & 7          & 131  & 4456  & -14.47 & 6          & 16   & 143   & -6.25  \\ [2mm]
\textit{Triple-zeta} &            &      &       &        &            &      &       &        &            &      &       &        &            &      &       &        &            &      &       &        &            &      &       &        \\ [2mm]
6-311G          & 30         & 86   & 847   & -22.40 & 28         & 66   & 432   & -20.01 & 24         & 73   & 534   & -24.41 & 30         & 69   & 481   & -19.21 & 50         & 167  & 4814  & -49.01 & 30         & 74   & 474   & -31.44 \\
6-311+G         & 30         & 90   & 841   & -26.39 & 28         & 70   & 523   & -23.44 & 25         & 75   & 538   & -27.19 & 29         & 72   & 567   & -22.64 & 56         & 171  & 4842  & -56.64 & 33         & 79   & 484   & -37.10 \\
6-311G*         & 12         & 20   & 119   & 8.31   & 8          & 23   & 235   & 1.14   & 10         & 30   & 432   & -8.93  & 8          & 18   & 94    & 3.63   & 12         & 134  & 4478  & -0.65  & 9          & 23   & 151   & 3.02   \\
6-311+G*        & 9          & 19   & 116   & 6.21   & 7          & 20   & 189   & -1.26  & 10         & 29   & 425   & -11.02 & 7          & 17   & 121   & 1.40   & 11         & 133  & 4439  & -6.87  & 8          & 24   & 190   & -1.60  \\
def2-TZVP       & 1          & 4    & 27    & -0.10  & 2          & 5    & 56    & 0.84   & 3          & 17   & 428   & -2.54  & 2          & 4    & 46    & 0.58   & 3          & 64   & 2177  & -4.17  & 1          & 5    & 49    & -0.18  \\
def2-TZVPP      & 1          & 2    & 14    & 0.42   & 1          & 3    & 23    & 1.53   & 4          & 16   & 428   & -2.97  & 1          & 3    & 35    & 1.30   & 2          & 64   & 2177  & -1.56  & 1          & 3    & 15    & 0.94   \\
def2-TZVPD      & 1          & 3    & 24    & -0.71  & 2          & 3    & 24    & 0.00   & 3          & 17   & 428   & -3.14  & 2          & 3    & 25    & -0.15  & 5          & 67   & 2267  & -7.71  & 2          & 5    & 39    & -1.92  \\
def2-TZVPPD     & 1          & 2    & 8     & 0.00   & 1          & 3    & 17    & 0.84   & 4          & 16   & 428   & -3.78  & 1          & 3    & 18    & 0.76   & 2          & 66   & 2267  & -3.60  & 1          & 2    & 15    & -0.22  \\
cc-pVTZ         & 2          & 4    & 45    & -0.67  & 2          & 7    & 91    & 0.42   & 5          & 18   & 437   & -6.08  & 2          & 5    & 39    & -0.07  & 2          & 80   & 2706  & -3.50  & 2          & 5    & 43    & -0.08  \\
aug-cc-pVTZ     & 2          & 4    & 27    & -1.86  & 2          & 4    & 40    & -1.45  & 7          & 19   & 434   & -8.34  & 2          & 4    & 37    & -1.70  & 5          & 74   & 2474  & -8.51  & 2          & 5    & 46    & -3.04  \\
pc-2            & 2          & 5    & 28    & 1.74   & 2          & 5    & 73    & 1.04   & 3          & 16   & 438   & -2.46  & 2          & 4    & 38    & 1.27   & 7          & 47   & 1553  & 5.35   & 3          & 6    & 21    & 2.92   \\
aug-pc-2        & 1          & 4    & 18    & 1.01   & 1          & 3    & 21    & 0.26   & 3          & 15   & 436   & -2.81  & 1          & 3    & 21    & 0.47   & 5          & 40   & 1320  & 2.55   & 2          & 5    & 35    & 1.25   \\
pcseg-2         & 2          & 4    & 25    & 1.50   & 1          & 5    & 62    & 0.92   & 3          & 16   & 432   & -2.54  & 2          & 4    & 39    & 1.17   & 6          & 68   & 2310  & 3.18   & 2          & 5    & 27    & 2.31   \\
aug-pcseg-2     & 2          & 4    & 20    & 0.97   & 1          & 3    & 26    & 0.29   & 3          & 15   & 432   & -2.65  & 1          & 4    & 24    & 0.54   & 5          & 318  & 2217  & -66.40 & 1          & 80   & 1092  & -6.78  \\
    \bottomrule
    \end{tabular}}
\end{sidewaystable}

Basis set incompleteness error (BSIE) is actually typically ignored in benchmarking studies despite the fact that in practice calculations are typically performed with only double- or triple-zeta basis sets. Therefore, here we quantify the difference between using a modest basis set (DZ or TZ) compared to a very large basis set (as specified above) for each level of theory considered. \Cref{tab:bsie} presents the median error (Med. Error), RMSE, and the mean signed error (MSE) related to the basis set incompleteness error (BSIE) for all basis set considered across the different level of theory classes featured in this study. The table shows that the double-zeta basis set choice matters as evidenced by the large variations in the median errors (between 7--44\,\cm{}) regardless of the level of theory class considered, with unpolarised basis sets delivering the worst performance overall (as shown in \Cref{fig:sf_comparison}). Similarly, triple-zeta Pople-style basis sets have a much poorer performance than other triple-zeta basis sets and should thus be avoided in future calculations. Note that the median error and RMSE associated with the BSIE for triple-zeta basis sets coupled with hybrid and double-hybrid functionals are very similar, showcasing their appropriate performance in harmonic frequency calculations when feasible computationally.

Comparison of \Cref{tab:lot_error} and \Cref{tab:bsie} shows that the error incurred in harmonic frequency calculations with the various approximations scales as follows: hybrid functional error $>$ double-zeta (DZ) basis set error $>$ anharmonicity error $\approx$ double hybrid (DH) functional error $>$ triple-zeta (TZ) basis set error. This ordering suggests that Hybrid/DZ and DH/TZ model chemistries are likely to represent the most efficient model chemistries in harmonic frequency calculations for a given computational time, with similar errors arising from the level of theory and basis set choice. The choice of the hybrid functional and the double-zeta basis set are very important in determining performance and should not be neglected.

\section{Assessing Model Chemistry Performance}
\label{sec:modechem_perfm}

\subsection{Scaling Factors}

\begin{figure}
    \centering
    \includegraphics[width=0.95\textwidth]{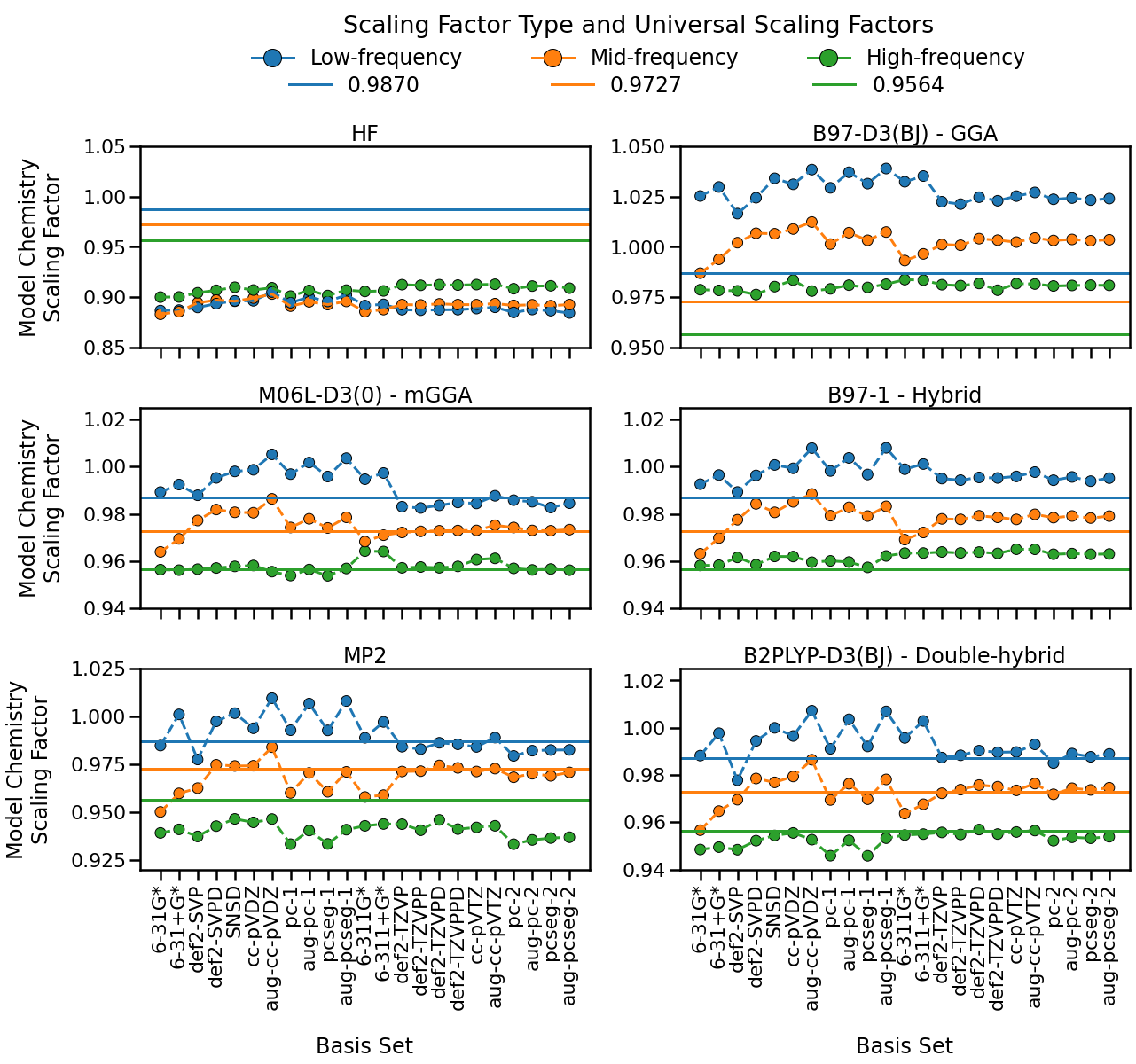}
    \caption{Low-, mid- and high-frequency scaling factors for a handful of model chemistries featured in this study. Data ponits are joined to aid readability. The three solid lines in the figure represent the universal frequency-range-specific scaling factors for the low- (0.9870), mid- (0.9727), and high-frequency (0.9564) regions.}
    \label{fig:sfs}
\end{figure}

\Cref{fig:sfs} shows the optimised low-, mid-, and high-frequency scaling factors for a selection of model chemistries, demonstrating the basis set convergence in each method; model chemistries of the same class had similar behaviour with figures and quantitative scaling factors for all model chemistries available in the supplementary material. The solid lines in the figure represent the universal scaling factors reported in \Cref{tab:frange_ccsdt} and the circles are the individual model-chemistry-specific scaling factors. Data points are joined in the figure to aid readability. Note the different scales for all plots in the figure.

The differences between the low, mid and high-frequency scaling factors in \Cref{fig:sfs} demonstrate the need for frequency-range-specific scaling factors. We can see that the scaling factor values generally decrease from the low- to the high-frequency region. This is not necessarily expected to some readers, but is rationalised by the fact that high frequencies are mostly stretches that have significant anharmonicity due to bond breaking, while the lower frequency bends and stretches are harmonically bound.  As showcased in \citet{21ZaMc}, scaling factors from the HF and B97-D3(BJ) (GGA functional) methods considerably differ from the universal values.

Though not representing a significant improvement in model chemistry performance (see following sections), the figure shows some variability in the scaling factor values across the double-zeta basis set space when comparing non-augmented and augmented basis sets with diffuse functions. For triple-zeta basis sets, however, there is very little variability with near identical scaling factors obtained regardless of the basis set choice or augmentation status; this implies these triple-zeta scaling factors are near the basis set limit scaling factors for the particular model chemistry, aligning with the very low basis set incompleteness error found for triple-zeta basis sets in the previous section (see \Cref{tab:bsie}).

\subsection{Median True Errors}

\begin{figure}
    \centering
    \includegraphics[width=1\textwidth]{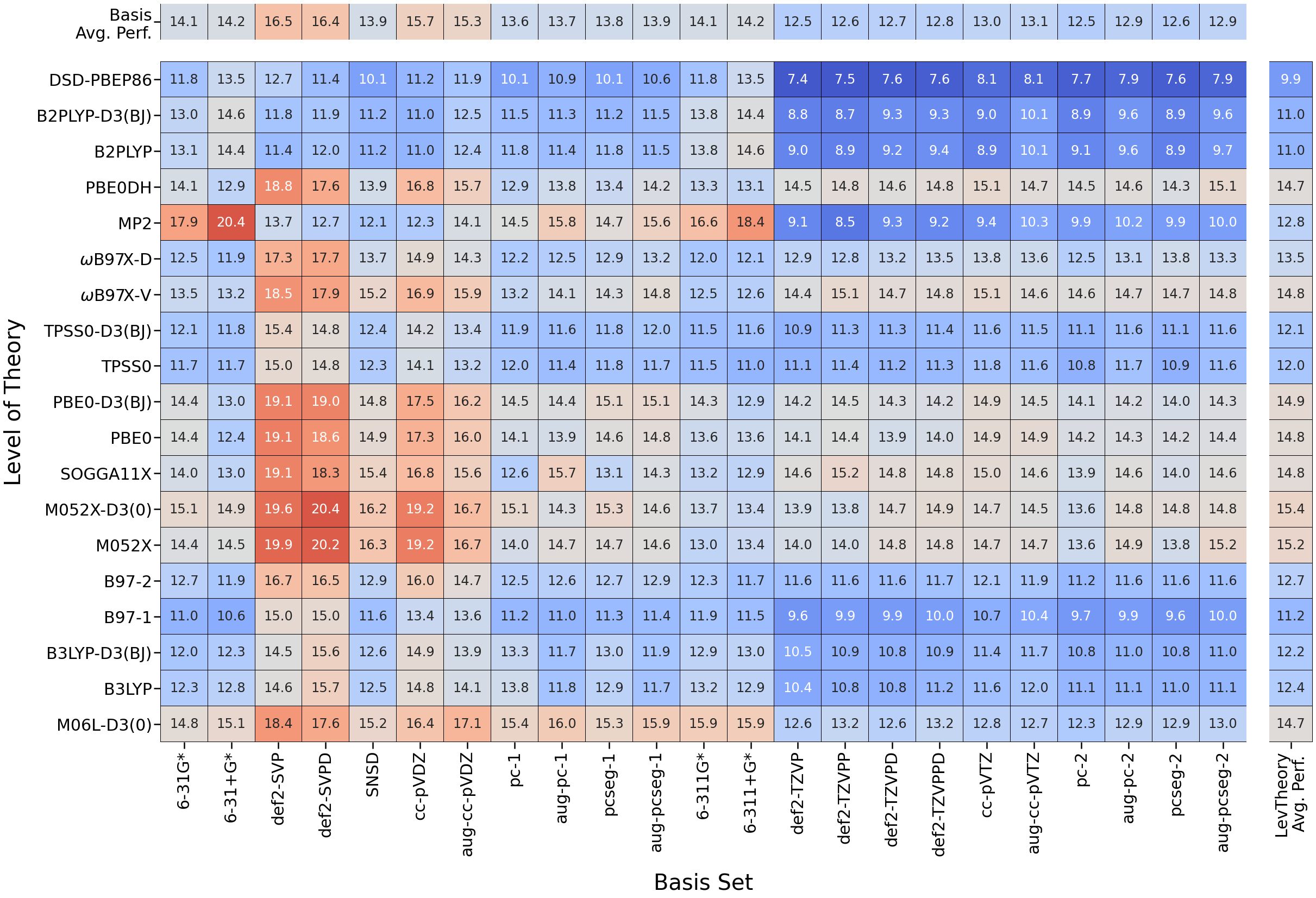}
    \caption{Median true errors (in \cm{}) for all model chemistries featured in this publication. Average performance for all basis sets and levels of theory are outlined in the top and right panels in the figure, respectively. Levels of theory are organised according to their class,i.e., mGGAs, hybrid functionals, MP2, and double-hybrid functionals, with basis sets organised in increasing size and zeta-quality.}
    \label{fig:allmc_perf}
\end{figure}

\Cref{fig:allmc_perf} presents the median true errors (in \cm{}) for all model chemistries considered, with average performance outlined in the top and right panels for all basis sets and levels of theory, respectively. Levels of theory and basis sets in the figure are organised approximately in increasing computational timings and complexity. As discussed in Section \ref{sec:comp_details}, we chose median errors over RMSEs (that are highly dependent on outliers) to assess typical performance.

Considering the choice of basis set, \Cref{fig:allmc_perf} clearly demonstrates that median true error trends follow the basis set incompleteness error trends shown in \Cref{tab:bsie}. Specifically, the choice of double-zeta basis sets is important, Pople triple-zeta basis sets perform poorly but other triple-zeta basis sets highlight for their superior performance with very similar median true errors amongst them. For double-zeta basis sets, 6-31(+)G* and (aug)-pc(seg)-1 variants significantly outperform the def2-SVP(D) and (aug-)cc-pVDZ basis sets. 

Considering the level of theory classes, \Cref{fig:allmc_perf} shows that, in general, double-hybrid functionals lead to improved performance especially when combined with a triple-zeta basis set, with errors less dependent on basis set. However, not all double-hybrids can claim such superior performance; PBE0DH functional has far poorer performance worse than many hybrid functionals. In the hybrid functionals space, the variability in performance is larger; B97-1 has the strongest overall performance, followed by TPSS0, B3LYP, B97-2, and $\omega$B97X-D (and their dispersion corrected versions). The Minnesota functionals (e.g. M05-2X) and PBE0 classes have the poorest performance. 

Top-performing choices from our approach correspond to the DSD-PBEP86 double-hybrid functional and a non-Pople triple-zeta basis set where median true errors between 7--8\,\cm{} can be achieved. For faster calculations, B97-1/6-31+G* has the lowest median error of 10.6 \cm{}. We defer explicit recommendations to \Cref{sec:recommendations} in order to consider not just median true error but outliers and model chemistry reliability.  

\subsection{Data Outliers}

\begin{figure}
    \centering
    \includegraphics[width=1\textwidth]{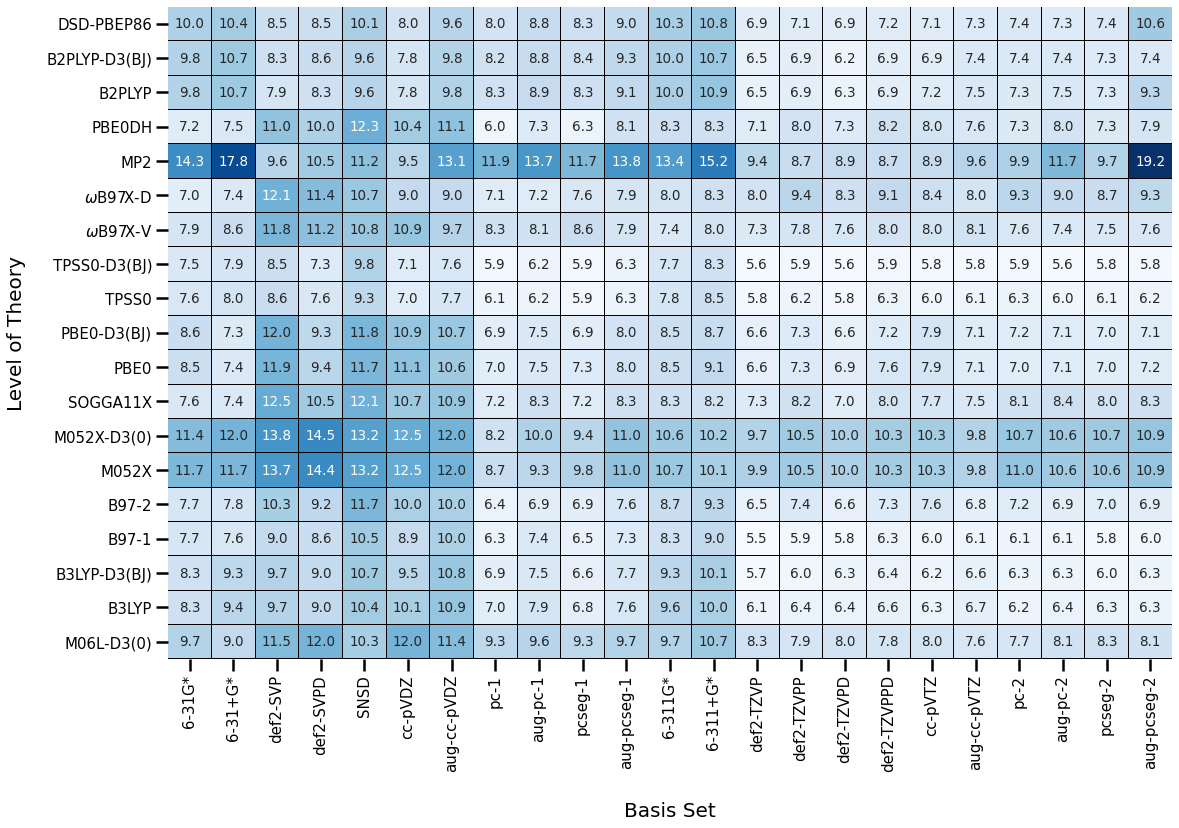}
    \caption{Percentage of outliers for all model chemistries featured in this publication. Levels of theory are organised according to their class, i.e., mGGAs, hybrid functionals, MP2, and double-hybrid functionals, with basis sets organised in increasing size and zeta-quality.}
    \label{fig:outs_heatmap}
\end{figure}

When predicting fundamental frequencies by scaling calculated raw harmonic frequencies, some vibrational modes will always display large frequency errors due to their inherently more (or less) anharmonic nature, making the scaling factor an insufficient model for quantitative predictions. It is important, however, to understand if there are particular model chemistries that lead to an increase in poor predictions. Here, we choose to define outliers as predicted data that differences in more than 50\,\cm{} from experimental values; at this point, the quantitative prediction has limited usefulness.

\Cref{fig:outs_heatmap} presents a heatmap of the percentage of outliers for all model chemistries featured in this study. The percentage of outliers generally decreases for more robust model chemistries, e.g., double-hybrid and hybrid functionals with triple-zeta basis sets, to a minimum of about 6\,\% (approximately 70 out of the 1,295 calculated frequencies) from inherently more anharmonic vibrational modes. We note that the MP2 method has a particularly high prevalence of outliers (up to 19\%), especially when coupled with double-zeta basis sets. 

\subsection{Model Chemistry Reliability}

To calculate all scaled harmonic frequencies for the model chemistries featured in this analysis, we followed a largely automated approach consisting on producing all corresponding initial geometries, input files, running the calculations, and parsing the harmonic frequencies from the output files. For the wide majority of jobs (more than 95\,\%), the calculations finished successfully without requiring manual intervention. However, there were some cases where additional requirements were needed to ensure the calculation's convergence and collection of data. In general, we found three main problems limiting our approach: (1) geometries failing to converge due to poor starting geometries (from RDkit predictions), (2) convergence to a transition state leading to imaginary frequencies, and (3) incorrect number of frequencies in the calculations for linear or highly symmetric molecules. 

\begin{figure}
    \centering
    \includegraphics[width=1\textwidth]{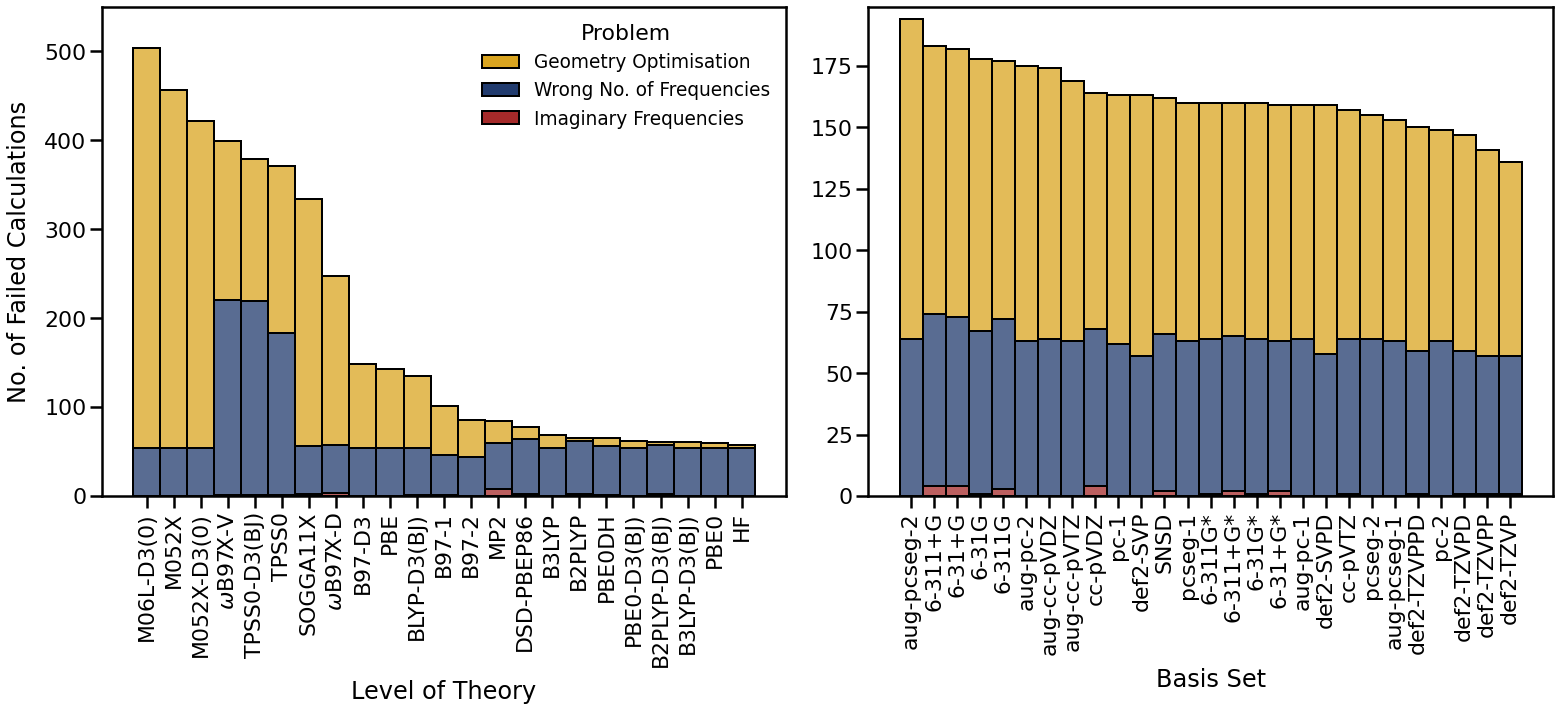}
    \caption{Stacked histograms of the number of failed calculations across all levels of theory (left) and basis sets (right) featured in this study. The colours represent the type of the underlying problem in the failed calculations.}
    \label{fig:failed_jobs}
\end{figure}

\Cref{fig:failed_jobs} presents a stacked histogram of the number of failed calculations across all levels of theory (left) and basis sets (right), respectively, hued by the type of problem encountered in the calculation. Note the difference in scale for both plots in the figure. It is evident that the level of theory choice had a much greater influence on the number of failed calculations than the basis set choice. HF is most reliable, with Becke and the PBE-derived functionals also quite reliable. The $\omega$ families have intermediate reliability while the heavily parameterised Minnesota functional (e.g. M05-2X) has the highest difficulty with convergence.

The figure shows that the convergence of the initial molecular geometry was the most widespread limitation in our approach. Indeed, some methods required tighter initial molecular geometries than those provided from our python approach (see Section \ref{sec:comp_details}) to converge. These methods mostly correspond to the M06L-D3(0), M052X (and its D3(0) version), $\omega$B97X-V, TPSS0 (and its D3(0) version), and SOGGA11X functionals. Conversely, double-hybrid and some hybrid functionals (PBE0 and its B3(BJ) version) were not as sensitive to the initial molecular geometries, with only a handful of unconverged calculations.

Linear and highly symmetric molecules represented a significant challenge for all levels of theory evaluated. The prevalence of imaginary frequencies  was largely attributed to this type of molecules, though this limitation was the least common in our approach. A significant issue arose with calculations displaying the wrong number of harmonic vibrational frequencies
 when dealing with these linear and highly symmetric molecules. In most cases, turning the symmetry off helped overcome such limitation; however, for some cases we were not able to fix the calculations and instead opted for identifying and importing the missing degenerate frequency manually. The left-most plot in the figure shows that for most functionals the number of calculations displaying the wrong number of frequencies is rather constant, as these corresponded particularly to the challenging
\ce{C3O2} and \ce{C2H4} molecules; however, $\omega$B97X-V and the TPSS0 family struggle the most with accurately predicting the right number of harmonic vibrational frequencies in linear and highly symmetric molecules.

While some of the limitations described in this section are likely to be less problematic when focusing on one molecule, the results from our automated calculations are nevertheless indicative of a method's reliability and useful to consider.

\subsection{Additional Considerations}

\subsubsection{Dispersion Corrections}

\begin{table}[h!]
    \centering
    \caption{Average scaling factors (SF) for the low- ($<$1,000\,\cm{}), mid- (1,000--2,000\,\cm{}), and high-frequency ($\geq$\,2,000\,\cm{}) ranges and median true errors (in \cm{}) for a handful of density functionals with and without empirical dispersion corrections across all double- and triple-zeta basis sets. Numbers in parenthesis are one standard deviation in the last digit of the reported value.}
    \label{tab:disp_methods}
    \scalebox{0.85}{
    \begin{tabular}{lccccccccc}
    \toprule
        \multirow{2}{*}{Level of Theory} & \mc{4}{c}{Double-zeta Basis Sets}  & & \mc{4}{c}{Triple-zeta Basis Sets} \\
        \cmidrule(r){2-5} \cmidrule(r){6-10}
         & \mc{1}{c}{SF Low}  & \mc{1}{c}{SF Mid} & \mc{1}{c}{SF High} & \mc{1}{c}{Med. Error} & & \mc{1}{c}{SF Low}  & \mc{1}{c}{SF Mid} & \mc{1}{c}{SF High} & \mc{1}{c}{Med. Error} \\
    \midrule
        M052X       & 0.97(1) & 0.957(6) & 0.939(2) & 21(8)   &   & 0.97(1) & 0.955(4) & 0.941(2)  & 19(7)      \\
        M052X-D3(0) & 0.97(1) & 0.957(6) & 0.939(1) & 21(7)   &   & 0.97(1) & 0.955(4) & 0.941(2)  & 19(7)      \\
                    &         &          &          &            &         &          &           &            \\
        B3LYP       & 1.00(1) & 0.980(7) & 0.961(2) & 23(6)   &   & 1.00(2) & 0.977(3) & 0.964(1)  & 20(7)      \\
        B3LYP-D3(BJ)  & 1.00(2) & 0.979(7) & 0.961(2) & 22(6)   &   & 1.00(1) & 0.977(3) & 0.963(1)  & 20(7)      \\
                    &         &          &          &            &         &          &           &            \\
        PBE0        & 0.99(1) & 0.976(6) & 0.951(2) & 20(6)   &   & 0.98(2) & 0.970(4) & 0.9556(8) & 18(6)      \\
        PBE0-D3(BJ)   & 0.99(1) & 0.970(7) & 0.951(2) & 20(6)   &   & 0.98(2) & 0.970(4) & 0.9555(8) & 18(6)      \\
                    &         &          &          &            &         &          &           &            \\
        B2PLYP      & 1.01(2) & 0.975(8) & 0.952(3) & 20(9)   &   & 1.00(2) & 0.974(5) & 0.956(3)  & 17(10)     \\
        B2PLYP-D3(BJ) & 1.01(2) & 0.975(8) & 0.952(3) & 20(9)   &   & 1.00(2) & 0.974(5) & 0.956(3)  & 17(10)     \\
    \bottomrule
    \end{tabular}}
\end{table}
At the forefront of computational quantum chemistry, including empirical dispersion corrections to the density functional choice has become a necessity to guarantee superior performance in the calculation of different chemical properties \cite{11Gr,11GrEhGo,16GrHaBr,17Go}. Here, for a handful of density functionals (see \Cref{tab:methods}) we have included both their original and dispersion-corrected versions to examine their performance when calculating harmonic vibrational frequencies.

\Cref{tab:disp_methods} presents the average scaling factors and median true errors (in \cm{}) for these functionals across all double- and triple-zeta basis sets. The same scaling factors and median errors are found with and without the empirical dispersion corrections, suggesting no overall improvement in the scaled harmonic frequencies. However, we expect the influence of including dispersion-corrections to the functionals to become more important in larger molecular systems where long-range interactions become more apparent.  

Dispersion-corrected functionals, especially those considered in this analysis, represent excellent and reliable options for the calculation of many different chemical properties \cite{17GoHaBa,17MaHe}, and thus we strongly encourage their use in harmonic frequency calculations to ensure an overall satisfactory model chemistry choice. We also base this recommendation on the fact that virtually identical computational timings were found when using the dispersion-corrected and dispersion-less functionals.

\subsubsection{Diffuse Functions}

\begin{sidewaystable}
    \centering
    \caption{Average scaling factors (SF) for the low- ($<$1,000\,\cm{}), mid- (1,000--2,000\,\cm{}), and high-frequency ($\geq$\,2,000\,\cm{}) ranges and median true errors (in \cm{}) for the augmented and non-augmented version of basis sets with diffuse functions across all mGGA, hybrid, and double-hybrid functionals. Numbers in parenthesis are one standard deviation in the last digit of the reported value.}
    \label{tab:diff_bases}
    \scalebox{0.83}{
    \begin{tabular}{lcccccccccccc}
    \toprule
\multirow{2}{*}{Basis   Set} & \multicolumn{4}{c}{mGGA}                     & \multicolumn{4}{c}{Hybrids}                  & \multicolumn{4}{c}{Double-hybirds}           \\
                                \cmidrule(r){2-5}                           \cmidrule(r){6-9}                               \cmidrule(r){10-13}
                             & SF Low   & SF Mid   & SF High   & Med. Error & SF Low   & SF Mid   & SF High   & Med. Error & SF Low   & SF Mid   & SF High   & Med. Error \\
    \midrule
\textit{Double-zeta }                 &          &          &           &            &          &          &           &            &          &          &           &   \\
                             &          &          &           &            &          &          &           &            &          &          &           &            \\
6-31G*                       & 0.989(2) & 0.964(1) & 0.9564(5) & 17(4)      & 0.972(2) & 0.949(1) & 0.9459(5) & 17(2)      & 0.977(2) & 0.95(1)  & 0.9428(5) & 18(3)      \\
6-31+G*                      & 0.993(2) & 0.969(1) & 0.9562(5) & 18(5)      & 0.976(2) & 0.956(1) & 0.9462(5) & 17(3)      & 0.985(2) & 0.958(1) & 0.9438(5) & 18(4)      \\[1mm]
def2-SVP                     & 0.988(2) & 0.977(2) & 0.9567(6) & 22(5)      & 0.97(2)  & 0.963(2) & 0.9491(6) & 19(4)      & 0.971(2) & 0.964(2) & 0.9443(6) & 16(4)      \\
def2-SVPD                    & 0.995(1) & 0.982(2) & 0.9572(5) & 23(7)      & 0.979(1) & 0.97(2)  & 0.949(5)  & 20(5)      & 0.985(1) & 0.972(2) & 0.9469(5) & 18(6)      \\[1mm]
cc-pVDZ                      & 0.998(2) & 0.98(1)  & 0.9582(6) & 24(7)      & 0.981(2) & 0.97(1)  & 0.9513(6) & 20(5)      & 0.986(2) & 0.972(1) & 0.9492(6) & 18(6)      \\
aug-cc-pVDZ                  & 1.005(2) & 0.987(2) & 0.9557(1) & 26(7)      & 0.989(2) & 0.974(2) & 0.9495(1) & 20(6)      & 0.997(2) & 0.98(2)  & 0.9485(1) & 21(7)      \\[1mm]
pc-1                         & 0.997(2) & 0.974(1) & 0.9538(5) & 19(5)      & 0.979(2) & 0.965(1) & 0.948(5)  & 16(4)      & 0.982(2) & 0.964(1) & 0.9419(5) & 15(4)      \\
aug-pc-1                     & 1.002(2) & 0.978(1) & 0.9564(5) & 22(6)      & 0.988(2) & 0.969(1) & 0.9501(5) & 17(5)      & 0.993(2) & 0.97(1)  & 0.9469(5) & 16(6)      \\[1mm]
pcseg-1                      & 0.996(2) & 0.974(1) & 0.9536(5) & 19(5)      & 0.979(2) & 0.965(1) & 0.9474(5) & 17(4)      & 0.983(2) & 0.964(1) & 0.9419(5) & 15(4)      \\
aug-pcseg-1                  & 1.003(2) & 0.979(1) & 0.9569(5) & 22(6)      & 0.989(2) & 0.969(1) & 0.9509(5) & 17(5)      & 0.996(2) & 0.971(1) & 0.9474(5) & 17(6)      \\
                             &          &          &           &            &          &          &           &            &          &          &           &            \\
\textit{Triple-zeta }                 &          &          &           &            &          &          &           &            &          &          &           &   \\
                             &          &          &           &            &          &          &           &            &          &          &           &            \\
6-311G*                      & 0.995(2) & 0.968(2) & 0.9642(5) & 18(5)      & 0.979(2) & 0.956(2) & 0.9515(5) & 16(3)      & 0.984(2) & 0.957(2) & 0.9489(5) & 16(3)      \\
6-311+G*                     & 0.997(2) & 0.971(1) & 0.964(5)  & 19(6)      & 0.981(2) & 0.958(1) & 0.9512(5) & 16(3)      & 0.991(2) & 0.961(1) & 0.9493(5) & 16(3)      \\[1mm]
def2-TZVP                    & 0.983(1) & 0.972(1) & 0.9572(5) & 16(5)      & 0.974(1) & 0.964(1) & 0.9525(5) & 16(3)      & 0.978(1) & 0.967(1) & 0.9509(5) & 13(4)      \\
def2-TZVPD                   & 0.983(2) & 0.973(1) & 0.9571(5) & 17(5)      & 0.975(2) & 0.965(1) & 0.9526(5) & 16(4)      & 0.979(2) & 0.969(1) & 0.9517(5) & 14(5)      \\[1mm]
def2-TZVPP                   & 0.982(1) & 0.973(1) & 0.9575(5) & 16(5)      & 0.973(1) & 0.964(1) & 0.952(5)  & 16(3)      & 0.977(1) & 0.967(1) & 0.9497(5) & 13(5)      \\
def2-TZVPPD                  & 0.985(2) & 0.973(1) & 0.9577(5) & 17(5)      & 0.974(2) & 0.964(1) & 0.9519(5) & 16(3)      & 0.979(2) & 0.969(1) & 0.9499(5) & 14(5)      \\[1mm]
cc-pVTZ                      & 0.984(2) & 0.973(1) & 0.9606(5) & 17(5)      & 0.975(2) & 0.963(1) & 0.9529(5) & 16(3)      & 0.979(2) & 0.967(1) & 0.9506(5) & 14(4)      \\
aug-cc-pVTZ                  & 0.988(2) & 0.975(1) & 0.9611(5) & 18(5)      & 0.977(2) & 0.966(1) & 0.9527(5) & 16(4)      & 0.982(2) & 0.97(1)  & 0.9513(5) & 14(5)      \\[1mm]
pc-2                         & 0.986(2) & 0.974(1) & 0.9569(4) & 16(5)      & 0.974(2) & 0.965(1) & 0.9515(4) & 16(3)      & 0.975(2) & 0.966(1) & 0.9474(4) & 14(4)      \\
aug-pc-2                     & 0.985(2) & 0.973(1) & 0.9563(5) & 17(5)      & 0.974(2) & 0.965(1) & 0.9516(5) & 16(4)      & 0.978(2) & 0.968(1) & 0.9485(5) & 14(4)      \\[1mm]
pcseg-2                      & 0.982(2) & 0.973(1) & 0.9566(6) & 17(5)      & 0.973(2) & 0.964(1) & 0.9512(6) & 15(4)      & 0.976(2) & 0.967(1) & 0.948(6)  & 14(5)      \\
aug-pcseg-2                  & 0.984(2) & 0.973(1) & 0.9562(6) & 17(5)      & 0.974(2) & 0.965(1) & 0.9515(6) & 16(4)      & 0.977(2) & 0.968(1) & 0.9487(6) & 14(4)      \\
    \bottomrule
    \end{tabular}}
\end{sidewaystable}

Augmenting the basis set with diffuse functions (on both hydrogen and non-hydrogen atoms) is essential to accurately predict dipole moments in computational quantum chemistry \cite{20ZaMc}, which are in turn essential in the calculation of vibrational intensities. Thus, generally we would strongly recommend using basis sets augmented with diffuse functions when superior vibrational intensity predictions are needed. We caution, however, that the addition of diffuse functions increases convergence issues, especially when using triple-zeta basis sets for systems where the diffuse basis functions' contribution to the overall electron density is negligible. If these  convergence issues arise, it is reasonable to retry the calculation without the diffuse functions and expect similar results in most cases.

It is important to know whether the addition of diffuse basis functions to the basis set changes the average scaling factor and median error; this data is detailed for each non-augmented/augmented basis set pair in \Cref{tab:diff_bases} averaged over all mGGA, hybrid and double-hybrid functionals considered. We find there is some variability in the scaling factor for the augmented vs unaugmented basis set  especially for double-zeta basis sets. No overall improvement in the median errors is found when augmenting the basis set with diffuse functions.

\section{Recommendations}
\label{sec:recommendations}

\begin{table}[h!]
    \centering
    \caption{Model chemistry recommendations for harmonic frequency calculations outlining the model chemistry class, optimised scaling factors, median true error performance (in \cm{}), percentage of outliers (\%\,Outs.), approximate reliability (Reliab.), and ranking for timings (Time). Numbers in parenthesis are one standard deviation in the last digit of the reported value.}
    \label{tab:recommendations}
    \scalebox{0.7}{
    \begin{tabular}{lcccccccc}
    \toprule
            \multirow{2}{*}{Model Chemistry$^{(a)}$} & \multirow{2}{*}{Class} & \multicolumn{3}{c}{Scaling Factors$^{(b)}$} & \multirow{2}{*}{Med. Error} & \multirow{2}{*}{\% Outs$^{(c)}$} & \multirow{2}{*}{Time$^{(d)}$} & \multirow{2}{*}{Reliab.$^{(e)}$} \\
            \cmidrule(r){3-5}
            &  & Low & Mid & High &  &  &  &  \\
    \midrule
            &  &  &  &  &  &  &  &  \\
            \multicolumn{9}{c}{\textbf{Very high-accuracy for frequencies and intensities}} \\
            &  &  &  &  &  &  &  &  \\
            \multicolumn{9}{l}{\textit{Constrain:} Triple-zeta basis sets augmented with diffuse functions (accurate dipole moments predictions and thus vibrational intensities)} \\
            \multicolumn{9}{p{23cm}}{\textit{Comments:} Dispersion-corrected DH functionals with superior performance in thermochemistry calculations. TZ basis sets with diffuse functions for superior vibrational intensity predictions. These model chemistries are not only more accurate but faster than other DH/TZ model chemistries.} \\
            &  &  &  &  &  &  &  &  \\
            DSD-PBEP86/def2-TZVPD    & DH/TZ      & 0.981(2)  & 0.972(1)  & 0.953(1)  & 7.6(7) & 7.0  & 4.5  & High  \\
            DSD-PBEP86/def2-TZVPPD   & DH/TZ      & 0.980(2)  & 0.971(1)  & 0.951(1)  & 7.6(7) & 7.2  & 5.0  & High  \\
            B2PLYP-D3(BJ)/def2-TZVPPD       & DH/TZ      & 0.990(2)  & 0.9750(6) & 0.9550(4) & 9.3(8) & 6.9  & 5.0  & High  \\
            B2PLYP-D3(BJ)/def2-TZVPD        & DH/TZ      & 0.990(2)  & 0.9757(5) & 0.9569(4) & 9.3(7) & 6.2  & 4.5  & High  \\
            &  &  &  &  &  &  &  &  \\
    \midrule
            &  &  &  &  &  &  &  &  \\
            \multicolumn{9}{c}{\textbf{Superior performance for frequencies and intensities}} \\
            &  &  &  &  &  &  &  &  \\
            \multicolumn{9}{l}{\textit{Constrain:} Hybrid functionals with augmented basis sets (better vibrational intensity predictions)}  \\
            \multicolumn{9}{p{23cm}}{\textit{Comments:} Hybrid functionals for faster calculations with satisfactory performance in thermochemical calculations and non-covalent interactions. Double- and triple-zeta bases augmented with diffuse functions for superior intensity predictions. Model chemistries ideal for large-scale calculations involving small- to medium-sized molecules.}                                                         \\
            &  &  &  &  &  &  &  &  \\
            B97-1/def2-TZVPD         & Hybrid/TZ  & 0.995(2) & 0.9792(7) & 0.9638(3) & 9.9(7)   & 5.8 & 3.0 & High   \\
            B97-1/def2-TZVPPD        & Hybrid/TZ  & 0.995(1) & 0.9785(7) & 0.9633(4) & 10.0(7)  & 6.3 & 3.0 & High   \\
            B97-1/6-31+G*            & Hybrid/DZ  & 0.996(2) & 0.9698(7) & 0.9583(3) & 10.6(8)  & 7.6 & 2.0 & High   \\
            TPSS0-D3(BJ)/def2-TZVPD         & Hybrid/TZ  & 0.970(2) & 0.9579(7) & 0.9476(3) & 11.3(8)  & 5.6 & 2.5 & Medium \\
            $\omega$B97X-D/6-31+G*          & Hybrid/DZ  & 0.970(2) & 0.953(1)  & 0.9475(5) & 12(1)    & 7.4 & 2.0 & Medium \\
            &  &  &  &  &  &  &  &  \\
    \midrule
            &  &  &  &  &  &  &  &  \\
            \multicolumn{9}{c}{\textbf{Routine/exploratory calculations}} \\
            &  &  &  &  &  &  &  &  \\
            \multicolumn{9}{l}{\textit{Constrain:} Hybrid and GGA functionals with unaugmented double-zeta basis sets} \\
            \multicolumn{9}{p{23cm}}{\textit{Comments:} Model chemistries ideal for routine and exploratory calculations where affordable   computational times are essential, e.g., larger molecular systems. Note that calculations with these model chemistries might require further manual assistance to allow proper convergence.} \\
            &  &  &  &  &  &  &  &  \\
            B97-1/6-31G*             & Hybrid/DZ  & 0.993(1) & 0.9631(6) & 0.9581(3) & 11(1)     & 7.7   & 1.5   & High      \\
            B97-1/pc-1               & Hybrid/DZ  & 0.998(2) & 0.9792(7) & 0.9601(3) & 11.2(7)   & 6.3   & 1.5   & High      \\
            B97-1/pcseg-1            & Hybrid/DZ  & 0.997(2) & 0.9791(7) & 0.957(1)  & 11.3(8)   & 6.5   & 1.5   & High      \\
            TPSS0-D3(BJ)/pcseg-1            & Hybrid/DZ  & 0.974(2) & 0.9555(7) & 0.9409(4) & 11.8(9)   & 5.9   & 1     & Low       \\
            TPSS0-D3(BJ)/pc-1               & Hybrid/DZ  & 0.974(2) & 0.9557(6) & 0.9412(4) & 11.9(9)   & 6.0   & 1     & Low       \\
            TPSS0-D3(BJ)/6-31G*             & Hybrid/DZ  & 0.968(2) & 0.9421(7) & 0.9394(3) & 12.1(9)   & 7.5   & 1     & Medium    \\
            $\omega$B97X-D/pc-1             & Hybrid/DZ  & 0.976(2) & 0.9642(8) & 0.9500(4) & 12(2)     & 7.1   & 1.5   & Medium    \\
            $\omega$B97X-D/6-31G*           & Hybrid/DZ  & 0.967(2) & 0.948(1)  & 0.9473(5) & 13(1)     & 7.0   & 1.5   & Medium    \\
            B97-D3(BJ)/def2-SVP             & GGA/DZ     & 1.017(2) & 1.002(1)  & 0.978(1)  & 14(1)     & 12.0  & 1.5   & High      \\
            PBE/6-31G*               & GGA/DZ     & 1.026(1) & 0.9951(8) & 0.9819(5) & 15(1)     & 11.0  & 1.5   & High      \\
            &  &  &  &  &  &  &  &  \\
    \bottomrule
    \end{tabular}}
        \begin{tablenotes}
        \tiny{\item[] (a) Note that diffuse functions can cause convergence issues for molecules where they are not needed to describe the electron density; in this case, removing the diffuse functions from the basis set should achieve very similar performance.
        \item (b) Frequency domains defined as follows: low-frequency ($<$\,1,000\,\cm{}), Mid-frequency (1,000--2,000\,\cm{}), and High-frequency ($\geq$\,2,000\,\cm{}).
        \item (c) Qualitative measure of the prevalence of outliers (see \Cref{fig:outs_heatmap}) as recorded from our calculations.
        \item (d) Approximate computational timings where 1 indicates very cheap and 5 very expensive calculations.
        \item (e) Qualitative metric of expected reliability based on the number of failed calculations in our approach. Note that this metric is only representative of the calculations performed on the VIBFREQ1295 database and it might change in other cases.}
        \end{tablenotes}
\end{table}

We strongly recommend the use of frequency-range specific (not global) and model-chemistry-specific (not universal) scaling factors when aiming for high-quality prediction of molecular fundamental frequencies through scaling of calculated harmonic vibrational frequencies within the double harmonic approximation. 

Taking into consideration the median true error performance, prevalence of outliers and the overall reliability for the model chemistries featured in this study, we present in \Cref{tab:recommendations} our model chemistry recommendations for these scaled harmonic frequency calculations. The table highlights the expected errors, some statistical metrics, as well as the most promising applications for each model chemistry featured. We categorised our recommendations into three main groups organised from the most to least-intensive computationally demanding: (1) very high-accuracy frequency and intensity predictions, (2) superior frequency and intensity predictions, and (3) routine or exploratory calculations.

For very high-accuracy frequency and intensity predictions, our recommendations focus on model chemistries belonging to the double-hybrid/triple-zeta basis set class. In particular, we recommend model chemistries merging the DSD-PBEP86 and B2PLYP-D3(BJ) double-hybrid functionals with the def2-TZVPD and def2-TZVPPD triple-zeta basis sets, given their outstanding performance in harmonic frequency calculations, with median true errors approaching the anharmonicity error at 7.5\,\cm{}. We expect these recommendations to be highly reliable in terms of their likelihood of unconverged calculations and display minimal outliers in routine calculations. Note that we chose the augmented version with diffuse functions of these basis sets to ensure accurate dipole moment calculations and thus superior transition vibrational intensity predictions. In principle, the aug-cc-pVTZ, aug-pc-2, and aug-pcseg-2 triple-zeta basis sets could also be recommended in this category due to their highly similar performances (see \Cref{fig:allmc_perf}). However, the def2-$n$ triple-zeta basis sets are about 25\,\% smaller than their Jensen and Dunning counterparts, making them almost 3 times faster as recorded from our calculations. We therefore warn the user about the significant increase in computational timings if the triple-zeta Dunning and Jensen basis sets are chosen for harmonic frequency calculations in this category.

Our second category in the table concerns model chemistries for superior frequency and intensity predictions. Here, we constrained our recommendations to hybrid functionals with double- and triple-zeta augmented basis sets to ensure faster yet reliable options; use of double-zeta basis sets will significantly reduce computational timings. These model chemistries in this category not only display median errors below 12\,\cm{}, but they also highlight for their moderate reliability and expected prevalence of outliers. The appealing accuracy and reasonable computational timings for these recommendations make them attractive options for large-scale harmonic frequency calculations, such as those seeking to produce approximate vibrational spectral data for thousands of molecules rapidly \cite{19SoPeSe,21ZaSyRo}. We note that the B97-1 functional, which is the strongly performing hybrid functional for this property, does not have dispersion corrections easily available in modern quantum chemistry packages currently; we strongly recommend these be added for future updates in order to improve vibrational frequency prediction accuracy for larger molecules. 

Finally, we also provide recommendations for routine and exploratory calculations, using less-demanding yet accurate model chemistries. These recommendations were mostly constrained to hybrid functionals with unaugmented double-zeta basis sets. Median errors below 12\,\cm{} can be observed in most cases at a surprisingly low computational cost. However, further manual assistance in the input files, e.g. stronger initial molecular geometries, loosening symmetry constrains, might be needed to ensure the proper convergence of calculations involving some of these model chemistries as highlighted in the reliability column in the table. We also included top-performing GGA/DZ model chemistries within these recommendations.

\begin{figure}
    \centering
    \includegraphics[width=0.75\textwidth]{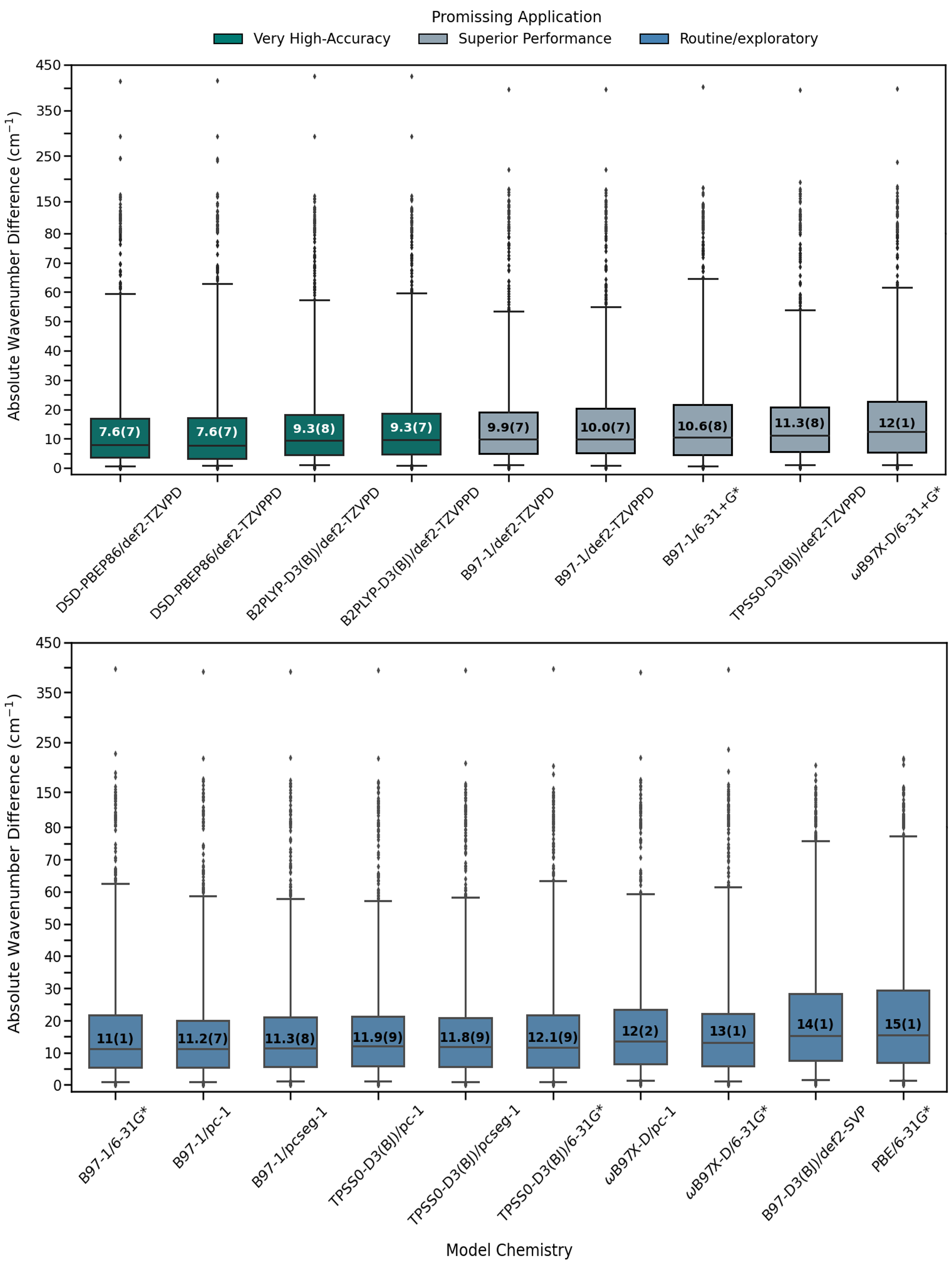}
    \caption{Absolute wavenumber difference (in \cm{}) between the scaled harmonic and experimental fundamental frequencies for our model chemistry recommendations: in green, model chemistries for very high-accuracy frequency and intensity predictions; in grey, model chemistries for superior frequency and intensity predictions; and, in blue, recommendations for routine and exploratory calculations. The numbers inside the boxes represent the median true error (in \cm{}) for the model chemistries considered. The bottom and top whiskers in the figure encapsulate 5 and 95\,\% of the data, respectively. Numbers in parenthesis are one standard deviation in the last digit of the reported value. Note that we changed the scale of the absolute  wavenumber difference axis for data points above 80\,\cm{} to allow readability of both the boxes and outliers for each model chemistry considered.}
    \label{fig:box_recommendations}
\end{figure}

Using a box-and-whisker representation, \Cref{fig:box_recommendations} presents the distribution of errors (absolute wavenumber difference) between the scaled harmonic and experimental fundamental frequencies for all model chemistry recommendations presented in \Cref{tab:recommendations} hued by their different potential applications. The box limits in the figure lie at 25\% and 75\% while the whisker limits lie at 5\% and 95\% of the data respectively. The fact that the 95\% whisker is usually more than three times the error of the 75\% upper box limit demonstrates that though there is the potential for vibrational frequencies to be very poorly represented within any harmonic approximation, in general the performance is far stronger than a RMSE value would predict. This distribution justifies our use of median errors throughout this manuscript when assessing model chemistry performance.

It is also important to note from \Cref{fig:box_recommendations} that, regardless of the model chemistry choice, large outliers are inherently present in routine scaled harmonic frequency calculations. In particular, our recommendations display outliers with deviations larger than 400\,\cm{} from experiment for problematic open-shell molecules and some low-frequency vibrational modes.

Our recommendations deliberately do not include the popular B3LYP/6-31G* model chemistry. Though its performance for harmonic vibrational frequency predictions is reasonable (median true error of $\sim$\,13\,\cm{}), there have been a large body of work demonstrating the model chemistry's poor performance for important properties, such as thermochemistry predictions \cite{12KrGoGr,17MaHe,17GoHaBa,19GoMe}, which are often calculated alongside vibrational frequencies. We strongly recommend the more robust $\omega$B97X-D/6-31G* model chemistry be used in preference to B3LYP/6-31G*, though for optimal performance in harmonic frequency calculations, B97-1 and TPSS0-D3(BJ) are better choices of functionals. Our routine and exploratory recommendations in \Cref{tab:recommendations} not only display an attractive performance in harmonic frequency calculations but represent solid options in the calculation of general thermochemistry properties \cite{17MaHe,17GoHaBa}. We strongly encourage the reader to use these model chemistry recommendations in preference to the outdated B3LYP/6-31G* model chemistry.

\section{Concluding Remarks and Future Directions}
\label{sec:conclusions}

Use of scaled harmonic vibrational frequency calculations to predict experimental fundamental frequencies is widespread across chemistry and thus reliable model chemistry recommendations (i.e., a level of theory and basis set pair), highlighting expected errors and limitations, are pivotal to ensure their appropriate use in computational quantum chemistry. Here, we have assessed the performance of over 600 model chemistries spanning 23 levels of theory and 27 basis sets using the VIBFREQ1295 database \cite{22ZaMc_VIBFREQ} containing 1,295 experimental fundamental frequencies and high-level \abinitio{} harmonic frequencies for 141 organic-like molecules. Our recommendations from this analysis represent solid model chemistry options with median true errors below 12\,\cm{} and approaching the lower-bound in performance set by the anharmonicity error at 7.5\,\cm{}. Not only we expect superior agreement between the scaled harmonic frequencies computed with these model chemistries and the experimental fundamental frequencies for a given molecule, but minimal prevalence of outliers, unconverged molecular geometries, and imaginary frequencies for most well-behaved organic-like molecules. We present these model chemistry recommendations in \Cref{tab:recommendations}, along with their scaling factor values, expected errors, and potential applications.

Scaled harmonic frequency calculations also play a crucial role in the calculation of thermodynamic corrections \cite{20SpGr,20PrGrGr,22Ka}, e.g., absolute molecular entropies and zero-point energies, and, though not explicitly evaluated here, we expect our recommendations to potentially be suitable for this application as well.

Noting the pivotal role of scaling factors in harmonic frequency calculations, our analysis of the different scaling factor types further confirms the importance of implementing frequency-range-specific scaling factors for each model chemistry individually, to achieve superior performance. The values for the frequency coverage outlined in \Cref{tab:frange_ccsdt} represent appropriate thresholds for distinguishing between the low- ($<$\,1,000\,\cm{}), mid- (1,000--2,000\,\cm{}), and high-frequency ($>$\,2,000\,\cm{}) ranges of the vibrational spectrum, grouping similar vibrational modes together \cite{22ZaMc_VIBFREQ}. Using three-group scaling substantially reduces errors over a global scaling approach and should always be used.

Two main sources of error contribute towards the deviation of the calculated harmonic frequencies from the experimental fundamental frequencies: (1) the lack of anharmonicity in the calculations, and (2) the model chemistry choice.  The lack of anharmonicity has the largest contribution towards this deviation (7.5\,\cm{} on average) and is irreducible within the harmonic approximation. Thus, the model chemistry choice should be properly selected to minimise model chemistry error as much as possible, as inappropriate level of theory and basis set combinations can easily double the expected errors.

We also examined the prevalence of outliers (i.e. very erroneous predictions) and the reliability of a given model chemistry (i.e. how many failed calculations). Our results show that large deviations from experiment (more than 50\,\cm{}) cannot be avoided in routine harmonic frequency calculations (as some frequencies are inherently more anharmonic than typical), but they can be minimised when selecting an appropriate model chemistry choice. Likewise, the model chemistry choice, in particular the level of theory choice, largely influences the amount of failed calculations due to unconverged initial molecular geometries, or the prevalence of spurious data such as imaginary frequencies. For linear and highly-symmetric molecules, all model chemistries considered struggled to predict the right number of harmonic vibrational frequencies. We therefore warn the user about this limitation and recommend proceeding with caution in calculations involving such molecules. Our recommendations in \Cref{tab:recommendations} not only represent accurate options in terms of their median true errors, but mitigate to their best extent these limitations.

The most natural and important follow-up to this work is the comprehensive benchmarking of model chemistry options in anharmonic vibrational frequency calculations, e.g., VSCF, VCI, and VPT2, which are becoming increasingly more affordable and popular. We expect that superior accuracy in the computed fundamental frequencies may be achieved when implementing anharmonic approaches (as anharmonicity error can be removed), but this needs to be demonstrated. We note recent publications have identified very large errors in some computed frequencies, e.g., 100s of wavenumbers for P-H stretches in \ce{PH3} \cite{21ZaSyRo,20BaCeFu}, resulting from problematic high-order derivatives for some functionals (including $\omega$B97X-D). If these issues can be reliably avoided, however, anharmonic calculations promise not only higher accuracy  but the prediction of frequencies and intensities of non-fundamental infrared spectral bands not present in harmonic frequency calculations \cite{86Bo,96JuGe,08BoCaMe,10ScLaBe,14BaBiBl,15BaBiPu,18BiBlPu} that could have non-negligible intensities, especially for higher temperature systems. 

\begin{acknowledgement}

This research was undertaken with the assistance of resources from the National Computational Infrastructure (NCI Australia), an NCRIS enabled capability supported by the Australian Government.

\end{acknowledgement}

\begin{suppinfo}

The data underlying this study are available in the published article and its online supplementary material. 

We provide the csv files containing all scaling factors for the model chemistries featured in this study, as well as the corresponding statistical metrics and scaled harmonic frequencies.

We also provide a PDF file describing further analysis in this study concerning the global scaling approach, further details into failed calculations, and samples of the input files used to run the calculations.  We encourage the reader to look into the supplementary material for a more detailed description of the files and data provided.

\end{suppinfo}


\bibliography{references_harmbench}

\providecommand{\latin}[1]{#1}
\makeatletter
\providecommand{\doi}
  {\begingroup\let\do\@makeother\dospecials
  \catcode`\{=1 \catcode`\}=2 \doi@aux}
\providecommand{\doi@aux}[1]{\endgroup\texttt{#1}}
\makeatother
\providecommand*\mcitethebibliography{\thebibliography}
\csname @ifundefined\endcsname{endmcitethebibliography}
  {\let\endmcitethebibliography\endthebibliography}{}
\begin{mcitethebibliography}{108}
\providecommand*\natexlab[1]{#1}
\providecommand*\mciteSetBstSublistMode[1]{}
\providecommand*\mciteSetBstMaxWidthForm[2]{}
\providecommand*\mciteBstWouldAddEndPuncttrue
  {\def\EndOfBibitem{\unskip.}}
\providecommand*\mciteBstWouldAddEndPunctfalse
  {\let\EndOfBibitem\relax}
\providecommand*\mciteSetBstMidEndSepPunct[3]{}
\providecommand*\mciteSetBstSublistLabelBeginEnd[3]{}
\providecommand*\EndOfBibitem{}
\mciteSetBstSublistMode{f}
\mciteSetBstMaxWidthForm{subitem}{(\alph{mcitesubitemcount})}
\mciteSetBstSublistLabelBeginEnd
  {\mcitemaxwidthsubitemform\space}
  {\relax}
  {\relax}

\bibitem[Hait and Head-Gordon(2018)Hait, and Head-Gordon]{18HaHe1}
Hait,~D.; Head-Gordon,~M. {How Accurate Is Density Functional Theory at
  Predicting Dipole Moments? An Assessment Using a New Database of 200
  Benchmark Values}. \emph{Journal of Chemical Theory and Computation}
  \textbf{2018}, \emph{14}, 1969--1981\relax
\mciteBstWouldAddEndPuncttrue
\mciteSetBstMidEndSepPunct{\mcitedefaultmidpunct}
{\mcitedefaultendpunct}{\mcitedefaultseppunct}\relax
\EndOfBibitem
\bibitem[Zapata and McKemmish(2020)Zapata, and McKemmish]{20ZaMc}
Zapata,~J.~C.; McKemmish,~L.~K. {Computation of Dipole Moments: A
  Recommendation on the Choice of the Basis Set and the Level of Theory}.
  \emph{The Journal of Physical Chemistry A} \textbf{2020}, \emph{124},
  7538--7548\relax
\mciteBstWouldAddEndPuncttrue
\mciteSetBstMidEndSepPunct{\mcitedefaultmidpunct}
{\mcitedefaultendpunct}{\mcitedefaultseppunct}\relax
\EndOfBibitem
\bibitem[Hait and Head-Gordon(2018)Hait, and Head-Gordon]{18HaHe2}
Hait,~D.; Head-Gordon,~M. {How accurate are static polarizability predictions
  from density functional theory? An assessment over 132 species at equilibrium
  geometry}. \emph{Physical Chemistry Chemical Physics} \textbf{2018},
  \emph{20}, 19800--19810\relax
\mciteBstWouldAddEndPuncttrue
\mciteSetBstMidEndSepPunct{\mcitedefaultmidpunct}
{\mcitedefaultendpunct}{\mcitedefaultseppunct}\relax
\EndOfBibitem
\bibitem[Mardirossian and Head-Gordon(2017)Mardirossian, and
  Head-Gordon]{17MaHe}
Mardirossian,~N.; Head-Gordon,~M. {Thirty years of density functional theory in
  computational chemistry: An overview and extensive assessment of 200 density
  functionals}. \emph{Molecular Physics} \textbf{2017}, \emph{115},
  2315--2372\relax
\mciteBstWouldAddEndPuncttrue
\mciteSetBstMidEndSepPunct{\mcitedefaultmidpunct}
{\mcitedefaultendpunct}{\mcitedefaultseppunct}\relax
\EndOfBibitem
\bibitem[Goerigk \latin{et~al.}(2017)Goerigk, Hansen, Bauer, Ehrlich, Najibi,
  and Grimme]{17GoHaBa}
Goerigk,~L.; Hansen,~A.; Bauer,~C.; Ehrlich,~S.; Najibi,~A.; Grimme,~S. {A look
  at the density functional theory zoo with the advanced GMTKN55 database for
  general main group thermochemistry, kinetics and noncovalent interactions}.
  \emph{Physical Chemistry Chemical Physics} \textbf{2017}, \emph{19},
  32184--32215\relax
\mciteBstWouldAddEndPuncttrue
\mciteSetBstMidEndSepPunct{\mcitedefaultmidpunct}
{\mcitedefaultendpunct}{\mcitedefaultseppunct}\relax
\EndOfBibitem
\bibitem[Gryn'ova \latin{et~al.}(2013)Gryn'ova, Marshall, Blanksby, and
  Coote]{13GrMaBl}
Gryn'ova,~G.; Marshall,~D.~L.; Blanksby,~S.~J.; Coote,~M.~L. {Switching radical
  stability by pH-induced orbital conversion}. \emph{Nature Chemistry}
  \textbf{2013}, \emph{5}, 474--481\relax
\mciteBstWouldAddEndPuncttrue
\mciteSetBstMidEndSepPunct{\mcitedefaultmidpunct}
{\mcitedefaultendpunct}{\mcitedefaultseppunct}\relax
\EndOfBibitem
\bibitem[Zhang \latin{et~al.}(2020)Zhang, Yu, Zhang, Jiang, Li, Hu, Li, Zhao,
  Wang, Xie, Zhang, Dai, Wu, Zhang, Jiang, Li, and Yang]{20ZhYuZh}
Zhang,~B. \latin{et~al.}  {Infrared spectroscopy of neutral water clusters at
  finite temperature: Evidence for a noncyclic pentamer}. \emph{Proceedings of
  the National Academy of Sciences} \textbf{2020}, \emph{117},
  15423--15428\relax
\mciteBstWouldAddEndPuncttrue
\mciteSetBstMidEndSepPunct{\mcitedefaultmidpunct}
{\mcitedefaultendpunct}{\mcitedefaultseppunct}\relax
\EndOfBibitem
\bibitem[Long \latin{et~al.}(2021)Long, Wang, Xia, He, Bao, and
  Truhlar]{21LoWaXi}
Long,~B.; Wang,~Y.; Xia,~Y.; He,~X.; Bao,~J.~L.; Truhlar,~D.~G. {Atmospheric
  Kinetics: Bimolecular Reactions of Carbonyl Oxide by a Triple-Level
  Strategy}. \emph{Journal of the American Chemical Society} \textbf{2021},
  \emph{143}, 8402--8413\relax
\mciteBstWouldAddEndPuncttrue
\mciteSetBstMidEndSepPunct{\mcitedefaultmidpunct}
{\mcitedefaultendpunct}{\mcitedefaultseppunct}\relax
\EndOfBibitem
\bibitem[Ning and Truhlar(2021)Ning, and Truhlar]{21NiTr}
Ning,~J.; Truhlar,~D.~G. {Spin-Orbit Coupling Changes the Identity of the
  Hyper-Open-Shell Ground State of Ce+, and the Bond Dissociation Energy of
  CeH+Proves to Be Challenging for Theory}. \emph{Journal of Chemical Theory
  and Computation} \textbf{2021}, \emph{17}, 1421--1434\relax
\mciteBstWouldAddEndPuncttrue
\mciteSetBstMidEndSepPunct{\mcitedefaultmidpunct}
{\mcitedefaultendpunct}{\mcitedefaultseppunct}\relax
\EndOfBibitem
\bibitem[Zhang \latin{et~al.}(2022)Zhang, Yu, Zhang, Jiang, Li, Hu, Li, Zhao,
  Wang, Xie, Zhang, Dai, Wu, Zhang, Jiang, Li, and Yang]{22LoMaLa}
Zhang,~B. \latin{et~al.}  {The stability of covalent dative bond significantly
  increases with increasing solvent polarity}. \emph{Nature Communications}
  \textbf{2022}, \emph{13}, 2107\relax
\mciteBstWouldAddEndPuncttrue
\mciteSetBstMidEndSepPunct{\mcitedefaultmidpunct}
{\mcitedefaultendpunct}{\mcitedefaultseppunct}\relax
\EndOfBibitem
\bibitem[Scott and Radom(1996)Scott, and Radom]{96ScRa}
Scott,~A.~P.; Radom,~L. {Harmonic vibrational frequencies: An evaluation of
  Hartree-Fock, M{\o}ller-Plesset, quadratic configuration interaction, density
  functional theory, and semiempirical scale factors}. \emph{Journal of
  Physical Chemistry} \textbf{1996}, \emph{100}, 16502--16513\relax
\mciteBstWouldAddEndPuncttrue
\mciteSetBstMidEndSepPunct{\mcitedefaultmidpunct}
{\mcitedefaultendpunct}{\mcitedefaultseppunct}\relax
\EndOfBibitem
\bibitem[Zapata~Trujillo and McKemmish(2021)Zapata~Trujillo, and
  McKemmish]{21ZaMc}
Zapata~Trujillo,~J.~C.; McKemmish,~L.~K. {Meta‐analysis of uniform scaling
  factors for harmonic frequency calculations}. \emph{WIREs Computational
  Molecular Science} \textbf{2021}, e1584\relax
\mciteBstWouldAddEndPuncttrue
\mciteSetBstMidEndSepPunct{\mcitedefaultmidpunct}
{\mcitedefaultendpunct}{\mcitedefaultseppunct}\relax
\EndOfBibitem
\bibitem[Pople \latin{et~al.}(1993)Pople, Scott, Wong, and Radom]{93PoScWo}
Pople,~J.~A.; Scott,~A.~P.; Wong,~M.~W.; Radom,~L. {Scaling Factors for
  Obtaining Fundamental Vibrational Frequencies and Zero‐Point Energies from
  HF/6–31G* and MP2/6–31G* Harmonic Frequencies}. \emph{Israel Journal of
  Chemistry} \textbf{1993}, \emph{33}, 345--350\relax
\mciteBstWouldAddEndPuncttrue
\mciteSetBstMidEndSepPunct{\mcitedefaultmidpunct}
{\mcitedefaultendpunct}{\mcitedefaultseppunct}\relax
\EndOfBibitem
\bibitem[H{\"{a}}ttig \latin{et~al.}(2010)H{\"{a}}ttig, Tew, and
  K{\"{o}}hn]{10HaTeKo}
H{\"{a}}ttig,~C.; Tew,~D.~P.; K{\"{o}}hn,~A. {Communications: Accurate and
  efficient approximations to explicitly correlated coupled-cluster singles and
  doubles, CCSD-F12}. \emph{Journal of Chemical Physics} \textbf{2010},
  \emph{132}, 231102\relax
\mciteBstWouldAddEndPuncttrue
\mciteSetBstMidEndSepPunct{\mcitedefaultmidpunct}
{\mcitedefaultendpunct}{\mcitedefaultseppunct}\relax
\EndOfBibitem
\bibitem[Peterson \latin{et~al.}(2008)Peterson, Adler, and Werner]{08PeAdWe}
Peterson,~K.~A.; Adler,~T.~B.; Werner,~H.~J. {Systematically convergent basis
  sets for explicitly correlated wavefunctions: The atoms H, He, B-Ne, and
  Al-Ar}. \emph{Journal of Chemical Physics} \textbf{2008}, \emph{128},
  084102\relax
\mciteBstWouldAddEndPuncttrue
\mciteSetBstMidEndSepPunct{\mcitedefaultmidpunct}
{\mcitedefaultendpunct}{\mcitedefaultseppunct}\relax
\EndOfBibitem
\bibitem[Hill and Peterson(2010)Hill, and Peterson]{10HiPe}
Hill,~J.~G.; Peterson,~K.~A. {Correlation consistent basis sets for explicitly
  correlated wavefunctions: Valence and core-valence basis sets for Li, Be, Na,
  and Mg}. \emph{Physical Chemistry Chemical Physics} \textbf{2010}, \emph{12},
  10460--10468\relax
\mciteBstWouldAddEndPuncttrue
\mciteSetBstMidEndSepPunct{\mcitedefaultmidpunct}
{\mcitedefaultendpunct}{\mcitedefaultseppunct}\relax
\EndOfBibitem
\bibitem[Zapata~Trujillo and McKemmish(2022)Zapata~Trujillo, and
  McKemmish]{22ZaMc_VIBFREQ}
Zapata~Trujillo,~J.~C.; McKemmish,~L.~K. {VIBFREQ1295: A New Database for
  Vibrational Frequency Calculations}. \emph{The Journal of Physical Chemistry
  A} \textbf{2022}, \emph{2022}, 4100--4122\relax
\mciteBstWouldAddEndPuncttrue
\mciteSetBstMidEndSepPunct{\mcitedefaultmidpunct}
{\mcitedefaultendpunct}{\mcitedefaultseppunct}\relax
\EndOfBibitem
\bibitem[Boussessi \latin{et~al.}(2020)Boussessi, Ceselin, Tasinato, and
  Barone]{20BoCeTa}
Boussessi,~R.; Ceselin,~G.; Tasinato,~N.; Barone,~V. {DFT meets the segmented
  polarization consistent basis sets: Performances in the computation of
  molecular structures, rotational and vibrational spectroscopic properties}.
  \emph{Journal of Molecular Structure} \textbf{2020}, \emph{1208},
  127886\relax
\mciteBstWouldAddEndPuncttrue
\mciteSetBstMidEndSepPunct{\mcitedefaultmidpunct}
{\mcitedefaultendpunct}{\mcitedefaultseppunct}\relax
\EndOfBibitem
\bibitem[Barone \latin{et~al.}(2020)Barone, Ceselin, Fus{\`{e}}, and
  Tasinato]{20BaCeFu}
Barone,~V.; Ceselin,~G.; Fus{\`{e}},~M.; Tasinato,~N. {Accuracy Meets
  Interpretability for Computational Spectroscopy by Means of Hybrid and
  Double-Hybrid Functionals}. \emph{Frontiers in Chemistry} \textbf{2020},
  \emph{8}, 584203\relax
\mciteBstWouldAddEndPuncttrue
\mciteSetBstMidEndSepPunct{\mcitedefaultmidpunct}
{\mcitedefaultendpunct}{\mcitedefaultseppunct}\relax
\EndOfBibitem
\bibitem[Goerigk and Mehta(2019)Goerigk, and Mehta]{19GoMe}
Goerigk,~L.; Mehta,~N. {A Trip to the Density Functional Theory Zoo: Warnings
  and Recommendations for the User}. \emph{Australian Journal of Chemistry}
  \textbf{2019}, \emph{72}, 563--573\relax
\mciteBstWouldAddEndPuncttrue
\mciteSetBstMidEndSepPunct{\mcitedefaultmidpunct}
{\mcitedefaultendpunct}{\mcitedefaultseppunct}\relax
\EndOfBibitem
\bibitem[Nagy and Jensen(2017)Nagy, and Jensen]{17NaJe}
Nagy,~B.; Jensen,~F. \emph{Reviews in Computational Quantum Chemistry}; John
  Wiley {\&} Sons, Ltd, 2017; pp 93--149\relax
\mciteBstWouldAddEndPuncttrue
\mciteSetBstMidEndSepPunct{\mcitedefaultmidpunct}
{\mcitedefaultendpunct}{\mcitedefaultseppunct}\relax
\EndOfBibitem
\bibitem[Haunschild \latin{et~al.}(2016)Haunschild, Barth, and Marx]{16HaBaMa}
Haunschild,~R.; Barth,~A.; Marx,~W. {Evolution of DFT studies in view of a
  scientometric perspective}. \emph{Journal of Cheminformatics} \textbf{2016},
  \emph{8}, 52\relax
\mciteBstWouldAddEndPuncttrue
\mciteSetBstMidEndSepPunct{\mcitedefaultmidpunct}
{\mcitedefaultendpunct}{\mcitedefaultseppunct}\relax
\EndOfBibitem
\bibitem[Haunschild \latin{et~al.}(2019)Haunschild, Barth, and
  French]{19HaBaFr}
Haunschild,~R.; Barth,~A.; French,~B. {A comprehensive analysis of the history
  of DFT based on the bibliometric method RPYS}. \emph{Journal of
  Cheminformatics} \textbf{2019}, \emph{11}, 72\relax
\mciteBstWouldAddEndPuncttrue
\mciteSetBstMidEndSepPunct{\mcitedefaultmidpunct}
{\mcitedefaultendpunct}{\mcitedefaultseppunct}\relax
\EndOfBibitem
\bibitem[Pritchard \latin{et~al.}(2019)Pritchard, Altarawy, Didier, Gibson, and
  Windus]{19PrAlDi}
Pritchard,~B.~P.; Altarawy,~D.; Didier,~B.; Gibson,~T.~D.; Windus,~T.~L. {New
  Basis Set Exchange: An Open, Up-to-Date Resource for the Molecular Sciences
  Community}. \emph{Journal of Chemical Information and Modeling}
  \textbf{2019}, \emph{59}, 4814--4820\relax
\mciteBstWouldAddEndPuncttrue
\mciteSetBstMidEndSepPunct{\mcitedefaultmidpunct}
{\mcitedefaultendpunct}{\mcitedefaultseppunct}\relax
\EndOfBibitem
\bibitem[Roothaan(1951)]{51Ro}
Roothaan,~C.~C. {New developments in molecular orbital theory}. \emph{Reviews
  of Modern Physics} \textbf{1951}, \emph{23}, 69--89\relax
\mciteBstWouldAddEndPuncttrue
\mciteSetBstMidEndSepPunct{\mcitedefaultmidpunct}
{\mcitedefaultendpunct}{\mcitedefaultseppunct}\relax
\EndOfBibitem
\bibitem[Grimme(2006)]{06Gr}
Grimme,~S. {Semiempirical GGA-type density functional constructed with a
  long-range dispersion correction}. \emph{Journal of Computational Chemistry}
  \textbf{2006}, \emph{27}, 1787--1799\relax
\mciteBstWouldAddEndPuncttrue
\mciteSetBstMidEndSepPunct{\mcitedefaultmidpunct}
{\mcitedefaultendpunct}{\mcitedefaultseppunct}\relax
\EndOfBibitem
\bibitem[Grimme \latin{et~al.}(2011)Grimme, Ehrlich, and Goerigk]{11GrEhGo}
Grimme,~S.; Ehrlich,~S.; Goerigk,~L. {Effect of the damping function in
  dispersion corrected density functional theory}. \emph{Journal of
  Computational Chemistry} \textbf{2011}, \emph{32}, 1456--1465\relax
\mciteBstWouldAddEndPuncttrue
\mciteSetBstMidEndSepPunct{\mcitedefaultmidpunct}
{\mcitedefaultendpunct}{\mcitedefaultseppunct}\relax
\EndOfBibitem
\bibitem[Becke(1988)]{88Be}
Becke,~A.~D. {Density-functional exchange-energy approximation with correct
  asymptotic behavior}. \emph{Physical Review A} \textbf{1988}, \emph{38},
  3098--3100\relax
\mciteBstWouldAddEndPuncttrue
\mciteSetBstMidEndSepPunct{\mcitedefaultmidpunct}
{\mcitedefaultendpunct}{\mcitedefaultseppunct}\relax
\EndOfBibitem
\bibitem[Lee \latin{et~al.}(1988)Lee, Yang, and Parr]{88LeYaPa}
Lee,~C.; Yang,~W.; Parr,~R.~G. {Development of the Colle-Salvetti
  correlation-energy formula into a functional of the electron density}.
  \emph{Physical Review B} \textbf{1988}, \emph{37}, 785--789\relax
\mciteBstWouldAddEndPuncttrue
\mciteSetBstMidEndSepPunct{\mcitedefaultmidpunct}
{\mcitedefaultendpunct}{\mcitedefaultseppunct}\relax
\EndOfBibitem
\bibitem[Miehlich \latin{et~al.}(1989)Miehlich, Savin, Stoll, and
  Preuss]{89MiSaSt}
Miehlich,~B.; Savin,~A.; Stoll,~H.; Preuss,~H. {Results obtained with the
  correlation energy density functionals of becke and Lee, Yang and Parr}.
  \emph{Chemical Physics Letters} \textbf{1989}, \emph{157}, 200--206\relax
\mciteBstWouldAddEndPuncttrue
\mciteSetBstMidEndSepPunct{\mcitedefaultmidpunct}
{\mcitedefaultendpunct}{\mcitedefaultseppunct}\relax
\EndOfBibitem
\bibitem[Perdew \latin{et~al.}(1996)Perdew, Burke, and Ernzerhof]{96PeBuEr}
Perdew,~J.~P.; Burke,~K.; Ernzerhof,~M. {Generalized gradient approximation
  made simple}. \emph{Physical Review Letters} \textbf{1996}, \emph{77},
  3865--3868\relax
\mciteBstWouldAddEndPuncttrue
\mciteSetBstMidEndSepPunct{\mcitedefaultmidpunct}
{\mcitedefaultendpunct}{\mcitedefaultseppunct}\relax
\EndOfBibitem
\bibitem[Zhao and Truhlar(2006)Zhao, and Truhlar]{06ZhTr}
Zhao,~Y.; Truhlar,~D.~G. {A new local density functional for main-group
  thermochemistry, transition metal bonding, thermochemical kinetics, and
  noncovalent interactions}. \emph{Journal of Chemical Physics} \textbf{2006},
  \emph{125}, 194101\relax
\mciteBstWouldAddEndPuncttrue
\mciteSetBstMidEndSepPunct{\mcitedefaultmidpunct}
{\mcitedefaultendpunct}{\mcitedefaultseppunct}\relax
\EndOfBibitem
\bibitem[Goerigk and Grimme(2011)Goerigk, and Grimme]{11GoGr}
Goerigk,~L.; Grimme,~S. {A thorough benchmark of density functional methods for
  general main group thermochemistry, kinetics, and noncovalent interactions}.
  \emph{Physical Chemistry Chemical Physics} \textbf{2011}, \emph{13},
  6670--6688\relax
\mciteBstWouldAddEndPuncttrue
\mciteSetBstMidEndSepPunct{\mcitedefaultmidpunct}
{\mcitedefaultendpunct}{\mcitedefaultseppunct}\relax
\EndOfBibitem
\bibitem[Becke(1993)]{93Be}
Becke,~A.~D. {Density-functional thermochemistry. III. The role of exact
  exchange}. \emph{The Journal of Chemical Physics} \textbf{1993}, \emph{98},
  5648--5652\relax
\mciteBstWouldAddEndPuncttrue
\mciteSetBstMidEndSepPunct{\mcitedefaultmidpunct}
{\mcitedefaultendpunct}{\mcitedefaultseppunct}\relax
\EndOfBibitem
\bibitem[Stephens \latin{et~al.}(1994)Stephens, Devlin, Chabalowski, and
  Frisch]{94StDeCh}
Stephens,~P.~J.; Devlin,~F.~J.; Chabalowski,~C.~F.; Frisch,~M.~J. {Ab Initio
  calculation of vibrational absorption and circular dichroism spectra using
  density functional force fields}. \emph{Journal of Physical
  Chemistry{\textregistered}} \textbf{1994}, \emph{98}, 11623--11627\relax
\mciteBstWouldAddEndPuncttrue
\mciteSetBstMidEndSepPunct{\mcitedefaultmidpunct}
{\mcitedefaultendpunct}{\mcitedefaultseppunct}\relax
\EndOfBibitem
\bibitem[Hamprecht \latin{et~al.}(1998)Hamprecht, Cohen, Tozer, and
  Handy]{98HaCoTo}
Hamprecht,~F.~A.; Cohen,~A.~J.; Tozer,~D.~J.; Handy,~N.~C. {Development and
  assessment of new exchange-correlation functionals}. \emph{Journal of
  Chemical Physics} \textbf{1998}, \emph{109}, 6264--6271\relax
\mciteBstWouldAddEndPuncttrue
\mciteSetBstMidEndSepPunct{\mcitedefaultmidpunct}
{\mcitedefaultendpunct}{\mcitedefaultseppunct}\relax
\EndOfBibitem
\bibitem[Wilson \latin{et~al.}(2001)Wilson, Bradley, and Tozer]{01WiBrTo}
Wilson,~P.~J.; Bradley,~T.~J.; Tozer,~D.~J. {Hybrid exchange-correlation
  functional determined from thermochemical data and ab initio potentials}.
  \emph{Journal of Chemical Physics} \textbf{2001}, \emph{115},
  9233--9242\relax
\mciteBstWouldAddEndPuncttrue
\mciteSetBstMidEndSepPunct{\mcitedefaultmidpunct}
{\mcitedefaultendpunct}{\mcitedefaultseppunct}\relax
\EndOfBibitem
\bibitem[Zhao \latin{et~al.}(2006)Zhao, Schultz, and Truhlar]{06ZhScTr}
Zhao,~Y.; Schultz,~N.~E.; Truhlar,~D.~G. {Design of density functionals by
  combining the method of constraint satisfaction with parametrization for
  thermochemistry, thermochemical kinetics, and noncovalent interactions}.
  \emph{Journal of Chemical Theory and Computation} \textbf{2006}, \emph{2},
  364--382\relax
\mciteBstWouldAddEndPuncttrue
\mciteSetBstMidEndSepPunct{\mcitedefaultmidpunct}
{\mcitedefaultendpunct}{\mcitedefaultseppunct}\relax
\EndOfBibitem
\bibitem[Peverati and Truhlar(2011)Peverati, and Truhlar]{11PeTr}
Peverati,~R.; Truhlar,~D.~G. {Communication: A global hybrid generalized
  gradient approximation to the exchange-correlation functional that satisfies
  the second-order density-gradient constraint and has broad applicability in
  chemistry}. \emph{Journal of Chemical Physics} \textbf{2011}, \emph{135},
  191102\relax
\mciteBstWouldAddEndPuncttrue
\mciteSetBstMidEndSepPunct{\mcitedefaultmidpunct}
{\mcitedefaultendpunct}{\mcitedefaultseppunct}\relax
\EndOfBibitem
\bibitem[Adamo and Barone(1999)Adamo, and Barone]{99AdBa}
Adamo,~C.; Barone,~V. {Toward reliable density functional methods without
  adjustable parameters: The PBE0 model}. \emph{Journal of Chemical Physics}
  \textbf{1999}, \emph{110}, 6158--6170\relax
\mciteBstWouldAddEndPuncttrue
\mciteSetBstMidEndSepPunct{\mcitedefaultmidpunct}
{\mcitedefaultendpunct}{\mcitedefaultseppunct}\relax
\EndOfBibitem
\bibitem[Ernzerhof and Scuseria(1999)Ernzerhof, and Scuseria]{99ErSc}
Ernzerhof,~M.; Scuseria,~G.~E. {Assessment of the Perdew-Burke-Ernzerhof
  exchange-correlation functional}. \emph{Journal of Chemical Physics}
  \textbf{1999}, \emph{110}, 5029--5036\relax
\mciteBstWouldAddEndPuncttrue
\mciteSetBstMidEndSepPunct{\mcitedefaultmidpunct}
{\mcitedefaultendpunct}{\mcitedefaultseppunct}\relax
\EndOfBibitem
\bibitem[Grimme(2005)]{05Gr}
Grimme,~S. {Accurate calculation of the heats of formation for large main group
  compounds with spin-component scaled MP2 methods}. \emph{Journal of Physical
  Chemistry A} \textbf{2005}, \emph{109}, 3067--3077\relax
\mciteBstWouldAddEndPuncttrue
\mciteSetBstMidEndSepPunct{\mcitedefaultmidpunct}
{\mcitedefaultendpunct}{\mcitedefaultseppunct}\relax
\EndOfBibitem
\bibitem[Chai and Head-Gordon(2008)Chai, and Head-Gordon]{08ChHe}
Chai,~J.~D.; Head-Gordon,~M. {Long-range corrected hybrid density functionals
  with damped atom-atom dispersion corrections}. \emph{Physical Chemistry
  Chemical Physics} \textbf{2008}, \emph{10}, 6615--6620\relax
\mciteBstWouldAddEndPuncttrue
\mciteSetBstMidEndSepPunct{\mcitedefaultmidpunct}
{\mcitedefaultendpunct}{\mcitedefaultseppunct}\relax
\EndOfBibitem
\bibitem[Mardirossian and Head-Gordon(2014)Mardirossian, and
  Head-Gordon]{14MaHe}
Mardirossian,~N.; Head-Gordon,~M. {{$\omega$}b97X-V: A 10-parameter,
  range-separated hybrid, generalized gradient approximation density functional
  with nonlocal correlation, designed by a survival-of-the-fittest strategy}.
  \emph{Physical Chemistry Chemical Physics} \textbf{2014}, \emph{16},
  9904--9924\relax
\mciteBstWouldAddEndPuncttrue
\mciteSetBstMidEndSepPunct{\mcitedefaultmidpunct}
{\mcitedefaultendpunct}{\mcitedefaultseppunct}\relax
\EndOfBibitem
\bibitem[Head-Gordon \latin{et~al.}(1988)Head-Gordon, Pople, and
  Frisch]{88HePoFr}
Head-Gordon,~M.; Pople,~J.~A.; Frisch,~M.~J. {MP2 energy evaluation by direct
  methods}. \emph{Chemical Physics Letters} \textbf{1988}, \emph{153},
  503--506\relax
\mciteBstWouldAddEndPuncttrue
\mciteSetBstMidEndSepPunct{\mcitedefaultmidpunct}
{\mcitedefaultendpunct}{\mcitedefaultseppunct}\relax
\EndOfBibitem
\bibitem[S{\ae}b{\o} and Alml{\"{o}}f(1989)S{\ae}b{\o}, and
  Alml{\"{o}}f]{89SaAl}
S{\ae}b{\o},~S.; Alml{\"{o}}f,~J. {Avoiding the integral storage bottleneck in
  LCAO calculations of electron correlation}. \emph{Chemical Physics Letters}
  \textbf{1989}, \emph{154}, 83--89\relax
\mciteBstWouldAddEndPuncttrue
\mciteSetBstMidEndSepPunct{\mcitedefaultmidpunct}
{\mcitedefaultendpunct}{\mcitedefaultseppunct}\relax
\EndOfBibitem
\bibitem[Frisch \latin{et~al.}(1990)Frisch, Head-Gordon, and Pople]{90FrHePo_1}
Frisch,~M.~J.; Head-Gordon,~M.; Pople,~J.~A. {A direct MP2 gradient method}.
  \emph{Chemical Physics Letters} \textbf{1990}, \emph{166}, 275--280\relax
\mciteBstWouldAddEndPuncttrue
\mciteSetBstMidEndSepPunct{\mcitedefaultmidpunct}
{\mcitedefaultendpunct}{\mcitedefaultseppunct}\relax
\EndOfBibitem
\bibitem[Frisch \latin{et~al.}(1990)Frisch, Head-Gordon, and Pople]{90FrHePo_2}
Frisch,~M.~J.; Head-Gordon,~M.; Pople,~J.~A. {Semi-direct algorithms for the
  MP2 energy and gradient}. \emph{Chemical Physics Letters} \textbf{1990},
  \emph{166}, 281--289\relax
\mciteBstWouldAddEndPuncttrue
\mciteSetBstMidEndSepPunct{\mcitedefaultmidpunct}
{\mcitedefaultendpunct}{\mcitedefaultseppunct}\relax
\EndOfBibitem
\bibitem[Head-Gordon and Head-Gordon(1994)Head-Gordon, and Head-Gordon]{94HeHe}
Head-Gordon,~M.; Head-Gordon,~T. {Analytic MP2 frequencies without fifth-order
  storage. Theory and application to bifurcated hydrogen bonds in the water
  hexamer}. \emph{Chemical Physics Letters} \textbf{1994}, \emph{220},
  122--128\relax
\mciteBstWouldAddEndPuncttrue
\mciteSetBstMidEndSepPunct{\mcitedefaultmidpunct}
{\mcitedefaultendpunct}{\mcitedefaultseppunct}\relax
\EndOfBibitem
\bibitem[Biczysko \latin{et~al.}(2010)Biczysko, Panek, Scalmani, Bloino, and
  Barone]{10BiPaSc}
Biczysko,~M.; Panek,~P.; Scalmani,~G.; Bloino,~J.; Barone,~V. {Harmonic and
  anharmonic vibrational frequency calculations with the double-hybrid B2PLYP
  method: Analytic second derivatives and benchmark studies}. \emph{Journal of
  Chemical Theory and Computation} \textbf{2010}, \emph{6}, 2115--2125\relax
\mciteBstWouldAddEndPuncttrue
\mciteSetBstMidEndSepPunct{\mcitedefaultmidpunct}
{\mcitedefaultendpunct}{\mcitedefaultseppunct}\relax
\EndOfBibitem
\bibitem[Kozuch and Martin(2011)Kozuch, and Martin]{11KoMa}
Kozuch,~S.; Martin,~J.~M. {DSD-PBEP86: In search of the best double-hybrid DFT
  with spin-component scaled MP2 and dispersion corrections}. \emph{Physical
  Chemistry Chemical Physics} \textbf{2011}, \emph{13}, 20104--20107\relax
\mciteBstWouldAddEndPuncttrue
\mciteSetBstMidEndSepPunct{\mcitedefaultmidpunct}
{\mcitedefaultendpunct}{\mcitedefaultseppunct}\relax
\EndOfBibitem
\bibitem[Br{\'{e}}mond and Adamo(2011)Br{\'{e}}mond, and Adamo]{11BrAd}
Br{\'{e}}mond,~E.; Adamo,~C. {Seeking for parameter-free double-hybrid
  functionals: The PBE0-DH model}. \emph{Journal of Chemical Physics}
  \textbf{2011}, \emph{135}, 24106\relax
\mciteBstWouldAddEndPuncttrue
\mciteSetBstMidEndSepPunct{\mcitedefaultmidpunct}
{\mcitedefaultendpunct}{\mcitedefaultseppunct}\relax
\EndOfBibitem
\bibitem[Ditchfield \latin{et~al.}(1971)Ditchfield, Hehre, and Pople]{71DiHePo}
Ditchfield,~R.; Hehre,~W.~J.; Pople,~J.~A. {Self‐Consistent
  Molecular‐Orbital Methods. IX. An Extended Gaussian‐Type Basis for
  Molecular‐Orbital Studies of Organic Molecules}. \emph{The Journal of
  Chemical Physics} \textbf{1971}, \emph{54}, 724--728\relax
\mciteBstWouldAddEndPuncttrue
\mciteSetBstMidEndSepPunct{\mcitedefaultmidpunct}
{\mcitedefaultendpunct}{\mcitedefaultseppunct}\relax
\EndOfBibitem
\bibitem[Hehre \latin{et~al.}(1972)Hehre, Ditchfield, and Pople]{72HeDiPo}
Hehre,~W.~J.; Ditchfield,~K.; Pople,~J.~A. {Self-consistent molecular orbital
  methods. XII. Further extensions of gaussian-type basis sets for use in
  molecular orbital studies of organic molecules}. \emph{The Journal of
  Chemical Physics} \textbf{1972}, \emph{56}, 2257--2261\relax
\mciteBstWouldAddEndPuncttrue
\mciteSetBstMidEndSepPunct{\mcitedefaultmidpunct}
{\mcitedefaultendpunct}{\mcitedefaultseppunct}\relax
\EndOfBibitem
\bibitem[Dill and Pople(1975)Dill, and Pople]{75DiPo}
Dill,~J.~D.; Pople,~J.~A. {Self‐consistent molecular orbital methods. XV.
  Extended Gaussian‐type basis sets for lithium, beryllium, and boron}.
  \emph{The Journal of Chemical Physics} \textbf{1975}, \emph{62},
  2921--2923\relax
\mciteBstWouldAddEndPuncttrue
\mciteSetBstMidEndSepPunct{\mcitedefaultmidpunct}
{\mcitedefaultendpunct}{\mcitedefaultseppunct}\relax
\EndOfBibitem
\bibitem[Binkley and Pople(1977)Binkley, and Pople]{77BiPo}
Binkley,~J.~S.; Pople,~J.~A. {Self‐consistent molecular orbital methods. XIX.
  Split‐valence Gaussian‐type basis sets for beryllium}. \emph{The Journal
  of Chemical Physics} \textbf{1977}, \emph{66}, 879--880\relax
\mciteBstWouldAddEndPuncttrue
\mciteSetBstMidEndSepPunct{\mcitedefaultmidpunct}
{\mcitedefaultendpunct}{\mcitedefaultseppunct}\relax
\EndOfBibitem
\bibitem[Francl \latin{et~al.}(1982)Francl, Pietro, Hehre, Binkley, Gordon,
  DeFrees, and Pople]{82FrPiHe}
Francl,~M.~M.; Pietro,~W.~J.; Hehre,~W.~J.; Binkley,~J.~S.; Gordon,~M.~S.;
  DeFrees,~D.~J.; Pople,~J.~A. {Self-consistent molecular orbital methods.
  XXIII. A polarization-type basis set for second-row elements}. \emph{The
  Journal of Chemical Physics} \textbf{1982}, \emph{77}, 3654--3665\relax
\mciteBstWouldAddEndPuncttrue
\mciteSetBstMidEndSepPunct{\mcitedefaultmidpunct}
{\mcitedefaultendpunct}{\mcitedefaultseppunct}\relax
\EndOfBibitem
\bibitem[Clark \latin{et~al.}(1983)Clark, Chandrasekhar, Spitznagel, and
  Schleyer]{83ClChSp}
Clark,~T.; Chandrasekhar,~J.; Spitznagel,~G.~W.; Schleyer,~P. V.~R. {Efficient
  diffuse function‐augmented basis sets for anion calculations. III. The
  3‐21+G basis set for first‐row elements, Li–F}. \emph{Journal of
  Computational Chemistry} \textbf{1983}, \emph{4}, 294--301\relax
\mciteBstWouldAddEndPuncttrue
\mciteSetBstMidEndSepPunct{\mcitedefaultmidpunct}
{\mcitedefaultendpunct}{\mcitedefaultseppunct}\relax
\EndOfBibitem
\bibitem[Spitznagel \latin{et~al.}(1987)Spitznagel, Clark, von
  Ragu{\'{e}}~Schleyer, and Hehre]{87SpClRa}
Spitznagel,~G.~W.; Clark,~T.; von Ragu{\'{e}}~Schleyer,~P.; Hehre,~W.~J. {An
  evaluation of the performance of diffuse function‐augmented basis sets for
  second row elements, Na‐Cl}. \emph{Journal of Computational Chemistry}
  \textbf{1987}, \emph{8}, 1109--1116\relax
\mciteBstWouldAddEndPuncttrue
\mciteSetBstMidEndSepPunct{\mcitedefaultmidpunct}
{\mcitedefaultendpunct}{\mcitedefaultseppunct}\relax
\EndOfBibitem
\bibitem[Gill \latin{et~al.}(1992)Gill, Johnson, Pople, and Frisch]{92GiJoPo}
Gill,~P.~M.; Johnson,~B.~G.; Pople,~J.~A.; Frisch,~M.~J. {The performance of
  the Becke-Lee-Yang-Parr (B-LYP) density functional theory with various basis
  sets}. \emph{Chemical Physics Letters} \textbf{1992}, \emph{197},
  499--505\relax
\mciteBstWouldAddEndPuncttrue
\mciteSetBstMidEndSepPunct{\mcitedefaultmidpunct}
{\mcitedefaultendpunct}{\mcitedefaultseppunct}\relax
\EndOfBibitem
\bibitem[Hariharan and Pople(1973)Hariharan, and Pople]{73HaPo}
Hariharan,~P.~C.; Pople,~J.~A. {The influence of polarization functions on
  molecular orbital hydrogenation energies}. \emph{Theoretica Chimica Acta}
  \textbf{1973}, \emph{28}, 213--222\relax
\mciteBstWouldAddEndPuncttrue
\mciteSetBstMidEndSepPunct{\mcitedefaultmidpunct}
{\mcitedefaultendpunct}{\mcitedefaultseppunct}\relax
\EndOfBibitem
\bibitem[Krishnan \latin{et~al.}(1980)Krishnan, Binkley, Seeger, and
  Pople]{80KrBiSe}
Krishnan,~R.; Binkley,~J.~S.; Seeger,~R.; Pople,~J.~A. {Self-consistent
  molecular orbital methods. XX. A basis set for correlated wave functions}.
  \emph{The Journal of Chemical Physics} \textbf{1980}, \emph{72},
  650--654\relax
\mciteBstWouldAddEndPuncttrue
\mciteSetBstMidEndSepPunct{\mcitedefaultmidpunct}
{\mcitedefaultendpunct}{\mcitedefaultseppunct}\relax
\EndOfBibitem
\bibitem[McLean and Chandler(1980)McLean, and Chandler]{80McCh}
McLean,~A.~D.; Chandler,~G.~S. {Contracted Gaussian basis sets for molecular
  calculations. I. Second row atoms, Z=11-18}. \emph{The Journal of Chemical
  Physics} \textbf{1980}, \emph{72}, 5639--5648\relax
\mciteBstWouldAddEndPuncttrue
\mciteSetBstMidEndSepPunct{\mcitedefaultmidpunct}
{\mcitedefaultendpunct}{\mcitedefaultseppunct}\relax
\EndOfBibitem
\bibitem[Weigend and Ahlrichs(2005)Weigend, and Ahlrichs]{05WeAh}
Weigend,~F.; Ahlrichs,~R. {Balanced basis sets of split valence, triple zeta
  valence and quadruple zeta valence quality for H to Rn: Design and assessment
  of accuracy}. \emph{Physical Chemistry Chemical Physics} \textbf{2005},
  \emph{7}, 3297--3305\relax
\mciteBstWouldAddEndPuncttrue
\mciteSetBstMidEndSepPunct{\mcitedefaultmidpunct}
{\mcitedefaultendpunct}{\mcitedefaultseppunct}\relax
\EndOfBibitem
\bibitem[Rappoport and Furche(2010)Rappoport, and Furche]{10RaFu}
Rappoport,~D.; Furche,~F. {Property-optimized Gaussian basis sets for molecular
  response calculations}. \emph{Journal of Chemical Physics} \textbf{2010},
  \emph{133}, 134105\relax
\mciteBstWouldAddEndPuncttrue
\mciteSetBstMidEndSepPunct{\mcitedefaultmidpunct}
{\mcitedefaultendpunct}{\mcitedefaultseppunct}\relax
\EndOfBibitem
\bibitem[Dunning(1989)]{89Du}
Dunning,~T.~H. {Gaussian basis sets for use in correlated molecular
  calculations. I. The atoms boron through neon and hydrogen}. \emph{The
  Journal of Chemical Physics} \textbf{1989}, \emph{90}, 1007--1023\relax
\mciteBstWouldAddEndPuncttrue
\mciteSetBstMidEndSepPunct{\mcitedefaultmidpunct}
{\mcitedefaultendpunct}{\mcitedefaultseppunct}\relax
\EndOfBibitem
\bibitem[Woon and Dunning(1993)Woon, and Dunning]{93WoDu}
Woon,~D.~E.; Dunning,~T.~H. {Gaussian basis sets for use in correlated
  molecular calculations. III. The atoms aluminum through argon}. \emph{The
  Journal of Chemical Physics} \textbf{1993}, \emph{98}, 1358--1371\relax
\mciteBstWouldAddEndPuncttrue
\mciteSetBstMidEndSepPunct{\mcitedefaultmidpunct}
{\mcitedefaultendpunct}{\mcitedefaultseppunct}\relax
\EndOfBibitem
\bibitem[Woon and Dunning(1994)Woon, and Dunning]{94WoDu}
Woon,~D.~E.; Dunning,~T.~H. {Gaussian basis sets for use in correlated
  molecular calculations. IV. Calculation of static electrical response
  properties}. \emph{The Journal of Chemical Physics} \textbf{1994},
  \emph{100}, 2975--2988\relax
\mciteBstWouldAddEndPuncttrue
\mciteSetBstMidEndSepPunct{\mcitedefaultmidpunct}
{\mcitedefaultendpunct}{\mcitedefaultseppunct}\relax
\EndOfBibitem
\bibitem[Prascher \latin{et~al.}(2011)Prascher, Woon, Peterson, Dunning, and
  Wilson]{11PrWoPe}
Prascher,~B.~P.; Woon,~D.~E.; Peterson,~K.~A.; Dunning,~T.~H.; Wilson,~A.~K.
  {Gaussian basis sets for use in correlated molecular calculations. VII.
  Valence, core-valence, and scalar relativistic basis sets for Li, Be, Na, and
  Mg}. \emph{Theoretical Chemistry Accounts} \textbf{2011}, \emph{128},
  69--82\relax
\mciteBstWouldAddEndPuncttrue
\mciteSetBstMidEndSepPunct{\mcitedefaultmidpunct}
{\mcitedefaultendpunct}{\mcitedefaultseppunct}\relax
\EndOfBibitem
\bibitem[Kendall \latin{et~al.}(1992)Kendall, Dunning, and Harrison]{92KeDuHa}
Kendall,~R.~A.; Dunning,~T.~H.; Harrison,~R.~J. {Electron affinities of the
  first-row atoms revisited. Systematic basis sets and wave functions}.
  \emph{The Journal of Chemical Physics} \textbf{1992}, \emph{96},
  6796--6806\relax
\mciteBstWouldAddEndPuncttrue
\mciteSetBstMidEndSepPunct{\mcitedefaultmidpunct}
{\mcitedefaultendpunct}{\mcitedefaultseppunct}\relax
\EndOfBibitem
\bibitem[Jensen(2001)]{01Je}
Jensen,~F. {Polarization consistent basis sets: Principles}. \emph{The Journal
  of Chemical Physics} \textbf{2001}, \emph{115}, 9113--9125\relax
\mciteBstWouldAddEndPuncttrue
\mciteSetBstMidEndSepPunct{\mcitedefaultmidpunct}
{\mcitedefaultendpunct}{\mcitedefaultseppunct}\relax
\EndOfBibitem
\bibitem[Jensen(2002)]{02Je}
Jensen,~F. {Polarization consistent basis sets: II. Estimating the Kohn-Sham
  basis set limit}. \emph{Journal of Chemical Physics} \textbf{2002},
  \emph{116}, 7372--7379\relax
\mciteBstWouldAddEndPuncttrue
\mciteSetBstMidEndSepPunct{\mcitedefaultmidpunct}
{\mcitedefaultendpunct}{\mcitedefaultseppunct}\relax
\EndOfBibitem
\bibitem[Jensen and Helgaker(2004)Jensen, and Helgaker]{04JeHe}
Jensen,~F.; Helgaker,~T. {Polarization consistent basis sets. V. The elements
  Si–Cl}. \emph{The Journal of Chemical Physics} \textbf{2004}, \emph{121},
  3463--3470\relax
\mciteBstWouldAddEndPuncttrue
\mciteSetBstMidEndSepPunct{\mcitedefaultmidpunct}
{\mcitedefaultendpunct}{\mcitedefaultseppunct}\relax
\EndOfBibitem
\bibitem[Jensen(2007)]{07Je}
Jensen,~F. {Polarization consistent basis sets. 4: The elements He, Li, Be, B,
  Ne, Na, Mg, Al, and Ar}. \emph{Journal of Physical Chemistry A}
  \textbf{2007}, \emph{111}, 11198--11204\relax
\mciteBstWouldAddEndPuncttrue
\mciteSetBstMidEndSepPunct{\mcitedefaultmidpunct}
{\mcitedefaultendpunct}{\mcitedefaultseppunct}\relax
\EndOfBibitem
\bibitem[Jensen(2002)]{02Je_diffuse}
Jensen,~F. {Polarization consistent basis sets. III. The importance of diffuse
  functions}. \emph{Journal of Chemical Physics} \textbf{2002}, \emph{117},
  9234--9240\relax
\mciteBstWouldAddEndPuncttrue
\mciteSetBstMidEndSepPunct{\mcitedefaultmidpunct}
{\mcitedefaultendpunct}{\mcitedefaultseppunct}\relax
\EndOfBibitem
\bibitem[Jensen(2014)]{14Je}
Jensen,~F. {Unifying General and Segmented Contracted Basis Sets. Segmented
  Polarization Consistent Basis Sets}. \emph{Journal of Chemical Theory and
  Computation} \textbf{2014}, \emph{10}, 1074--1085\relax
\mciteBstWouldAddEndPuncttrue
\mciteSetBstMidEndSepPunct{\mcitedefaultmidpunct}
{\mcitedefaultendpunct}{\mcitedefaultseppunct}\relax
\EndOfBibitem
\bibitem[Barone \latin{et~al.}(2013)Barone, Biczysko, Bloino, Egidi, and
  Puzzarini]{13BaBiBl}
Barone,~V.; Biczysko,~M.; Bloino,~J.; Egidi,~F.; Puzzarini,~C. {Accurate
  structure, thermodynamics, and spectroscopy of medium-sized radicals by
  hybrid coupled cluster/density functional theory approaches: The case of
  phenyl radical}. \emph{Journal of Chemical Physics} \textbf{2013},
  \emph{138}, 234303\relax
\mciteBstWouldAddEndPuncttrue
\mciteSetBstMidEndSepPunct{\mcitedefaultmidpunct}
{\mcitedefaultendpunct}{\mcitedefaultseppunct}\relax
\EndOfBibitem
\bibitem[Landrum(2010)]{10La}
Landrum,~G. {RDKit: Open-source cheminformatics}. 2010;
  \url{https://www.rdkit.org/}\relax
\mciteBstWouldAddEndPuncttrue
\mciteSetBstMidEndSepPunct{\mcitedefaultmidpunct}
{\mcitedefaultendpunct}{\mcitedefaultseppunct}\relax
\EndOfBibitem
\bibitem[Weser(2017)]{17We}
Weser,~O. {An efficient and general library for the definition and use of
  internal coordinates in large molecular systems}. Ph.D.\ thesis, Georg August
  Universit{\"{a}}t G{\"{o}}ttingen, 2017\relax
\mciteBstWouldAddEndPuncttrue
\mciteSetBstMidEndSepPunct{\mcitedefaultmidpunct}
{\mcitedefaultendpunct}{\mcitedefaultseppunct}\relax
\EndOfBibitem
\bibitem[Frisch \latin{et~al.}(2016)Frisch, Trucks, Schlegel, Scuseria, Robb,
  Cheeseman, Scalmani, Barone, Petersson, Nakatsuji, Li, Caricato, Marenich,
  Bloino, Janesko, Gomperts, Mennucci, Hratchian, Ortiz, Izmaylov, Sonnenberg,
  Williams-Young, Ding, Lipparini, Egidi, Goings, Peng, Petrone, Henderson,
  Ranasinghe, Zakrzewski, Gao, Rega, Zheng, Liang, Hada, Ehara, Toyota, Fukuda,
  Hasegawa, Ishida, Nakajima, Honda, Kitao, Nakai, Vreven, Throssell,
  Montgomery, Peralta, Ogliaro, Bearpark, Heyd, Brothers, Kudin, Staroverov,
  Keith, Kobayashi, Normand, Raghavachari, Rendell, Burant, Iyengar, Tomasi,
  Cossi, Millam, Klene, Adamo, Cammi, Ochterski, Martin, Morokuma, Farkas,
  Foresman, and Fox]{g16}
Frisch,~M.~J. \latin{et~al.}  {Gaussian16 Revision C.01}. 2016\relax
\mciteBstWouldAddEndPuncttrue
\mciteSetBstMidEndSepPunct{\mcitedefaultmidpunct}
{\mcitedefaultendpunct}{\mcitedefaultseppunct}\relax
\EndOfBibitem
\bibitem[Shao \latin{et~al.}(2015)Shao, Gan, Epifanovsky, Gilbert, Wormit,
  Kussmann, Lange, Behn, Deng, Feng, Ghosh, Goldey, Horn, Jacobson, Kaliman,
  Khaliullin, Kus̈, Landau, Liu, Proynov, Rhee, Richard, Rohrdanz, Steele,
  Sundstrom, Woodcock, Zimmerman, Zuev, Albrecht, Alguire, Austin, Beran,
  Bernard, Berquist, Brandhorst, Bravaya, Brown, Casanova, Chang, Chen, Chien,
  Closser, Crittenden, Diedenhofen, Distasio, Do, Dutoi, Edgar, Fatehi,
  Fusti-Molnar, Ghysels, Golubeva-Zadorozhnaya, Gomes, Hanson-Heine, Harbach,
  Hauser, Hohenstein, Holden, Jagau, Ji, Kaduk, Khistyaev, Kim, Kim, King,
  Klunzinger, Kosenkov, Kowalczyk, Krauter, Lao, Laurent, Lawler, Levchenko,
  Lin, Liu, Livshits, Lochan, Luenser, Manohar, Manzer, Mao, Mardirossian,
  Marenich, Maurer, Mayhall, Neuscamman, Oana, Olivares-Amaya, Oneill,
  Parkhill, Perrine, Peverati, Prociuk, Rehn, Rosta, Russ, Sharada, Sharma,
  Small, Sodt, Stein, St{\"{u}}ck, Su, Thom, Tsuchimochi, Vanovschi, Vogt,
  Vydrov, Wang, Watson, Wenzel, White, Williams, Yang, Yeganeh, Yost, You,
  Zhang, Zhang, Zhao, Brooks, Chan, Chipman, Cramer, Goddard, Gordon, Hehre,
  Klamt, Schaefer, Schmidt, Sherrill, Truhlar, Warshel, Xu, Aspuru-Guzik, Baer,
  Bell, Besley, Chai, Dreuw, Dunietz, Furlani, Gwaltney, Hsu, Jung, Kong,
  Lambrecht, Liang, Ochsenfeld, Rassolov, Slipchenko, Subotnik, Van~Voorhis,
  Herbert, Krylov, Gill, and Head-Gordon]{15ShGaEp}
Shao,~Y. \latin{et~al.}  {Advances in molecular quantum chemistry contained in
  the Q-Chem 4 program package}. \emph{Molecular Physics} \textbf{2015},
  \emph{113}, 184--215\relax
\mciteBstWouldAddEndPuncttrue
\mciteSetBstMidEndSepPunct{\mcitedefaultmidpunct}
{\mcitedefaultendpunct}{\mcitedefaultseppunct}\relax
\EndOfBibitem
\bibitem[Halls \latin{et~al.}(2001)Halls, Velkovski, and Schlegel]{01HaVeSc}
Halls,~M.~D.; Velkovski,~J.; Schlegel,~H.~B. {Harmonic frequency scaling
  factors for Hartree-Fock, S-VWN, B-LYP, B3-LYP, B3-PW91 and MP2 with the
  Sadlej pVTZ electric property basis set}. \emph{Theoretical Chemistry
  Accounts} \textbf{2001}, \emph{105}, 413--421\relax
\mciteBstWouldAddEndPuncttrue
\mciteSetBstMidEndSepPunct{\mcitedefaultmidpunct}
{\mcitedefaultendpunct}{\mcitedefaultseppunct}\relax
\EndOfBibitem
\bibitem[Sinha \latin{et~al.}(2004)Sinha, Boesch, Gu, Wheeler, and
  Wilson]{04SiBoGu}
Sinha,~P.; Boesch,~S.~E.; Gu,~C.; Wheeler,~R.~A.; Wilson,~A.~K. {Harmonic
  vibrational frequencies: Scaling factors for HF, B3LYP, and MP2 methods in
  combination with correlation consistent basis sets}. \emph{Journal of
  Physical Chemistry A} \textbf{2004}, \emph{108}, 9213--9217\relax
\mciteBstWouldAddEndPuncttrue
\mciteSetBstMidEndSepPunct{\mcitedefaultmidpunct}
{\mcitedefaultendpunct}{\mcitedefaultseppunct}\relax
\EndOfBibitem
\bibitem[Andersson and Uvdal(2005)Andersson, and Uvdal]{05AnUv}
Andersson,~M.~P.; Uvdal,~P. {New scale factors for harmonic vibrational
  frequencies using the B3LYP density functional method with the
  triple-{$\zeta$} basis Set 6-311+G(d,p)}. \emph{Journal of Physical Chemistry
  A} \textbf{2005}, \emph{109}, 2937--2941\relax
\mciteBstWouldAddEndPuncttrue
\mciteSetBstMidEndSepPunct{\mcitedefaultmidpunct}
{\mcitedefaultendpunct}{\mcitedefaultseppunct}\relax
\EndOfBibitem
\bibitem[Merrick \latin{et~al.}(2007)Merrick, Moran, and Radom]{07MeMoRa}
Merrick,~J.~P.; Moran,~D.; Radom,~L. {An evaluation of harmonic vibrational
  frequency scale factors}. \emph{Journal of Physical Chemistry A}
  \textbf{2007}, \emph{111}, 11683--11700\relax
\mciteBstWouldAddEndPuncttrue
\mciteSetBstMidEndSepPunct{\mcitedefaultmidpunct}
{\mcitedefaultendpunct}{\mcitedefaultseppunct}\relax
\EndOfBibitem
\bibitem[Andrade \latin{et~al.}(2008)Andrade, Gon{\c{c}}alves, and
  Jorge]{08AnGoJo}
Andrade,~S.~G.; Gon{\c{c}}alves,~L.~C.; Jorge,~F.~E. {Scaling factors for
  fundamental vibrational frequencies and zero-point energies obtained from HF,
  MP2, and DFT/DZP and TZP harmonic frequencies}. \emph{Journal of Molecular
  Structure: THEOCHEM} \textbf{2008}, \emph{864}, 20--25\relax
\mciteBstWouldAddEndPuncttrue
\mciteSetBstMidEndSepPunct{\mcitedefaultmidpunct}
{\mcitedefaultendpunct}{\mcitedefaultseppunct}\relax
\EndOfBibitem
\bibitem[Laury \latin{et~al.}(2011)Laury, Boesch, Haken, Sinha, Wheeler, and
  Wilson]{11LaBoHa}
Laury,~M.~L.; Boesch,~S.~E.; Haken,~I.; Sinha,~P.; Wheeler,~R.~A.;
  Wilson,~A.~K. {Harmonic Vibrational Frequencies: Scale Factors for Pure,
  Hybrid, Hybrid Meta, and Double-Hybrid Functionals in Conjunction with
  Correlation Consistent Basis Sets}. \emph{Journal of computational chemistry}
  \textbf{2011}, \emph{32}, 2339--2347\relax
\mciteBstWouldAddEndPuncttrue
\mciteSetBstMidEndSepPunct{\mcitedefaultmidpunct}
{\mcitedefaultendpunct}{\mcitedefaultseppunct}\relax
\EndOfBibitem
\bibitem[Laury \latin{et~al.}(2012)Laury, Carlson, and Wilson]{12LaCaWi}
Laury,~M.~L.; Carlson,~M.~J.; Wilson,~A.~K. {Vibrational frequency scale
  factors for density functional theory and the polarization consistent basis
  sets}. \emph{Journal of Computational Chemistry} \textbf{2012}, \emph{33},
  2380--2387\relax
\mciteBstWouldAddEndPuncttrue
\mciteSetBstMidEndSepPunct{\mcitedefaultmidpunct}
{\mcitedefaultendpunct}{\mcitedefaultseppunct}\relax
\EndOfBibitem
\bibitem[Chan and Radom(2016)Chan, and Radom]{16ChRa}
Chan,~B.; Radom,~L. {Frequency Scale Factors for Some Double-Hybrid Density
  Functional Theory Procedures: Accurate Thermochemical Components for
  High-Level Composite Protocols}. \emph{Journal of Chemical Theory and
  Computation} \textbf{2016}, \emph{12}, 3774--3780\relax
\mciteBstWouldAddEndPuncttrue
\mciteSetBstMidEndSepPunct{\mcitedefaultmidpunct}
{\mcitedefaultendpunct}{\mcitedefaultseppunct}\relax
\EndOfBibitem
\bibitem[Chan(2017)]{17Cha}
Chan,~B. {Use of Low-Cost Quantum Chemistry Procedures for Geometry
  Optimization and Vibrational Frequency Calculations: Determination of
  Frequency Scale Factors and Application to Reactions of Large Systems}.
  \emph{Journal of Chemical Theory and Computation} \textbf{2017}, \emph{13},
  6052--6060\relax
\mciteBstWouldAddEndPuncttrue
\mciteSetBstMidEndSepPunct{\mcitedefaultmidpunct}
{\mcitedefaultendpunct}{\mcitedefaultseppunct}\relax
\EndOfBibitem
\bibitem[Pople(1965)]{65Po}
Pople,~J.~A. {Two-dimensional chart of quantum chemistry}. \emph{The Journal of
  Chemical Physics} \textbf{1965}, \emph{43}, 229--230\relax
\mciteBstWouldAddEndPuncttrue
\mciteSetBstMidEndSepPunct{\mcitedefaultmidpunct}
{\mcitedefaultendpunct}{\mcitedefaultseppunct}\relax
\EndOfBibitem
\bibitem[Grimme(2011)]{11Gr}
Grimme,~S. {Density functional theory with London dispersion corrections}.
  \emph{Wiley Interdisciplinary Reviews: Computational Molecular Science}
  \textbf{2011}, \emph{1}, 211--228\relax
\mciteBstWouldAddEndPuncttrue
\mciteSetBstMidEndSepPunct{\mcitedefaultmidpunct}
{\mcitedefaultendpunct}{\mcitedefaultseppunct}\relax
\EndOfBibitem
\bibitem[Grimme \latin{et~al.}(2016)Grimme, Hansen, Brandenburg, and
  Bannwarth]{16GrHaBr}
Grimme,~S.; Hansen,~A.; Brandenburg,~J.~G.; Bannwarth,~C. {Dispersion-Corrected
  Mean-Field Electronic Structure Methods}. \emph{Chemical Reviews}
  \textbf{2016}, \emph{116}, 5105--5154\relax
\mciteBstWouldAddEndPuncttrue
\mciteSetBstMidEndSepPunct{\mcitedefaultmidpunct}
{\mcitedefaultendpunct}{\mcitedefaultseppunct}\relax
\EndOfBibitem
\bibitem[Goerigk(2017)]{17Go}
Goerigk,~L. \emph{Non-Covalent Interactions in Quantum Chemistry and Physics:
  Theory and Applications}; Elsevier, 2017; pp 195--219\relax
\mciteBstWouldAddEndPuncttrue
\mciteSetBstMidEndSepPunct{\mcitedefaultmidpunct}
{\mcitedefaultendpunct}{\mcitedefaultseppunct}\relax
\EndOfBibitem
\bibitem[Sousa-Silva \latin{et~al.}(2019)Sousa-Silva, Petkowski, and
  Seager]{19SoPeSe}
Sousa-Silva,~C.; Petkowski,~J.~J.; Seager,~S. {Molecular Simulations for the
  Spectroscopic Detection of Atmospheric Gases}. \emph{Physical} \textbf{2019},
  \emph{21}, 18970--18987\relax
\mciteBstWouldAddEndPuncttrue
\mciteSetBstMidEndSepPunct{\mcitedefaultmidpunct}
{\mcitedefaultendpunct}{\mcitedefaultseppunct}\relax
\EndOfBibitem
\bibitem[Zapata~Trujillo \latin{et~al.}(2021)Zapata~Trujillo, Syme, Rowell,
  Burns, Clark, Gorman, Jacob, Kapodistrias, Kedziora, Lempriere, Medcraft,
  O'Sullivan, Robertson, Soares, Steller, Teece, Tremblay, Sousa-Silva, and
  McKemmish]{21ZaSyRo}
Zapata~Trujillo,~J.~C. \latin{et~al.}  {Computational Infrared Spectroscopy of
  958 Phosphorus-Bearing Molecules}. \emph{Frontiers in Astronomy and Space
  Sciences} \textbf{2021}, \emph{8}, 639068\relax
\mciteBstWouldAddEndPuncttrue
\mciteSetBstMidEndSepPunct{\mcitedefaultmidpunct}
{\mcitedefaultendpunct}{\mcitedefaultseppunct}\relax
\EndOfBibitem
\bibitem[Kruse \latin{et~al.}(2012)Kruse, Goerigk, and Grimme]{12KrGoGr}
Kruse,~H.; Goerigk,~L.; Grimme,~S. {Why the standard B3LYP/6-31G* model
  chemistry should not be used in DFT calculations of molecular
  thermochemistry: Understanding and correcting the problem}. \emph{Journal of
  Organic Chemistry} \textbf{2012}, \emph{77}, 10824--10834\relax
\mciteBstWouldAddEndPuncttrue
\mciteSetBstMidEndSepPunct{\mcitedefaultmidpunct}
{\mcitedefaultendpunct}{\mcitedefaultseppunct}\relax
\EndOfBibitem
\bibitem[Spicher and Grimme(2020)Spicher, and Grimme]{20SpGr}
Spicher,~S.; Grimme,~S. {Efficient Computation of Free Energy Contributions for
  Association Reactions of Large Molecules}. \emph{Journal of Physical
  Chemistry Letters} \textbf{2020}, \emph{11}, 6606--6611\relax
\mciteBstWouldAddEndPuncttrue
\mciteSetBstMidEndSepPunct{\mcitedefaultmidpunct}
{\mcitedefaultendpunct}{\mcitedefaultseppunct}\relax
\EndOfBibitem
\bibitem[Pracht \latin{et~al.}(2020)Pracht, Grant, and Grimme]{20PrGrGr}
Pracht,~P.; Grant,~D.~F.; Grimme,~S. {Comprehensive Assessment of GFN
  Tight-Binding and Composite Density Functional Theory Methods for Calculating
  Gas-Phase Infrared Spectra}. \emph{Journal of Chemical Theory and
  Computation} \textbf{2020}, \emph{16}, 7044--7060\relax
\mciteBstWouldAddEndPuncttrue
\mciteSetBstMidEndSepPunct{\mcitedefaultmidpunct}
{\mcitedefaultendpunct}{\mcitedefaultseppunct}\relax
\EndOfBibitem
\bibitem[Karton(2022)]{22Ka}
Karton,~A. \emph{Annual Reports in Computational Chemistry}; Elsevier, 2022;
  Vol.~18; pp 123--166\relax
\mciteBstWouldAddEndPuncttrue
\mciteSetBstMidEndSepPunct{\mcitedefaultmidpunct}
{\mcitedefaultendpunct}{\mcitedefaultseppunct}\relax
\EndOfBibitem
\bibitem[Bowman(1986)]{86Bo}
Bowman,~J.~M. {The Self-Consistent-Field Approach to Polyatomic Vibrations}.
  \emph{Accounts of Chemical Research} \textbf{1986}, \emph{19}, 202--208\relax
\mciteBstWouldAddEndPuncttrue
\mciteSetBstMidEndSepPunct{\mcitedefaultmidpunct}
{\mcitedefaultendpunct}{\mcitedefaultseppunct}\relax
\EndOfBibitem
\bibitem[Jung and Gerber(1996)Jung, and Gerber]{96JuGe}
Jung,~J.~O.; Gerber,~R.~B. {Vibrational wave functions and spectroscopy of
  (H2O)n, n=2,3,4,5: Vibrational self-consistent field with correlation
  corrections}. \emph{Journal of Chemical Physics} \textbf{1996}, \emph{105},
  10332--10348\relax
\mciteBstWouldAddEndPuncttrue
\mciteSetBstMidEndSepPunct{\mcitedefaultmidpunct}
{\mcitedefaultendpunct}{\mcitedefaultseppunct}\relax
\EndOfBibitem
\bibitem[Bowman \latin{et~al.}(2008)Bowman, Carrington, and Meyer]{08BoCaMe}
Bowman,~J.~M.; Carrington,~T.; Meyer,~H.-D. {Variational quantum approaches for
  computing vibrational energies of polyatomic molecules}. \emph{Molecular
  Physics} \textbf{2008}, \emph{106}, 2145--2182\relax
\mciteBstWouldAddEndPuncttrue
\mciteSetBstMidEndSepPunct{\mcitedefaultmidpunct}
{\mcitedefaultendpunct}{\mcitedefaultseppunct}\relax
\EndOfBibitem
\bibitem[Scribano \latin{et~al.}(2010)Scribano, Lauvergnat, and
  Benoit]{10ScLaBe}
Scribano,~Y.; Lauvergnat,~D.~M.; Benoit,~D.~M. {Fast vibrational configuration
  interaction using generalized curvilinear coordinates and self-consistent
  basis}. \emph{Journal of Chemical Physics} \textbf{2010}, \emph{133},
  1--13\relax
\mciteBstWouldAddEndPuncttrue
\mciteSetBstMidEndSepPunct{\mcitedefaultmidpunct}
{\mcitedefaultendpunct}{\mcitedefaultseppunct}\relax
\EndOfBibitem
\bibitem[Barone \latin{et~al.}(2014)Barone, Biczysko, and Bloino]{14BaBiBl}
Barone,~V.; Biczysko,~M.; Bloino,~J. {Fully anharmonic IR and Raman spectra of
  medium-size molecular systems: Accuracy and interpretation}. \emph{Physical
  Chemistry Chemical Physics} \textbf{2014}, \emph{16}, 1759--1787\relax
\mciteBstWouldAddEndPuncttrue
\mciteSetBstMidEndSepPunct{\mcitedefaultmidpunct}
{\mcitedefaultendpunct}{\mcitedefaultseppunct}\relax
\EndOfBibitem
\bibitem[Barone \latin{et~al.}(2015)Barone, Biczysko, and Puzzarini]{15BaBiPu}
Barone,~V.; Biczysko,~M.; Puzzarini,~C. {Quantum chemistry meets spectroscopy
  for astrochemistry: Increasing complexity toward prebiotic molecules}.
  \emph{Accounts of Chemical Research} \textbf{2015}, \emph{48},
  1413--1422\relax
\mciteBstWouldAddEndPuncttrue
\mciteSetBstMidEndSepPunct{\mcitedefaultmidpunct}
{\mcitedefaultendpunct}{\mcitedefaultseppunct}\relax
\EndOfBibitem
\bibitem[Biczysko \latin{et~al.}(2018)Biczysko, Bloino, and
  Puzzarini]{18BiBlPu}
Biczysko,~M.; Bloino,~J.; Puzzarini,~C. {Computational challenges in
  Astrochemistry}. \emph{Wiley Interdisciplinary Reviews: Computational
  Molecular Science} \textbf{2018}, \emph{8}, 1349\relax
\mciteBstWouldAddEndPuncttrue
\mciteSetBstMidEndSepPunct{\mcitedefaultmidpunct}
{\mcitedefaultendpunct}{\mcitedefaultseppunct}\relax
\EndOfBibitem
\end{mcitethebibliography}

\end{document}